\documentclass[preprint,3p,12pt]{elsarticle}
\usepackage{mathrsfs}
\usepackage{amsmath}
\usepackage{stmaryrd}
\usepackage{bbding}
\usepackage{dcolumn}
\usepackage{graphicx}
\usepackage{amsfonts}
\usepackage{amssymb}
\usepackage{psfrag}
\usepackage{wrapfig}
\usepackage{subfigure}
\usepackage{makeidx}
\usepackage{bm}
\usepackage{epsf}
\usepackage{epsfig}
\usepackage{setspace}
\usepackage{graphicx}
\usepackage{epstopdf}
\usepackage{psfrag}
\usepackage{subfigure}
\usepackage{color}
\usepackage{enumerate}

\usepackage{tikz}
\usetikzlibrary{positioning,calc}
\usetikzlibrary{arrows,shapes}
\tikzstyle{block} = [rectangle,rounded corners,thin,align=center,fill=green!20,draw=black!20]
\tikzstyle{line} = [-latex]

\begin{document}

\title{A velocity-space adaptive unified gas kinetic scheme for continuum and rarefied flows}

\author[PKU]{Tianbai Xiao}
\ead{xiaotianbai@pku.edu.cn}

\author[HKUST,Shenzhen]{Kun Xu\corref{cor1}}
\ead{makxu@ust.hk}

\author[PKU]{Qingdong Cai}
\ead{caiqd@pku.edu.cn}

\address[PKU]{Department of Mechanics and Engineering Science, College of Engineering, Peking University, Beijing 100871, China}
\address[HKUST]{Department of Mathematics, Department of Mechanical and Aerospace Engineering, Hong Kong University of Science and Technology, Clear Water Bay, Kowloon, Hong Kong}
\address[Shenzhen]{Shenzhen Research Institute, Hong Kong University of Science and Technology, Shenzhen 518057, China}
\cortext[cor1]{Corresponding author}

\begin{abstract}

The compressible flow has intrinsically multiple scale nature due to the large variations of gas density and characteristic scale of flow structure, especially in hypersonic and reentry problems.
It is challenging to construct an accurate and efficient numerical algorithm to capture non-equilibrium flow physics across different regimes.
In this paper, a unified gas kinetic scheme with adaptive velocity space (AUGKS) for multiscale flow transport will be developed.
In near-equilibrium flow regions, particle distribution function is close to the Chapman-Enskog expansion and can be
formulated with a continuous velocity space, where only macroscopic conservative variables are updated.
With the emerging of non-equilibrium effects, the AUGKS automatically switches to a discrete velocity space to follow the evolution of particle distribution function.
Based on the Chapman-Enskog expansion, a criterion is proposed in this paper to quantify the intensity of non-equilibrium effects and is used for the continuous-discrete velocity space transformation.
Following the scale-dependent local evolution solution, the AUGKS presents the discretized gas dynamic equations directly on the cell size and time step scales, i.e., the so-called direct modeling method.
As a result, the scheme is able to capture the cross-scale flow physics from particle transport to hydrodynamic wave propagation, and provides a continuous variation of solutions from the Boltzmann to the Navier-Stokes.
Under the unified framework, different from conventional DSMC-NS hybrid method, the AUGKS does not need a buffer zone to match up kinetic and hydrodynamic solutions.
Instead, a continuous and discrete particle velocity space is naturally connected, which is feasible for the numerical simulations with unsteadiness or complex geometries.
Compared with the asymptotic preserving (AP) methods which solves kinetic equations uniformly over the entire flow field with discretized velocity space, the current velocity-space adaptive unified scheme speeds up the computation and reduces the memory requirement in multiscale flow problems, and maintains the equivalent accuracy.
The AUGKS provides an effective tool for non-equilibrium flow simulations.

\end{abstract}


\maketitle

\section{Introduction}

The gaseous flow shows a diverse set of behaviors on different characteristic scales.
For example, within the mean free path and collision time of gas molecules, particles travel freely during most of time with rare intermolecular collisions, leading to peculiar non-equilibrium flow dynamics.
However, on a much macroscopic level, the accumulating effect of collisions results in an equalization of local temperature and velocity, where the moderate non-equilibrium effects can be well described by viscous transport, heat conduction and mass diffusion, i.e., the so called transport phenomena \cite{chapman1970mathematical}.
From microscopic particle transport to macroscopic fluid motion, there is a continuous variation of flow dynamics.
Generally, different flow regimes can be categorized qualitatively according to the Knudsen number, which is defined as the ratio of the molecular mean free path to a characteristic physical length scale.
With the variation of $Kn$, the whole flow domain can be qualitatively divided into continuum ($Kn<0.001$), slip ($0.001<Kn<0.1$), transition ($0.1<Kn<10$) and free molecular regimes ($Kn>10$) \cite{tsien1946superaerodynamics}.
When $Kn$ is large, the particle transport and collision can be depicted separately in the Boltzmann equation.
In another limit with extremely small $Kn$, the Navier-Stokes equations are routinely used to describe macroscopic flow.

The traditional computational fluid dynamics targets to get numerical solutions of the corresponding governing equations.
For example, the most widely used numerical methods for the Boltzmann equation are the direct Boltzmann solvers \cite{aristov2012direct} and the direct simulation Monte Carlo (DSMC) method \cite{bird1994molecular}.
In the former methodology, a discretized particle velocity space is constructed and the particle distribution function is updated from transport and collision terms respectively.
On the other hand, the DSMC method mimics the same physical process while the distribution function is now represented by a large amount of test particles and the collision term is calculated statistically.
Due to the splitting treatment of particle transport and collision, the mesh size and time step should be restricted by the mean free path and collision time, and the computational cost is proportional to the amount of discretized velocity points or test particles used in the simulation.
Meanwhile, the compressible Navier-Stokes solvers are mostly based on the Riemann solvers for inviscid flux and the central difference method for viscous terms.
The macroscopic flow variables are followed in the simulation.
Compared with the kinetic methods, the computational cost of continuum flow solvers is much reduced.

The rapid developments of aerospace industry face new challenges for accurate and efficient simulation of complex flows. 
For example, when a shuttle reenters into the atmosphere, the surrounding gas becomes denser and denser from the rarefied upper atmosphere to lower continuum region, and thus the vehicle sees cross-scale flow physics during the landing process.
Besides, localized non-equilibrium flow structures often emerge around the vehicle in hypersonic cruise as a result of the geometric effect, such as shock, rarefaction wave, boundary layer and wake turbulence.
It is natural to couple different numerical methods in different domains to calculate aerodynamic force and heat efficiently.
Therefore, hybrid algorithms which combine continuum and kinetic approaches have been developed to simulate multiple scale flows where continuum and rarefied flow physics coexist in a single flow simulation \cite{bourgat1996coupling,tiwari1998coupling,boyd2011hybrid,wijesinghe2004three,schwartzentruber2006hybrid,schwartzentruber2007modular,dimarco2008hybrid,burt2009hybrid,degond2006macroscopic}.
In these numerical schemes, the main flow structure of the flow field is simulated by the continuum methods efficiently, with highly dissipative non-equilibrium regions being resolved by the kinetic methods.
Due to the complicated fivefold collision integral in the Boltzmann equation, the prevailing kinetic solver used for hybridization currently is the DSMC method.
In the calculation, a dynamic parameter is needed to determine the breakdown point of continuum description and decompose the flow domain.
It is easy to implement the hybrid method into a parallel simulation since the physical domain has already been splitted into blocks on different computational nodes.

In kinetic theory, the Chapman-Enskog expansion \cite{cercignani1988boltzmann} bridges the NS and Boltzmann solutions.
Although this successive expansion is mathematically attractive, there is little information provided about the intrinsic scale for the validation of macroscopic equations, such as the specific fluid element in the NS modeling.
The success of the mathematical derivation of high-order equations with inclusion of the so-called Burnett or super-Burnett terms is limited due to the lack of specified modeling scales for these extended hydrodynamic equations.
Besides, due to the uncertainty in choosing the length scale in the definition of Knudsen number, 
it becomes rather tough to predict a universal breakdown criterion for the Chapman-Enskog expansion and the use of the NS solutions, although
it is defined empirically that the NS equations are valid when ${Kn}\leq0.01$.
In addition, on particle mean free path and collision time scales, the kinetic method provides much more degrees of freedom to describe gas dynamics, while in the buffer zone these information needs to be shrunk to the mean field variables on a macroscopic level, such as density, momentum, energy, stress and heat flux.
The inherent incompatibility between the particle-based and PDE-based methods leads to considerable difficulties in the hybridization.
Usually a buffer zone is delicately defined for the information exchange between kinetic and continuum solutions, which may become more difficult in the simulation of unsteady flows.

In recent years, the unified gas kinetic scheme (UGKS) has been developed for cross-scale flows \cite{xu2010unified, xubook}.
Based on the direct modeling on the mesh size scale, the time-evolving flux function from the kinetic equation provides a smooth transition from particle transport to hydrodynamic wave propagation.
The UGKS is an asymptotic preserving (AP) scheme,
which preserves the discrete analogy of the Chapman-Enskog expansion and the time step in the simulation is not restricted by the collision time \cite{jin2010asymptotic}.
However, for the UGKS the memory requirement and computational cost due to the discretized velocity space limits its wide applications in aerospace industry.
In this paper, we develop an adaptive unified gas kinetic scheme (AUGKS), where the continuous and discrete velocity space is transformed dynamically in a single framework.
In the near-equilibrium region, the Chapman-Enskog expansion is used for the construction of distribution function with a continuous velocity space, and thus only macroscopic conservative variables are stored and updated in the simulation.
With increasing non-equilibrium effects,
the AUGKS tracks the evolution of distribution function directly with a discrete velocity space.
Based on the Chapman-Enskog expansion, a criterion to switch continuous-discrete velocity space in the simulation is proposed and validated through numerical experiments.
Compared with the original UGKS method, the current adaptive scheme frees the memory requirement in near-equilibrium flow regimes and speeds up the computation, but provides equivalent physical solutions.
Due to the use of particle distribution function in the whole flow field, the AUGKS avoids defining an interface condition for domain decomposition in order to connect fluid and kinetic representations.
In other words, no buffer zone is needed in the AUGKS.

This paper is organized as following.
Section 2 is a brief introduction of the fundamental kinetic theory.
Section 3 presents the numerical implementation of the adaptive unified gas kinetic scheme and proposes a switching criterion of the velocity space transformation.
Section 4 includes numerical examples to demonstrate the performance of the current scheme. 
The last section is the conclusion.

\section{Gas kinetic modeling}

The gas kinetic theory describes the time-space evolution of particle distribution function $f(\mathbf{x},\mathbf{u},t)$, where $\mathbf{x}\in \mathcal{R}^3$ is space variable and  $\mathbf{u}\in \mathcal{R}^3$ is particle velocity.
In the absence of external force field, the Boltzmann equation of a monatomic dilute gas writes,
\begin{equation}
f_t+\mathbf{u}\cdot\nabla_{\mathbf x} f = Q(f,f) = \int_{\mathcal{R}^3} \int_{\mathcal{S}^2} \left[ f(\mathbf{u}')f(\mathbf{u}_1')-f(\mathbf{u})f(\mathbf{u}_1) \right] B(\cos \theta,g) d\mathbf{\Omega} d\mathbf{u}_1,
\label{eqn:boltzmann equation}
\end{equation}
where $\mathbf{u},\mathbf{u_1}$ are the pre-collision velocities of two classes of molecules, and $\mathbf{u}',\mathbf{u_1}'$ are the corresponding post-collision velocities.
The collision kernel $B(\cos \theta,g)$ measures the strength of collisions in different directions, where $\theta$ is the deflection angle and $g=|\mathbf g|=|\mathbf{u}-\mathbf{u_1}|$ is the magnitude of relative pre-collision velocity.
The $\mathbf{\Omega}$ is the unit vector along the relative post-collision velocity $\mathbf{u}'-\mathbf{u_1}'$,
and the deflection angle $\theta$ satisfies $\cos \theta = \mathbf \Omega \cdot \mathbf g/g$.
The conservation of momentum and energy leads the following relations,
\begin{equation}
\begin{aligned}
&\mathbf{u}'=\frac{\mathbf{u}+\mathbf{u_1}}{2}+\frac{|\mathbf{u}-\mathbf{u_1}|}{2}\mathbf{\Omega}=\mathbf{u}+\frac{g\mathbf{\Omega}-\mathbf{g}}{2}, \\
&\mathbf{u_1}'=\frac{\mathbf{u}+\mathbf{u_1}}{2}-\frac{|\mathbf{u}-\mathbf{u_1}|}{2}\mathbf{\Omega}=\mathbf{u}_1-\frac{g\mathbf{\Omega}-\mathbf{g}}{2}.
\end{aligned}
\end{equation}

Due to the complicated fivefold integration in the Boltzmann collision operator, some simplified kinetic models have been constructed, such as the Shakhov \cite{shakhov1968generalization}.
In this model, the Boltzmann collision operator $Q(f,f)$ is replaced with a relaxation operator $S(f)$, which writes,
\begin{equation}
\begin{aligned}
&f_t+\mathbf{u}\cdot\nabla_{\mathbf x} f =S(f)=\frac{f^+-f}{\tau}, \\
&f^+=\rho \left( \frac{\lambda}{\pi} \right)^{\frac{3}{2}} e^{-\lambda(\mathbf u-\mathbf U)^2}
\left[1+(1-{Pr} )\mathbf{c} \cdot \mathbf{q} \left(\frac{\mathbf{c}^2}{RT}-5\right)/(5pRT)\right],
\end{aligned}
\label{eqn:shakhov equation}
\end{equation}
where $\tau=\mu/p$ is the collision time.
The macroscopic density, velocity, temperature and heat flux are marked with $\rho,\mathbf U,T,\mathbf q$.
The $\mathbf c=\mathbf u-\mathbf U$ is particle peculiar velocity, $Pr$ is the Prandtl number, $R$ is the gas constant, and $\lambda={\rho}/{(2p)}$.
In this paper, the numerical simulations will be conducted by either the full Boltzmann or the Shakhov collision terms.

The macroscopic conservative flow variables are related with the moments of particle distribution function via
\begin{equation*}
\textbf{W} =\left(
\begin{matrix}
\rho \\
\rho \mathbf U \\
\rho E
\end{matrix}
\right)=\int f\psi d\mathbf u,
\end{equation*}
and the collision terms satisfy the compatibility condition,
\begin{equation*}
\int Q(f,f_1)\psi d\mathbf u=\int S(f)\psi d\mathbf u=0,
\end{equation*}
where $\psi=\left(1,\mathbf u,\frac{1}{2} \mathbf u^2 \right)^T$ is a vector of collision invariants. 
Here we rewrite the collision terms $Q(f,f)$ and $S(f)$ into a general form $Q(f)$.

With a local constant collision time $\tau$, the integral solution of Eq.(\ref{eqn:shakhov equation}) can be constructed along the characteristic line,
\begin{equation}
f(\mathbf x,t,\mathbf u)=\frac{1}{\tau}\int_{t^0}^t f^+(\mathbf x',t',\mathbf u)e^{-(t-t')/\tau}dt' +e^{-t/\tau}f_0(\mathbf x^0,0,\mathbf u),
\label{eqn:integral solution}
\end{equation}
where $\mathbf x'=-\mathbf u(t-t')$ is the particle trajectory, and $f_0$ is the gas distribution function at the initial time step $t=t^0$.
Based on the above evolving solution, the corresponding discretized gas dynamic equations on the cell size and time step scales can be constructed in the gas kinetic scheme.

\section{Adaptive unified gas kinetic scheme}

In this section, we will present the principle and numerical implementation of the velocity-space adaptive unified gas kinetic scheme (AUGKS). 
The original gas kinetic scheme with continuous and discrete particle velocity space will be introduced first.
The detailed coupling of continuum and kinetic treatments and the switching criterion for velocity space transformation will be discussed.
For simplicity, the following introduction is based on two-dimensional case, while the extension to three dimension is straightforward.

\subsection{Unified gas kinetic scheme with discrete velocity space}

With the notation of cell averaged quantities in the control volume,
\begin{equation*}
\begin{aligned}
&\mathbf W_{x_i,y_j,t^n}=\mathbf W_{i,j}^n=\frac{1}{\Omega_{i,j}(\mathbf x)} \int_{\Omega_{i,j}} f(x,y,t^n)dxdy ,\\
&f_{x_i,y_j,t^n,u_l,v_m}=f_{i,j,l,m}^n=\frac{1}{\Omega_{i,j}(\mathbf x)\Omega_{l,m}(\mathbf u)} \int_{\Omega_{i,j}} \int_{\Omega_{l,m}}f(x,y,t^n,u,v) dxdydudv,
\end{aligned}
\end{equation*}
where $\Omega_{i,j}$ and $\Omega_{l,m}$ are the cell area in the discretized physical and velocity space,
the macroscopic conservative variables and the particle distribution function are updated in the UGKS,
\begin{equation}
\textbf{W}_{i,j}^{n+1}=\textbf{W}_{i,j}^n+\frac{1}{\Omega_{i,j}}\int_{t^n}^{t^{n+1}}\sum_{r} {\mathbf{F}}_r \Delta L_r dt,
\label{eqn:macro update}
\end{equation}
\begin{equation}
\begin{aligned}
f_{i,j,l,m}^{n+1}=&f_{i,j,l,m}^n+\frac{1}{\Omega_{i,j}}\int_{t^n}^{t^{n+1}} \sum_{r} u_r f_r \Delta L_r dt+ \int_{t^n}^{t^{n+1}}\int_{\Omega_{i,j}}Q(f) dt,
\end{aligned}
\label{eqn:distribution update}
\end{equation}
where $\textbf{F}_r$ is the flux of conservative variables, $f_r$ is the time-dependent gas distribution function at cell interface and $\Delta L_r$ is the cell interface length.

In the UGKS, the flux function $\mathbf F_r$ is evaluated through the particle distribution function at the cell interface, which is constructed from the evolving solution of the Shakhov equation.
If the interface location is simplified with $x_{i+1/2}=0$, $y_j=0$, and $t^n=0$, with a local constant collision time $\tau$, the integral solution in Eq.(\ref{eqn:integral solution}) can be written as,
\begin{equation}
\begin{aligned}
f(0,0,t,u_l,v_m,\xi)=&\frac{1}{\tau}\int_{0}^t f^+(x',y',t',u_l,v_m,\xi)e^{-(t-t')/\tau}dt' \\
&+e^{-t/\tau}f_0(x^0,y^0,0,u_l,v_m,\xi),
\end{aligned}
\label{eqn:ugks integral solution}
\end{equation}
where $x'=-u_l(t-t')$ and $y'=v_m(t-t')$ are the particle trajectories, and $x^0,y^0$ are the initial locations for the particle which passes through the cell interface at time $t$.
Here $f_0$ is the gas distribution function at the beginning of $n$-th time step.
The internal degree of freedom $\xi$ denotes the random motion in $z$ direction.
The above scale-dependent evolving solution provides a multiple scale evolution model,
where the contributions from both equilibrium hydrodynamic and non-equilibrium kinetic flow physics are included uniformly.

In the detailed numerical scheme, to the second order accuracy, the initial gas distribution function $f_0$ is reconstructed as
\begin{equation}
f_0(x,y,0,u_l,v_m,\xi)=\left\{
\begin{aligned}
&f_{i+1/2,j,l,m}^L+\sigma_{i,j,l,m}x+\theta_{i,j,l,m}y, \quad x\le 0, \\
&f_{i+1/2,j,l,m}^R+\sigma_{i+1,j,l,m}x+\theta_{i+1,j,l,m}y, \quad x> 0,
\end{aligned}
\right.
\label{eqn:f0 construct}
\end{equation}
where $f_{i+1/2,j,l,m}^L$ and $f_{i+1/2,j,l,m}^R$ are the reconstructed initial distribution functions at the left and right hand sides of a cell interface,
and $\sigma$ and $\theta$ are the slopes of distribution function along $x$ and $y$ directions.
In addition, the equilibrium distribution function around a cell interface is constructed as
\begin{equation}
f^+=f^+_0\left[1+(1-H[x])\bar{a}^Lx+H[x]\bar{a}^Rx+\bar by+\bar{A}t\right],
\label{eqn:g0 construct}
\end{equation}
where $f^+_0$ is the equilibrium distribution at $(x=0,y=0,t=0)$. 
The coefficients $(\bar a^{L,R},\bar b,\bar A)$ can be evaluated from the spatial distribution of conservative variables on both sides of the cell interface and the compatibility condition \cite{Xiao2017well}.
After all coefficients are determined, the time dependent interface distribution function becomes
\begin{equation}
\begin{aligned}
f(0,0,t,u_l,v_m,\xi)=&\left(1-e^{-t/\tau}\right) f^+_0\\
&+\left(\tau(-1+e^{-t/\tau})+te^{-t/\tau}\right)\bar a^{L,R}u_lf^+_0 \\
&+\left(\tau(-1+e^{-t/\tau})+te^{-t/\tau}\right)\bar bv_mf^+_0 \\
&+\tau \left(t/\tau-1+e^{-t/\tau}\right)\bar{A}f^+_0 \\
&+e^{-t/\tau}\left[\left(f_{i+1/2,j,l,m}^L-u_lt\sigma_{i,j,l,m}-v_mt\theta_{i,j,l,m}\right)H\left[u_l\right] \right.\\
&\left.+\left(f_{i+1/2,j,l,m}^R-u_lt\sigma_{i+1,j,l,m}-v_mt\theta_{i+1,j,l,m}\right)(1-H\left[u_l\right])\right] \\
=&\widetilde f^+_{i+1/2,j,k,l}+\widetilde f_{i+1/2,j,k,l},
\end{aligned}
\label{eqn:interface distribution}
\end{equation}
where $\widetilde f^+_{i+1/2,j,k,l}$ is related to equilibrium state integration and $\widetilde f_{i+1/2,j,k,l}$ is related to the initial distribution.
With the variation of the ratio between evolving time $t$ (i.e., the time step in the computation) and collision time $\tau$, the interface distribution function above provides self-conditioned multiple scale flow physics across different flow regimes.
After the interface distribution function is determined,
the corresponding fluxes of conservative variables are evaluated through
\begin{equation*}
{\textbf{F}}_r=\int_{\Omega_{l,m}} u_l f(0,0,t,u_l,v_m,\xi) \psi dudvd\xi.
\end{equation*}

Inside each control volume, the collision term $Q(f)$ is to be determined for the update of particle distribution function in Eq.(\ref{eqn:distribution update}).
In the unified scheme, the numerical treatments of $Q(f)$ for the exact Boltzmann collision term or the Shakhov model are given in \cite{mouhot2006fast,wu2013deterministic,wu2014solving,liu2016unified}.

\subsection{Near-continuum gas kinetic scheme with continuous velocity space}
In continuum flow with intensive intermolecular collisions, the particle distribution function is close to local equilibrium state, and thus the Navier-Stokes equations are valid to describe macroscopic fluid motion.
In this case, it is straightforward to apply the Chapman-Enskog expansion to construct the corresponding distribution function at every time step, and thus only macroscopic conservative variables need to be stored and updated.
Following this procedure, in the gas kinetic scheme with continuous particle velocity space \cite{xu2001gas}, the gas distribution function $f_0$ in Eq.(\ref{eqn:ugks integral solution}) at the beginning of each time step is expanded as following,
\begin{equation}
f_0=\left\{
\begin{aligned}
&f_0^{+(l)} \left(1+a^lx+b^lx-\tau\left(a^lu+b^lv+A^l\right)\right), \quad x\le 0 \\
&f_0^{+(r)}\left(1+a^rx+b^rx-\tau\left(a^ru+b^rv+A^r\right)\right), \quad x> 0
\end{aligned}
\right.
\end{equation}
where $f_0^{+(l)}$ and $f_0^{+(r)}$ are the corresponding equilibrium state at left and right sides of the interface, and
the coefficients $(a^{l,r},b^{l,r},A^{l,r})$ are related with spatial distribution of macroscopic variables.
The equilibrium distribution function $f^+$ around a cell interface is constructed in the same way in Eq.(\ref{eqn:g0 construct}), and the corresponding interface distribution function becomes,
\begin{equation}
\begin{aligned}
f(0,0,t,u,v,\xi)=&\left(1-e^{-t/\tau}\right)f_0^+\\
&+\left(\tau(-1+e^{-t/\tau})+te^{-t/\tau}\right)(\bar a^{L,R}u+\bar bv)f^+_0 \\
&+\tau \left(t/\tau-1+e^{-t/\tau}\right)\bar{A}f^+_0 \\
&+e^{-t/\tau}\left\{
[1-(t+\tau)(ua^l+vb^l)]H[u]f_0^{+(l)} \right. \\
&\left. + [1-(t+\tau)(ua^r+vb^r)][1-H[u]]f_0^{+(r)} \right\} \\
&+e^{-t/\tau} \left[ -\tau A^l H[u] f_0^{+(l)} - \tau A^r (1-H[u]) f_0^{+(r)} \right].
\end{aligned}
\label{eqn:gks interface distribution}
\end{equation}
The interface distribution function here is in the continuous form of particle velocity $(u,v)$, and the fluxes for macroscopic variables can be obtained through the analytical integration of Gaussian distribution.

\subsection{Adaptive unified gas kinetic scheme}

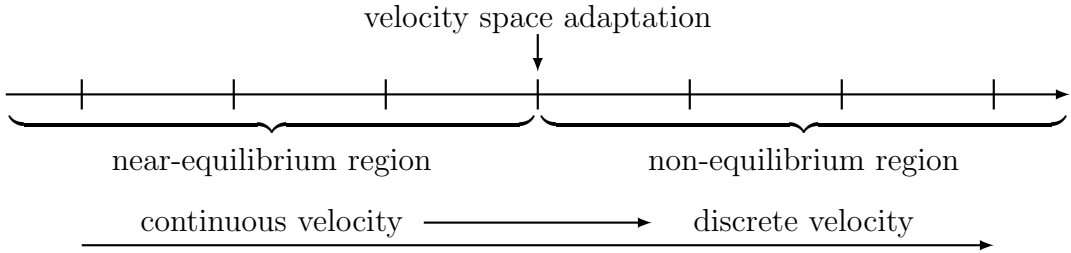
\begin{figure}[htb!]
	\centering
	{
		\begin{tikzpicture}[thick]
		\node[rectangle] (bx) {};
		\node[rotate = 0] at (-3.5, -0.4) {$\underbrace{\hspace{6.9cm}}$};
		\node[rotate = 0] at (+3.5, -0.4) {$\underbrace{\hspace{6.9cm}}$};
		\node at ($(bx)+(+0,+1.0)$) {velocity space adaptation};
		\draw[line] ($(bx)+(0,+0.8)$) -- ($(bx)+(+0,+0.3)$);
		\node at ($(bx)+(-3.5,-0.9)$) {near-equilibrium region};
		\node at ($(bx)+(+3.5,-0.9)$) {non-equilibrium region};
		\draw ($(bx)+(0,-0.2)$) -- ($(bx)+(0,+0.2)$);
		\draw ($(bx)+(-2,-0.2)$) -- ($(bx)+(-2,+0.2)$);
		\draw ($(bx)+(-4,-0.2)$) -- ($(bx)+(-4,+0.2)$);
		\draw ($(bx)+(-6,-0.2)$) -- ($(bx)+(-6,+0.2)$);
		\draw ($(bx)+(+2,-0.2)$) -- ($(bx)+(+2,+0.2)$);
		\draw ($(bx)+(+4,-0.2)$) -- ($(bx)+(+4,+0.2)$);
		\draw ($(bx)+(+6,-0.2)$) -- ($(bx)+(+6,+0.2)$);
		\draw[line] ($(bx)+(-7,+0)$) -- ($(bx)+(+7,+0)$);
		\draw[line] ($(bx)+(-6,-2)$) -- ($(bx)+(+6,-2)$);
		\node at ($(bx)+(-3.5,-1.7)$) {continuous velocity};
		\node at ($(bx)+(+3.5,-1.7)$) {discrete velocity};
		\draw[line] ($(bx)+(-1.5,-1.7)$) -- ($(bx)+(+1.5,-1.7)$);
		\end{tikzpicture}
	}
	\caption{Schematic of the adaptive scheme for multiscale flow.}
	\label{pic:hybrid}
\end{figure}

In a multiscale flow problem, to overcome the computational deficiency and memory burden from a large amount of discretized velocity points, 
it is feasible to combine both continuum and rarefied flow solvers into a single framework with dynamic continuous-discrete velocity transformation.
As shown in Fig.\ref{pic:hybrid}, in near-equilibrium flow regions, the particle distribution function (PDF) is formulated with a continuous velocity space based on the Chapman-Enskog expansion, and only macroscopic flow variables are updated.
In non-equilibrium regions, the AUGKS switches to a discretized velocity space to follow the evolution of particle distribution function.
The continuous and discrete velocity space are connected with an adaptation interface, at which the continuous solution of distribution function is sorted into discretized velocity space.

In the detailed numerical scheme, the macroscopic conservative variables are updated in Eq.(\ref{eqn:macro update}), while in non-equilibrium regions the particle distribution function is updated in Eq.(\ref{eqn:distribution update}).
Near the adaptation interface, at every time step $t^n$, if there is no recorded discretized distribution function at $t^{n-1}$ in the current "non-equilibrium" cell $(i,j)$, a local discretized velocity mesh will be generated, where the particle distribution function at velocity point $(l,m)$ is given by the discrete Chapman-Enskog expansion, 
\begin{equation}
f_{i,j,l,m}=f^+_{0(i,j,l,m)}[1-\tau(au_l+bv_m+A)],
\label{eqn:reconstruct distribution}
\end{equation}
where the coefficients $a,b,A$ are related with the spatial distribution of macroscopic variables, and can be determined in the same way in Sec. 3.2.
In the current scheme, the velocity mesh is generated within
\begin{equation}
u\in[U-3\sqrt{RT},U+3\sqrt{RT}],\ v\in[V-3\sqrt{RT},V+3\sqrt{RT}],
\end{equation}
where $(U,V)$ is macroscopic flow velocity, $T$ is temperature and $R$ is the gas constant.

To update the discretized distribution function in the adjacent cell next to the adaptation interface, the interface distribution function from the continuous GKS solution in Eq.(\ref{eqn:gks interface distribution}) is rewritten into the following discrete form,
\begin{equation}
\begin{aligned}
f(0,0,t,u_l,v_m,\xi)=&\left(1-e^{-t/\tau}\right)f_0^+\\
&+\left(\tau(-1+e^{-t/\tau})+te^{-t/\tau}\right)(\bar a^{L,R}u_l+\bar bv_m)f^+_0 \\
&+\tau \left(t/\tau-1+e^{-t/\tau}\right)\bar{A}f^+_0 \\
&+e^{-t/\tau}\left\{
[1-(t+\tau)(u_la^l+v_mb^l)]H[u_l]f_0^{+(l)} \right. \\
&\left. + [1-(t+\tau)(u_la^r+v_mb^r)][1-H[u_l]]f_0^{+(r)} \right\} \\
&+e^{-t/\tau} \left[ -\tau A^l H[u_l] f_0^{+(l)} - \tau A^r (1-H[u_l]) f_0^{+(r)} \right],
\end{aligned}
\label{eqn:transition interface distribution}
\end{equation}
and then it can be used to determine the fluxes of macroscopic variables and particle distribution function.
In this way, the fluxes at adaptation interface are fully determined and can be used to update the macroscopic variables in Eq.(\ref{eqn:macro update}) and particle distribution function in Eq.(\ref{eqn:distribution update}).

In the current scheme, the time step is determined by the CFL condition,
\begin{equation}
\Delta t = \mathrm{CFL} \frac{\Delta x \Delta y}{u_{max}\Delta y+v_{max}\Delta x},
\label{eqn:time step}
\end{equation}
where CFL is the CFL number, and $(u_{max},v_{max})$ is the largest particle velocity in $x$ and $y$ directions.

\subsection{Switching criterion of velocity space}

The accuracy and efficiency of the current adaptive scheme are based on a proper choice of location of velocity space adaptation.
The transition of discrete to continuous velocity space must be located in the region where the Navier-Stokes solutions provided by the GKS with a continuous velocity space are still valid.
For this issue, many empirical parameters for the breakdown of continuum description have been proposed.
Bird \cite{bird1994molecular} proposed a parameter $P=D(ln \rho)/Dt /\nu$ for the DSMC simulation of expansion flows, where $\rho$ is gas density and $\nu$ is collision frequency, and the breakdown value of $P$ for translational equilibrium is $0.05$.
Boyd et al. \cite{boyd1995predicting}
extended the above concept to a gradient-length-local Knudsen number $Kn_{GLL}=\ell |\nabla Q|/{Q}$, where $\ell$ is the local mean free path and $Q$ is the macroscopic flow quantity of interest, with the critical value $0.05$.
Considering the terms in the Chapman-Enskog distribution function, Garcia et al. [30] proposed a breakdown parameter based on dimensionless stress and heat flux $B=max(|\tau^*|,|q^*|)$, with the switching criterion of $0.1$.

Since the particle distribution function takes the Chapman-Enskog expansion form in the evolution process of the continuous GKS solver, here we propose an alternative switching criterion of particle velocity space directly from the Chapman-Enskog expansion.
For brevity, one-dimensional case will be used for illustration.
When there is no discontinuity inside the flow field, the particle distribution function in the Chapman-Enskog form writes,
\begin{equation}
f=f_0^+ [1-\tau (au+A)],
\end{equation}
where the coefficients $a$ and $A$ are the space and time variations of a distribution function, which can be expanded based on the collision invariants,
\begin{equation*}
\begin{aligned}
&a=a_1+a_2u+a_3\frac{1}{2}u^2=a_\alpha \psi_\alpha, \\
&A=A_1+A_2u+A_3\frac{1}{2}u^2=A_\alpha \psi_\alpha.
\end{aligned}
\end{equation*}
In the current one-dimensional case, the spatial slope $a$  is determined from the following relation,
\begin{equation*}
\frac{\partial \mathbf W}{\partial x}=\int a f^+_0 \psi d\mathbf u=M\mathbf a,
\end{equation*}
where $\psi$ is a vector of collision invariants, $M_{\alpha,\beta}=\int f^+_0 \psi_\alpha\alpha \psi_\alpha\beta du$ and $\mathbf a=(a_1,a_2,a_3)^T$.
The solution $\mathbf a$ can be written as,
\begin{equation*}
\begin{aligned}
&a_3=4\frac{\lambda^2}{(K+1)\rho} \left[ 2\frac{\partial (\rho E)}{\partial x}-2U\frac{\partial (\rho U)}{\partial x}+\frac{\partial \rho}{\partial x} \left( U^2-\frac{K+1}{2\lambda}\right) \right], \\
&a_2=2\frac{\lambda}{\rho} \left[ 2\frac{\partial (\rho U)}{\partial x}-U \frac{\partial \rho}{\partial x} \right]-Ua_3, \\
&a_1=\frac{1}{\rho} \frac{\partial \rho }{\partial x}-Ua_2-\frac{1}{2} \left[ U^2+\frac{K+1}{2\lambda} \right] a_3. \\
\end{aligned}
\end{equation*}
The time derivative $A$ is related to the temporal variation of conservative flow variables respectively, and can be evaluated from a successive procedure in the Chapman-Enskog expansion with the help of the Euler equations,
\begin{equation*}
\frac{\partial \mathbf W}{\partial t}=\int A f^+_0 \psi d\mathbf u=-\int u \frac{\partial f_0^+}{\partial x} \psi d\mathbf u.
\end{equation*}

Generally, the Navier-Stokes solutions can be applied when the Chapman-Enskog expansion is a proper approximation of the distribution function in near-equilibrium regime.
Therefore, based on the dimensionless collision time and space-time variations, a switching criterion for the velocity space transformation can be defined as
\begin{equation}
B=\tau \mathrm{max} (|  {\hat{a}}|, | {\hat A}|),
\label{eqn:velocity adaptation}
\end{equation}
where dimensionless variables are defined as,
\begin{equation}
\hat{\tau}=\frac{\tau (2RT_0)^{1/2}}{L_0} , {\hat{a}}=a {L_0}, {\hat{A}}=\frac{ A L_0}{(2RT_0)^{1/2}},
\end{equation}
where $R$ is gas constant, $L_0$ and $T_0$ are reference length and temperature.
The current switching criterion $B$ for particle velocity space will be tested in numerical experiments.

\subsection{Summary}

The numerical algorithm of the adaptive unified gas kinetic scheme is as following. 
In the AUGKS, we follow the evolution of both both conservative variables and particle distribution function.
In near-equilibrium flow regions, the particle distribution function is formulated by the Chapman-Enskog expansion in a continuous velocity space, and macroscopic flow variables are updated in Eq.(\ref{eqn:macro update}).
For non-equilibrium flows, the particle distribution function is updated explicitly in Eq.(\ref{eqn:distribution update}).
The scale-dependent flux function is determined by the particle distribution function at the interface, which comes from the integral solution of kinetic model equation in Eq.(\ref{eqn:integral solution}).
In each time step, the domain of continuous and discrete velocity space is specified by Eq.(\ref{eqn:velocity adaptation}), and the corresponding interface fluxes are provided by Eq.(\ref{eqn:interface distribution}), Eq.(\ref{eqn:transition interface distribution}) and Eq.(\ref{eqn:gks interface distribution}).
The detailed numerical procedures for AUGKS are given in Fig. \ref{pic:numerical procedure}.

\tikzstyle{round} = [rectangle, rounded corners, minimum width=2cm, minimum height=1cm, text centered, draw = black, fill = yellow!40]
\tikzstyle{start} = [trapezium, trapezium left angle=70, trapezium right angle=110, minimum width=2cm, minimum height=1cm, text centered, draw=black, fill = blue!40]
\tikzstyle{process} = [rectangle, minimum width=3cm, minimum height=1cm, text centered, draw=black, fill = yellow!50]
\tikzstyle{start} = [diamond, aspect = 3, text centered, draw=black, fill = green!15]
\tikzstyle{arrow} = [->,>=stealth]

\begin{figure}[htb!]
	\centering
	\begin{tikzpicture}[node distance=1.5cm]
	\node (start) [start] {Start};
	\node (timestep) [round, below of=start] {Calculate time step by Eq.(\ref{eqn:time step})};
	\node (adapt) [round, below of=timestep] {Evaluate velocity adaptation by Eq.(\ref{eqn:velocity adaptation})};
	\node (continuous) [round, below of=adapt] {Continuous velocity};
	\node (discrete) [round, below of=adapt, right of=adapt, xshift= 4cm] {Discrete velocity};
	\node (gks flux) [round, below of=continuous, left of=continuous, xshift=-1.2cm] {Calculate flux by Eq.(\ref{eqn:gks interface distribution})};
	\node (transition flux) [round, below of=continuous, right of=continuous, xshift=1cm] {Calculate flux by Eq.(\ref{eqn:transition interface distribution})};
	\node (ugks flux) [round, below of=discrete, right of=discrete, xshift=1.2cm] {Calculate flux by Eq.(\ref{eqn:interface distribution})};
	\node (deconstruct) [round, below of=gks flux] {destroy recorded PDF};
	\node (construct) [round, below of=transition flux, right of=transition flux, xshift=1cm] {Construct unrecorded PDF by Eq.(\ref{eqn:reconstruct distribution})};
	\node (updatemacro2) [round, below of=construct] {Update conservative variables by Eq.(\ref{eqn:macro update})};
	\node (pre) [round, below of=updatemacro2] {Calculate equilibrium distribution $f^{+(n+1)}$ and collision time $\tau^{n+1}$};
	\node (updatemicro) [round, below of=pre] {Update PDF by Eq.(\ref{eqn:distribution update})};
	\node (updatemacro1) [round, left of=updatemicro, xshift=-6.2cm] {Update conservative variables by Eq.(\ref{eqn:macro update})};
	\node (end) [start, below of=updatemacro1] {end};
	
	\draw [arrow](start) -- (timestep);
	\draw [arrow](timestep) -- (adapt);
	\draw [arrow](adapt) -- (continuous);
	\draw [arrow](adapt) -- (discrete);
	\draw [arrow](continuous) -- (gks flux);
	\draw [arrow](continuous) -- (transition flux);
	\draw [arrow](discrete) -- (transition flux);
	\draw [arrow](discrete) -- (ugks flux);
	\draw [arrow](ugks flux) -- (construct);
	\draw [arrow](transition flux) -- (construct);
	\draw [arrow](gks flux) -- (deconstruct);
	\draw [arrow](construct) -- (updatemacro2);
	\draw [arrow](updatemacro2) -- (pre);
	\draw [arrow](pre) -- (updatemicro);
	\draw [arrow](deconstruct) -- (updatemacro1);
	\draw [arrow](updatemacro1) -- (end);
	\draw [arrow](end) -- ($(end.west) - (3,0)$) node[anchor=north] {} |- (start);
	\draw [arrow](updatemicro) -- ($(updatemicro.south) - (0,1)$) node[anchor=west] {} |- (end);
	\end{tikzpicture}
	\caption{Numerical algorithm of AUGKS.}
	\label{pic:numerical procedure}
\end{figure}
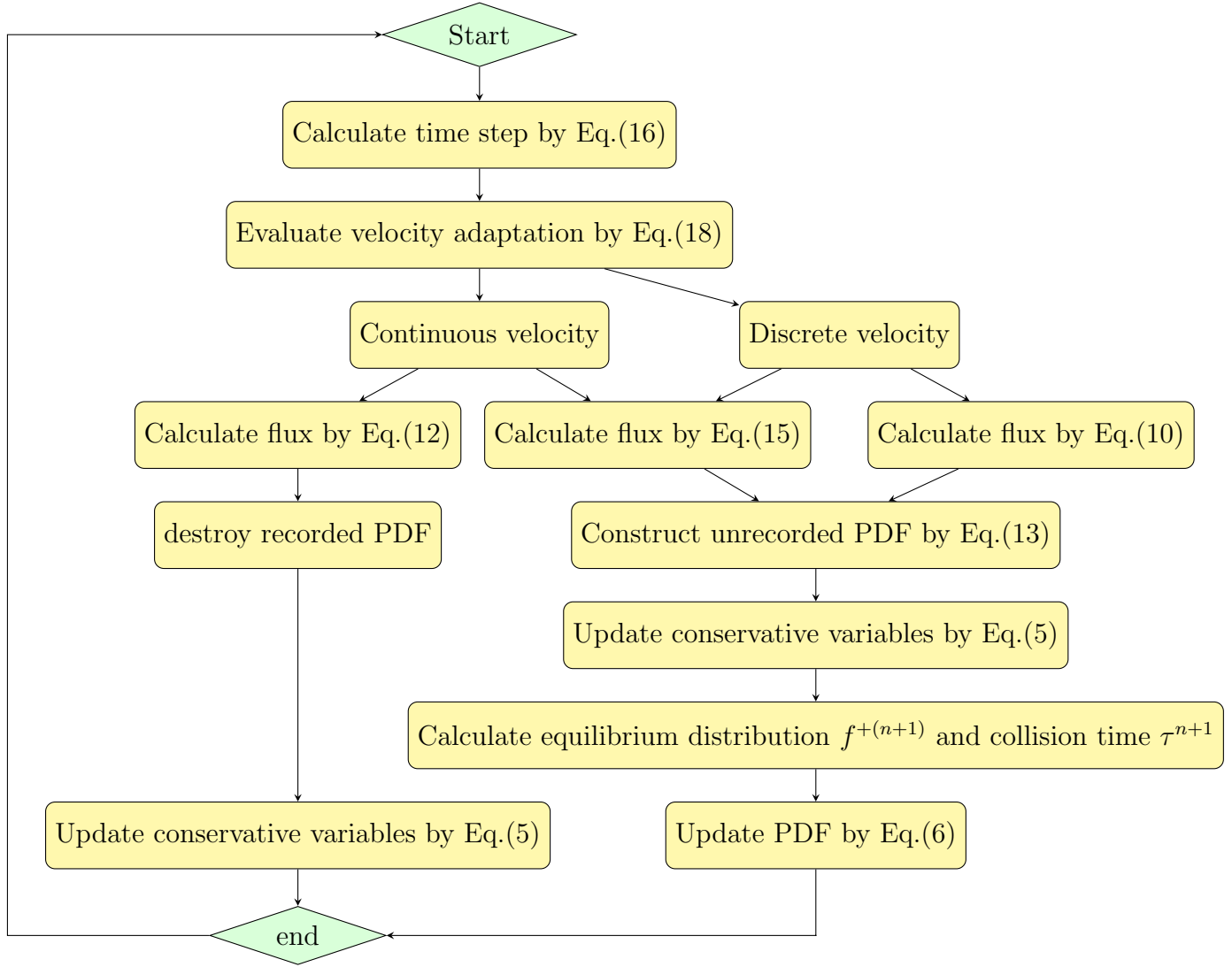

\section{Numerical experiments}

In this section, we are going to present some numerical experiments to test the performance of the current AUGKS.
In order to demonstrate multiscale performance of the algorithm, simulations from continuum Euler and Navier-Stokes to free molecule flow are presented.
The following dimensionless flow variables are used in the calculations, 
\begin{equation*}
\begin{aligned}
& \hat{x}=\frac{x}{L_0}, \hat{y}=\frac{y}{L_0}, \hat{\rho}=\frac{\rho}{\rho_0}, \hat{T}=\frac{T}{T_0}, \hat{u}_i=\frac{u_i}{(2RT_0)^{1/2}}, \\ &\hat{U}_i=\frac{U_i}{(2RT_0)^{1/2}}, \hat{f}=\frac{f}{\rho_0 (2RT_0)^{3/2}}, \hat{P}_{ij}=\frac{P_{ij}}{\rho_0 (2RT_0)}, \hat{q}_i=\frac{q_i}{\rho_0 (2RT_0)^{3/2}}, 
\end{aligned}
\end{equation*}
where $u_i$ is the particle velocity, $U_i$ is the macroscopic flow velocity, $P_{ij}$ is the stress tensor, $q_i$ is the heat flux.
The subscript zero represents the reference state.
For simplicity, the hat notation to denote dimensionless variables will be removed henceforth.
Argon gas is used in the simulation with the variable hard sphere (VHS) molecule model, and the dynamic viscosity is related with the Knudsen number in the reference state via
\begin{equation*}
\mu_{ref}=\frac{5(\alpha+1)(\alpha+2)\sqrt \pi}{4\alpha (5-2\omega)(7-2\omega)}Kn_{ref}.
\end{equation*}
In this simulation, we choose $\alpha=1.0$ and $\omega=0.5$ to recover a hard sphere gas,
and the viscosity varies with temperature through
\begin{equation*}
\mu=\mu_{ref} \left( \frac{T}{T_{ref}} \right)^\theta,
\end{equation*}
where $T_{ref}$ is the reference temperature and $\theta=0.81$ is the index of viscosity coefficient.

\subsection{Shock tube problem}

\begin{figure}[htb!]
	\centering
	\subfigure[Density]{
		\includegraphics[width=5cm]{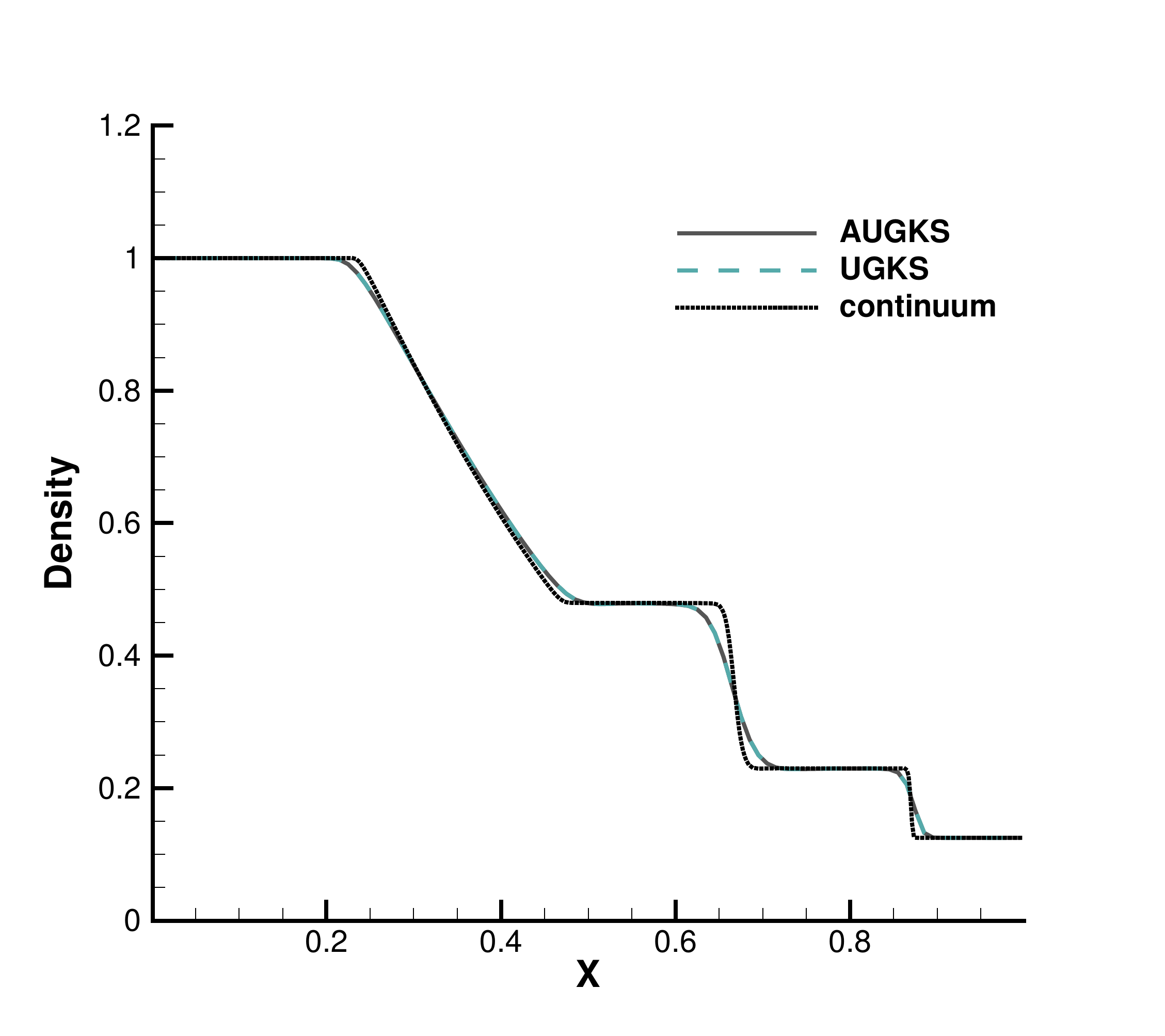}
	}
	\subfigure[Velocity]{
		\includegraphics[width=5cm]{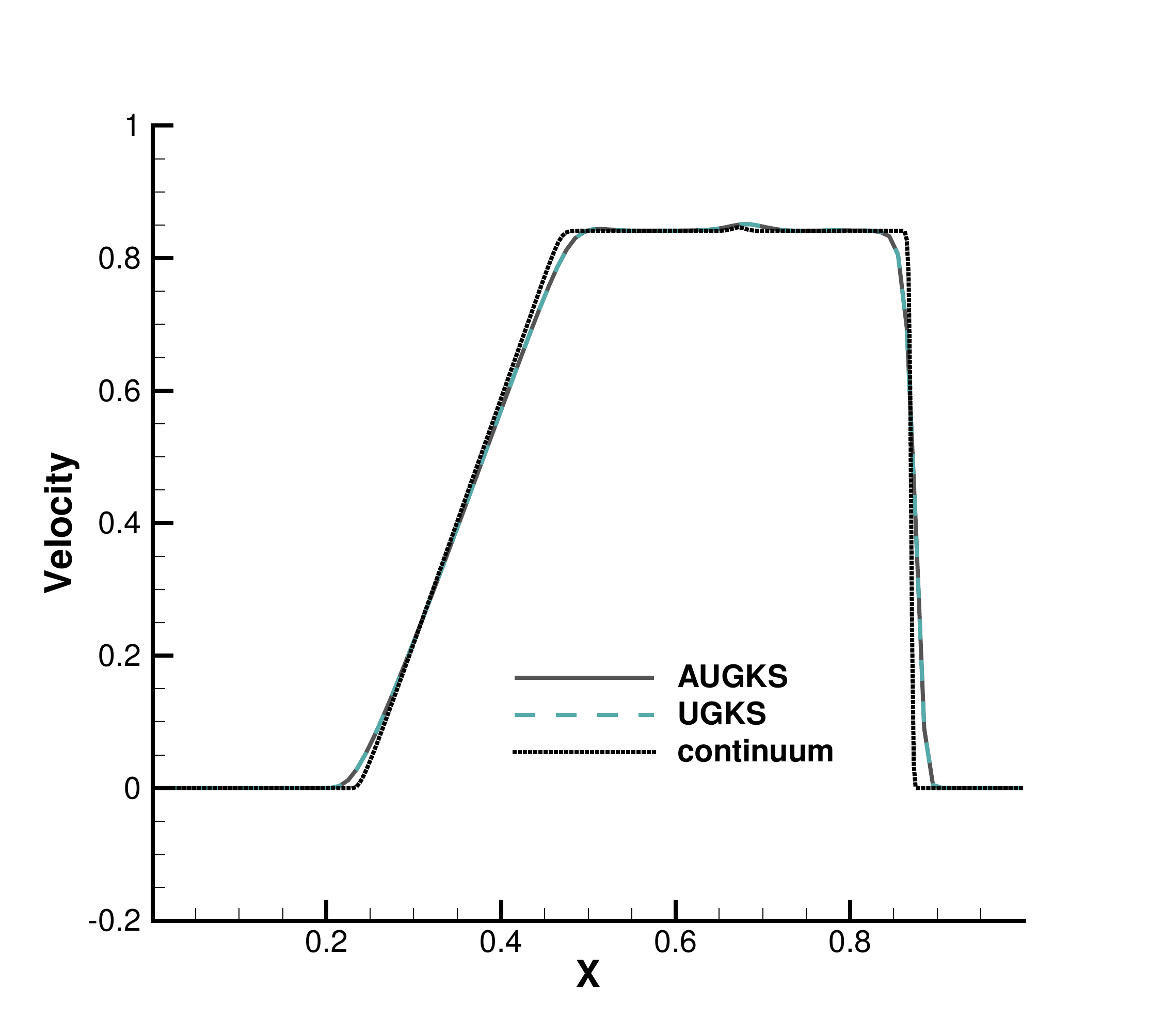}
	}
	\subfigure[Temperature]{
		\includegraphics[width=5cm]{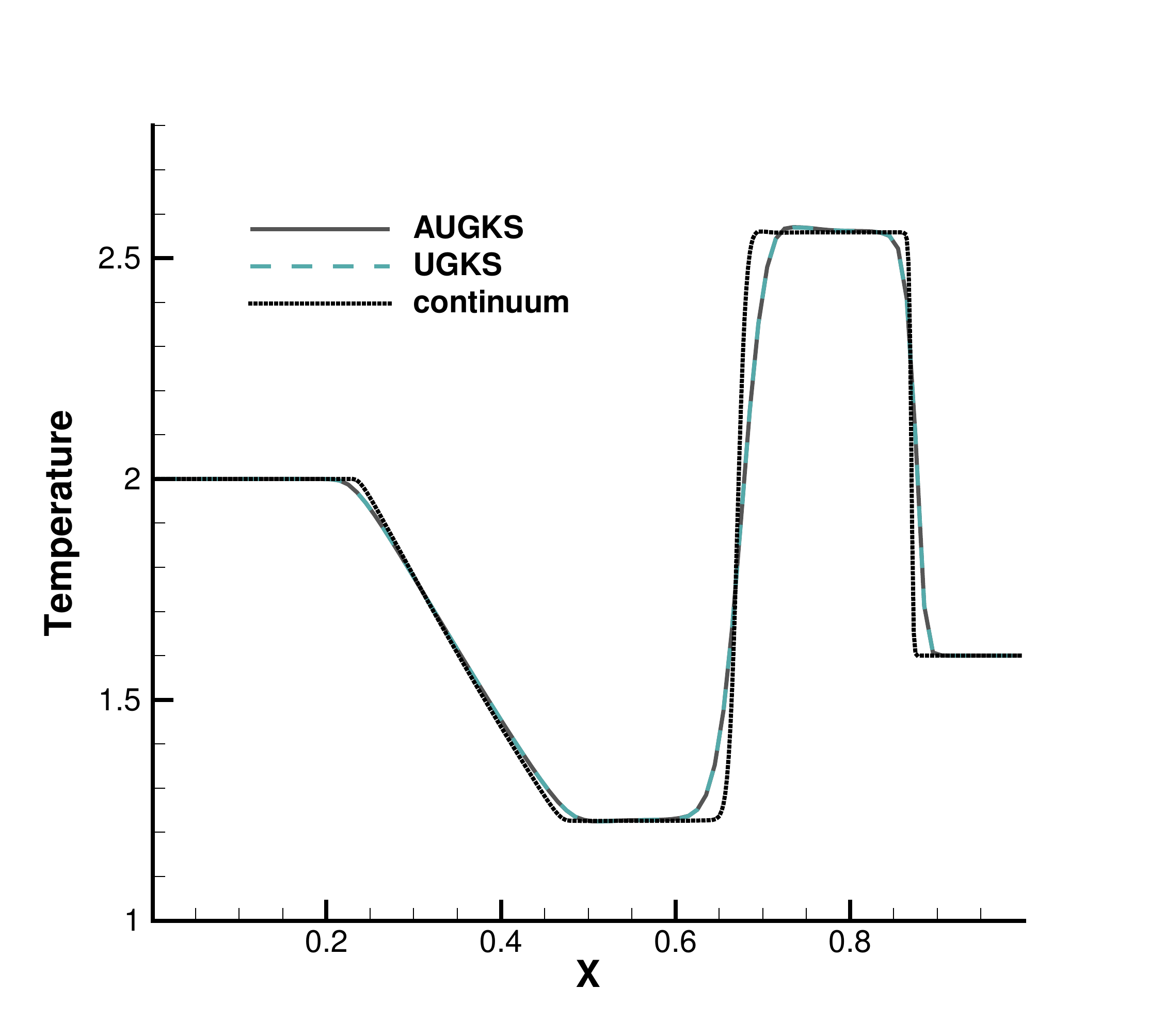}
	}
	\caption{Sod shock tube at $t=0.2$ with reference Knudsen number $Kn=0.0001$.}
	\label{pic:sod 1}
\end{figure}

\begin{figure}[htb!]
	\centering
	\subfigure[Density]{
		\includegraphics[width=5cm]{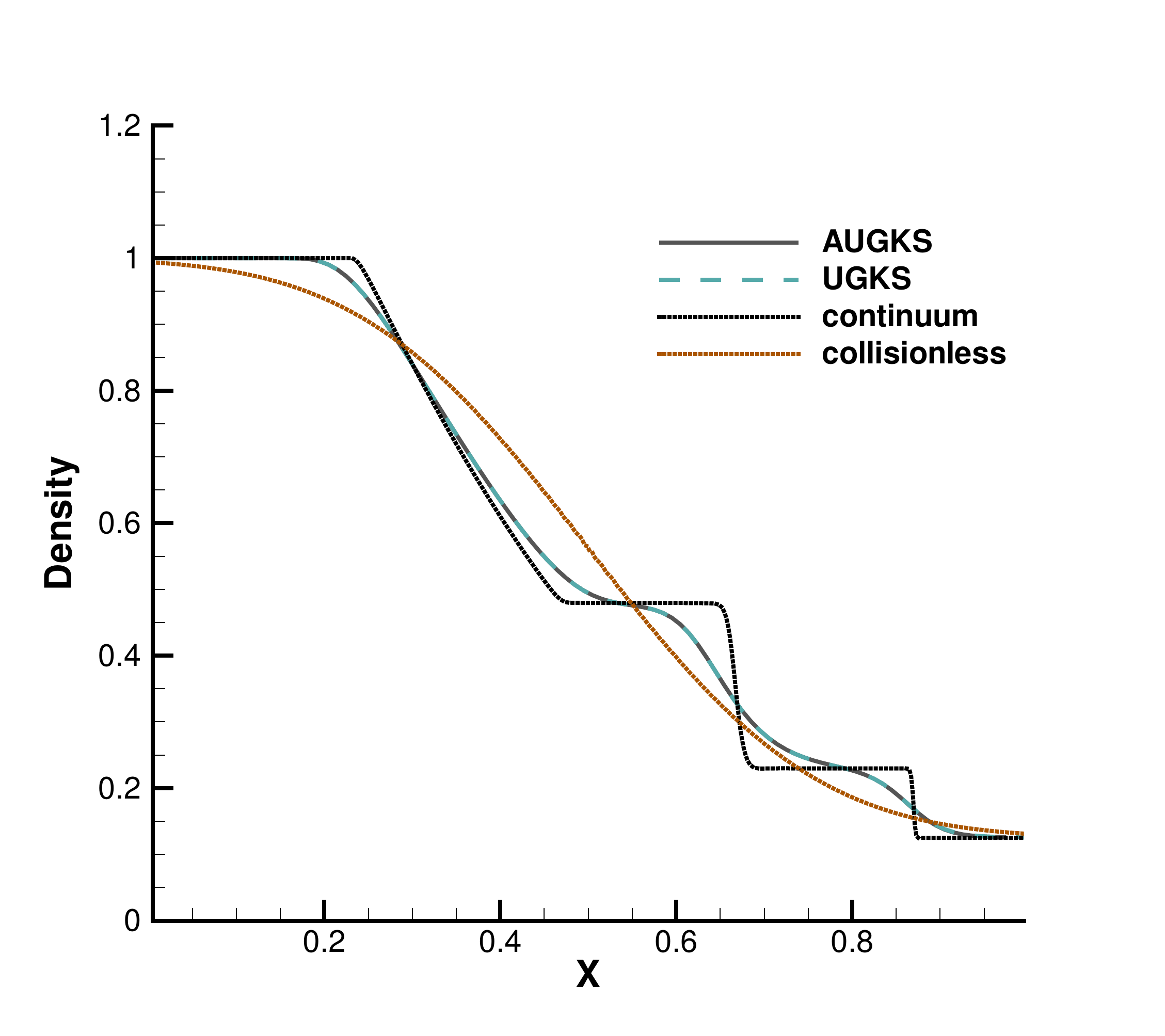}
	}
	\subfigure[Velocity]{
		\includegraphics[width=5cm]{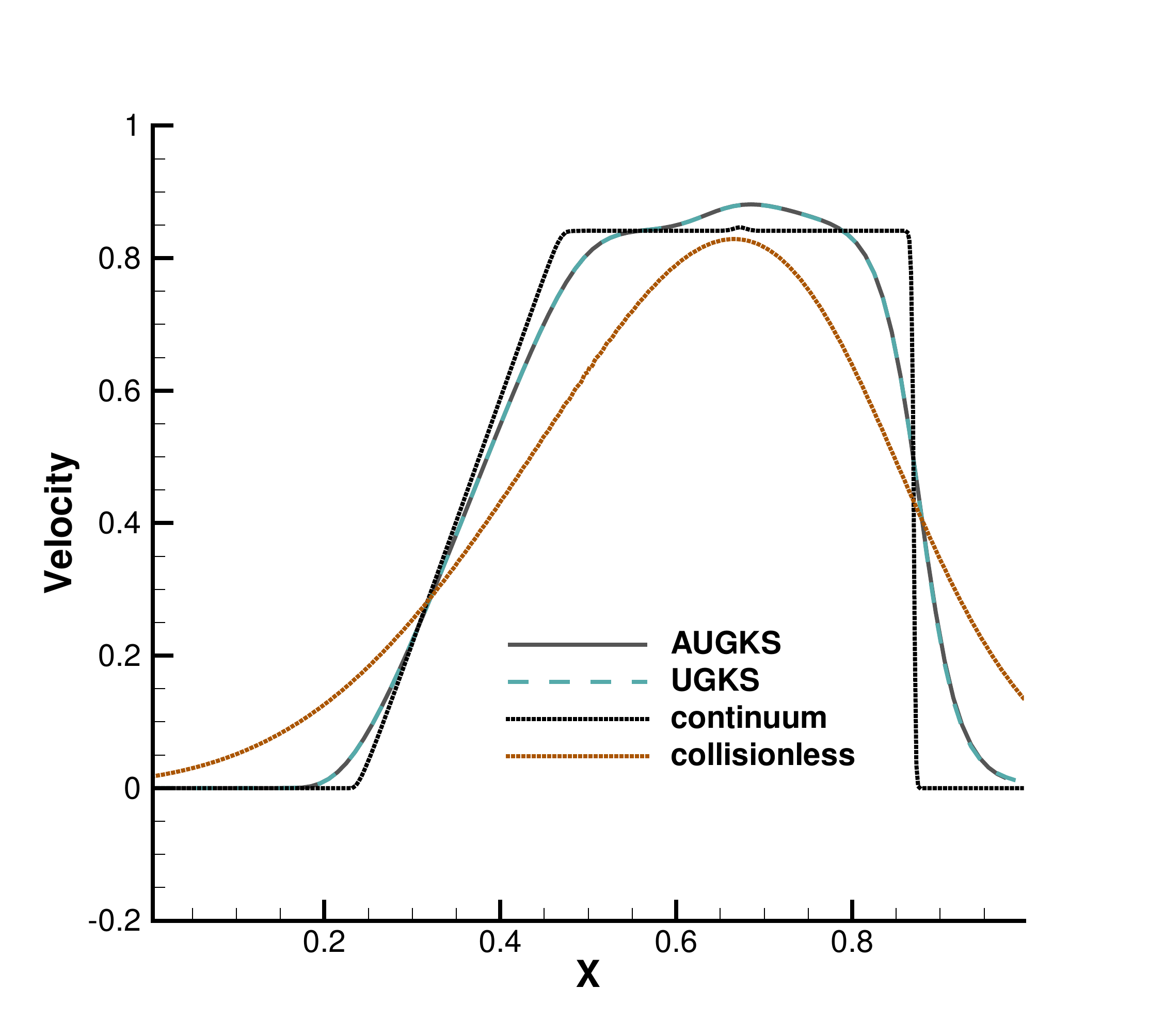}
	}
	\subfigure[Temperature]{
		\includegraphics[width=5cm]{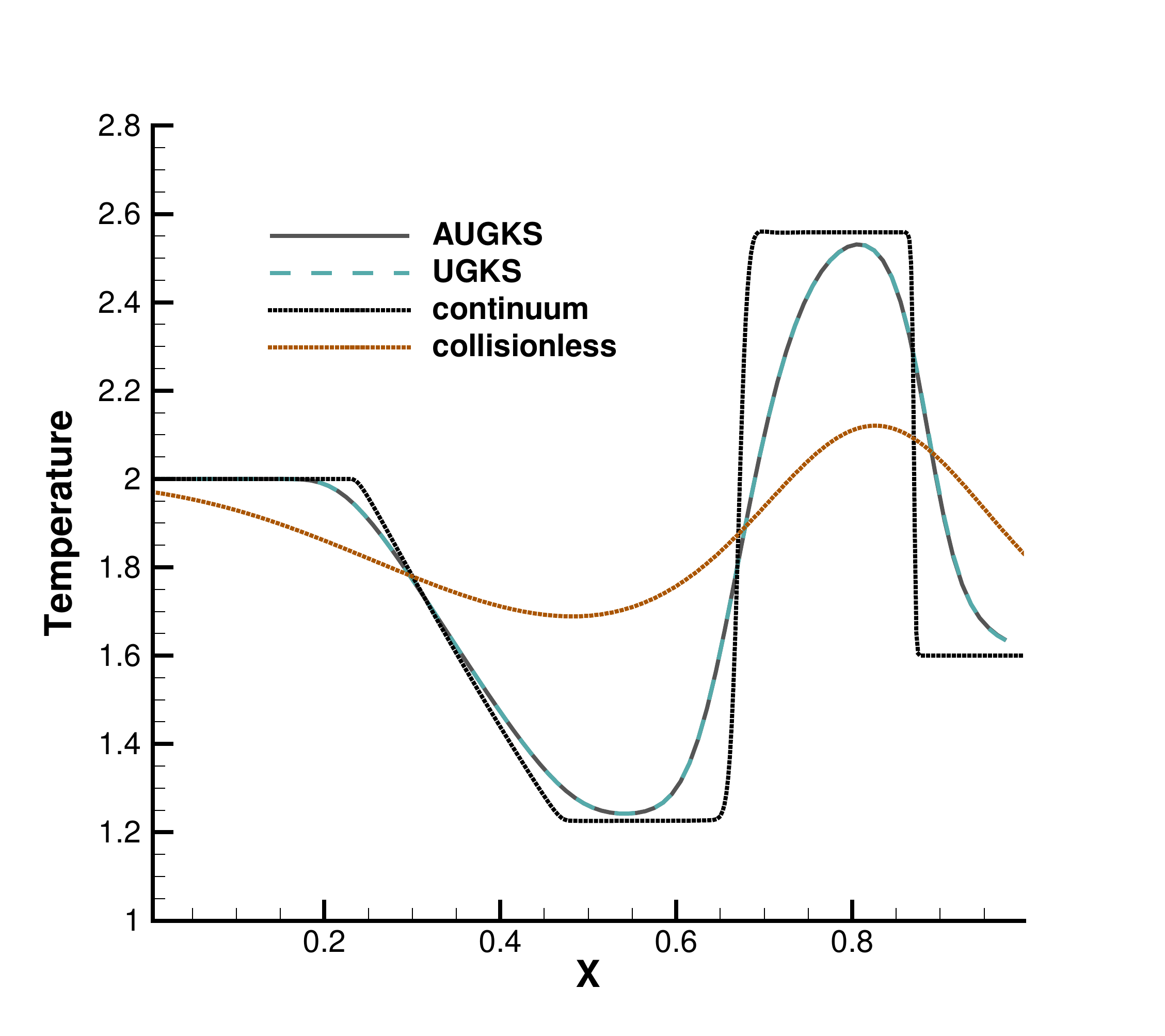}
	}
	\caption{Sod shock tube at $t=0.2$ with reference Knudsen number $Kn=0.001$.}
	\label{pic:sod 2}
\end{figure}

\begin{figure}[htb!]
	\centering
	\subfigure[Density]{
		\includegraphics[width=5cm]{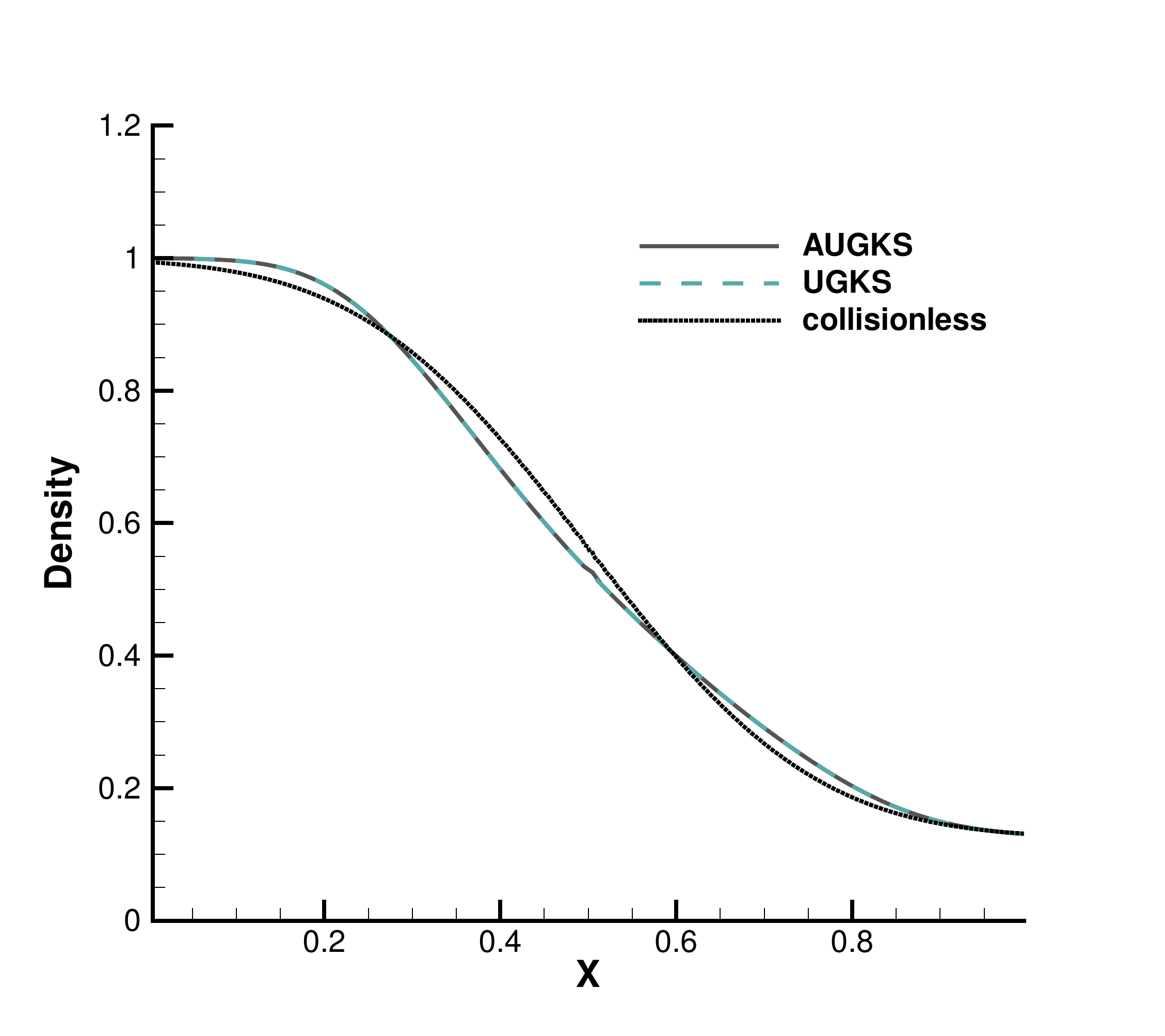}
	}
	\subfigure[Velocity]{
		\includegraphics[width=5cm]{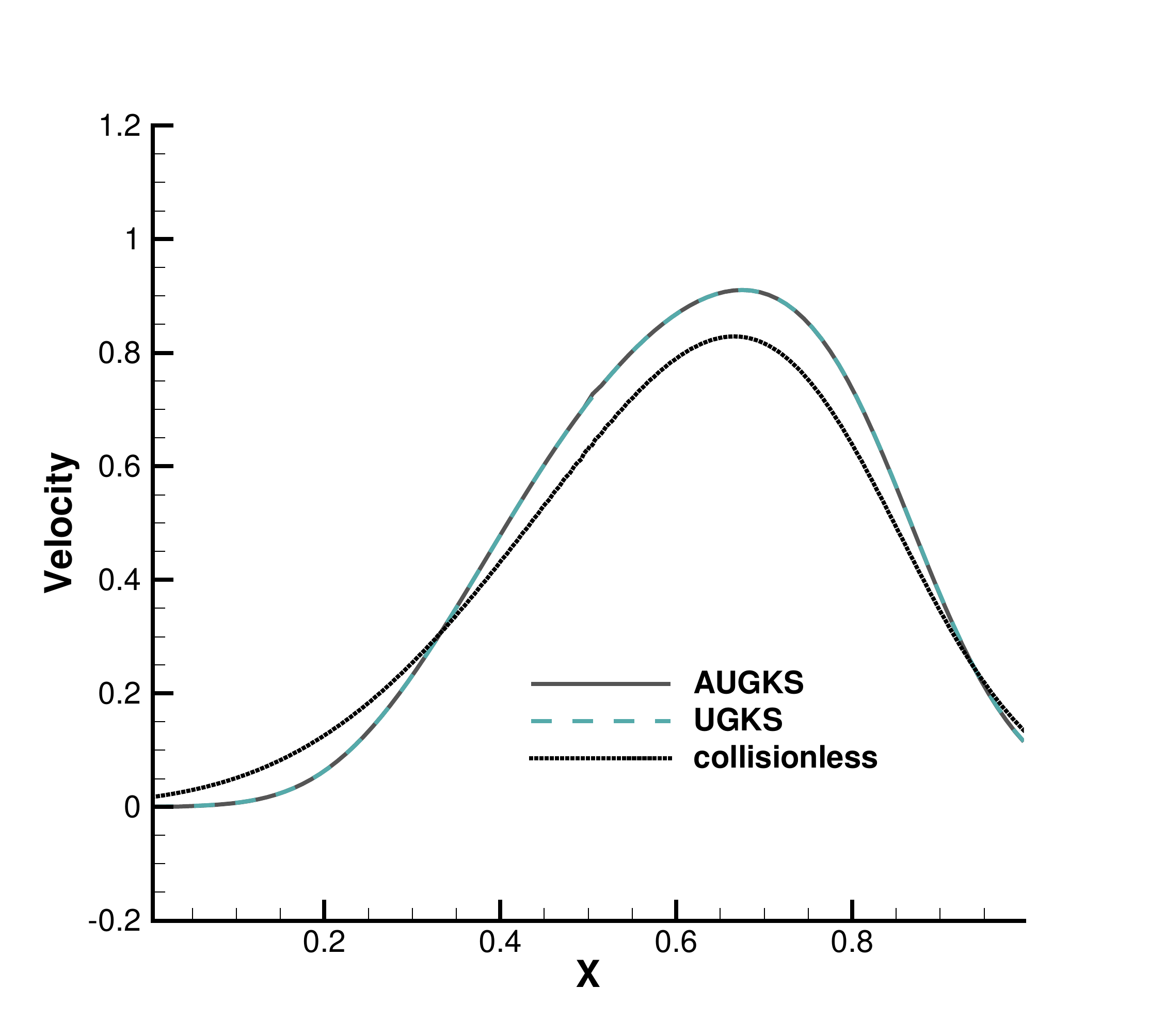}
	}
	\subfigure[Temperature]{
		\includegraphics[width=5cm]{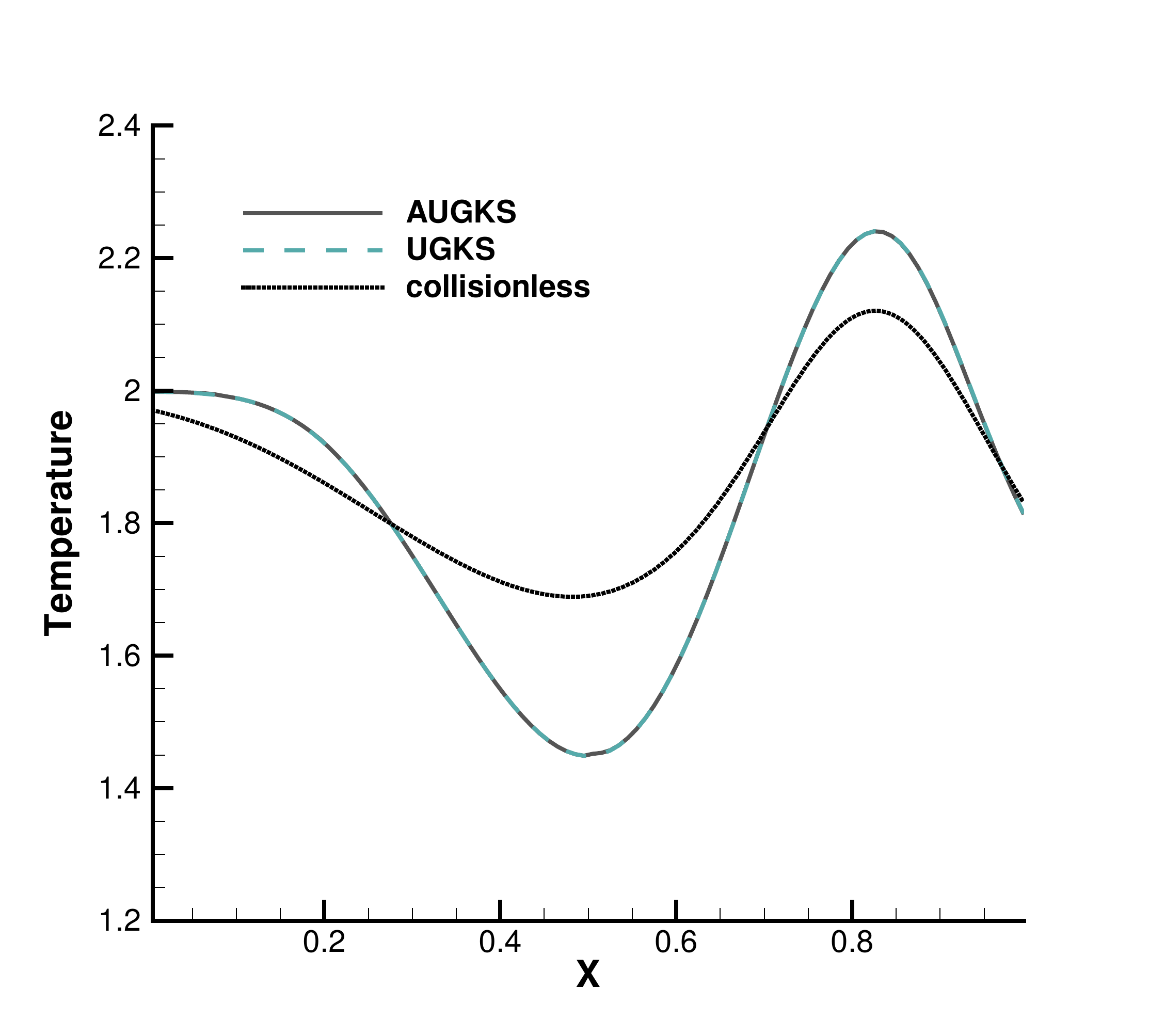}
	}
	\caption{Sod shock tube at $t=0.2$ with reference Knudsen number $Kn=0.01$.}
	\label{pic:sod 3}
\end{figure}

\begin{figure}[htb!]
	\centering
	\subfigure[Density]{
		\includegraphics[width=5cm]{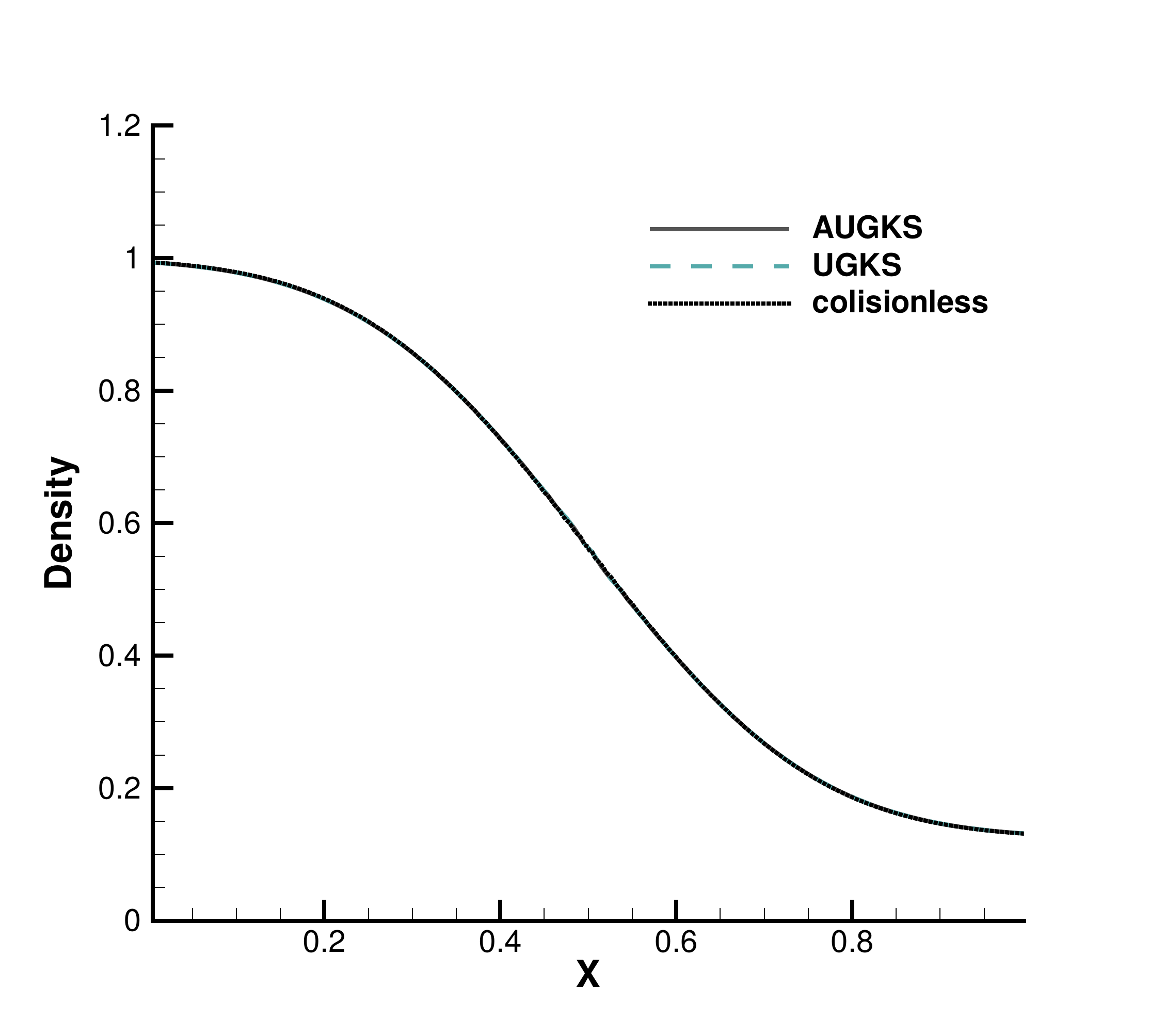}
	}
	\subfigure[Velocity]{
		\includegraphics[width=5cm]{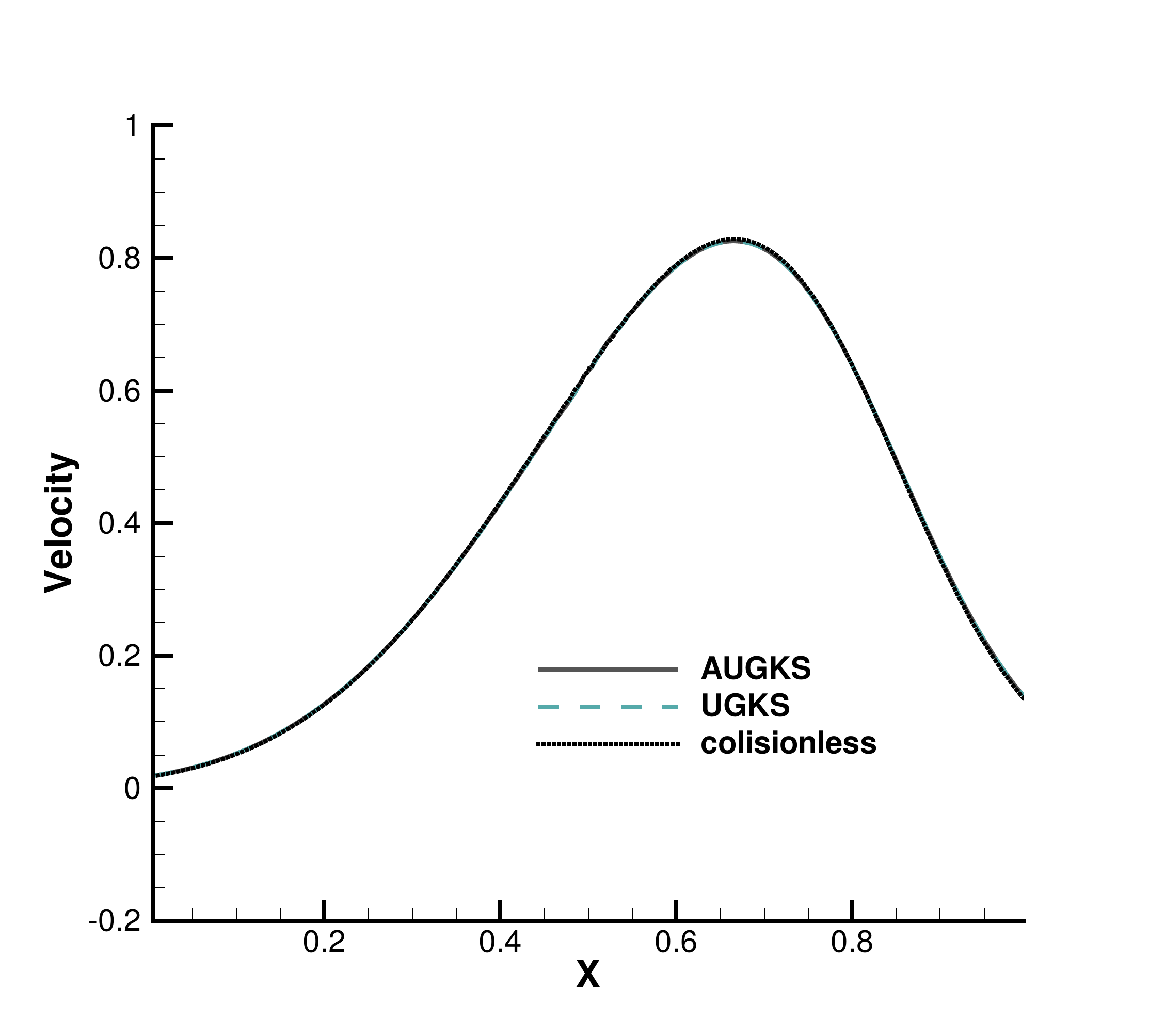}
	}
	\subfigure[Temperature]{
		\includegraphics[width=5cm]{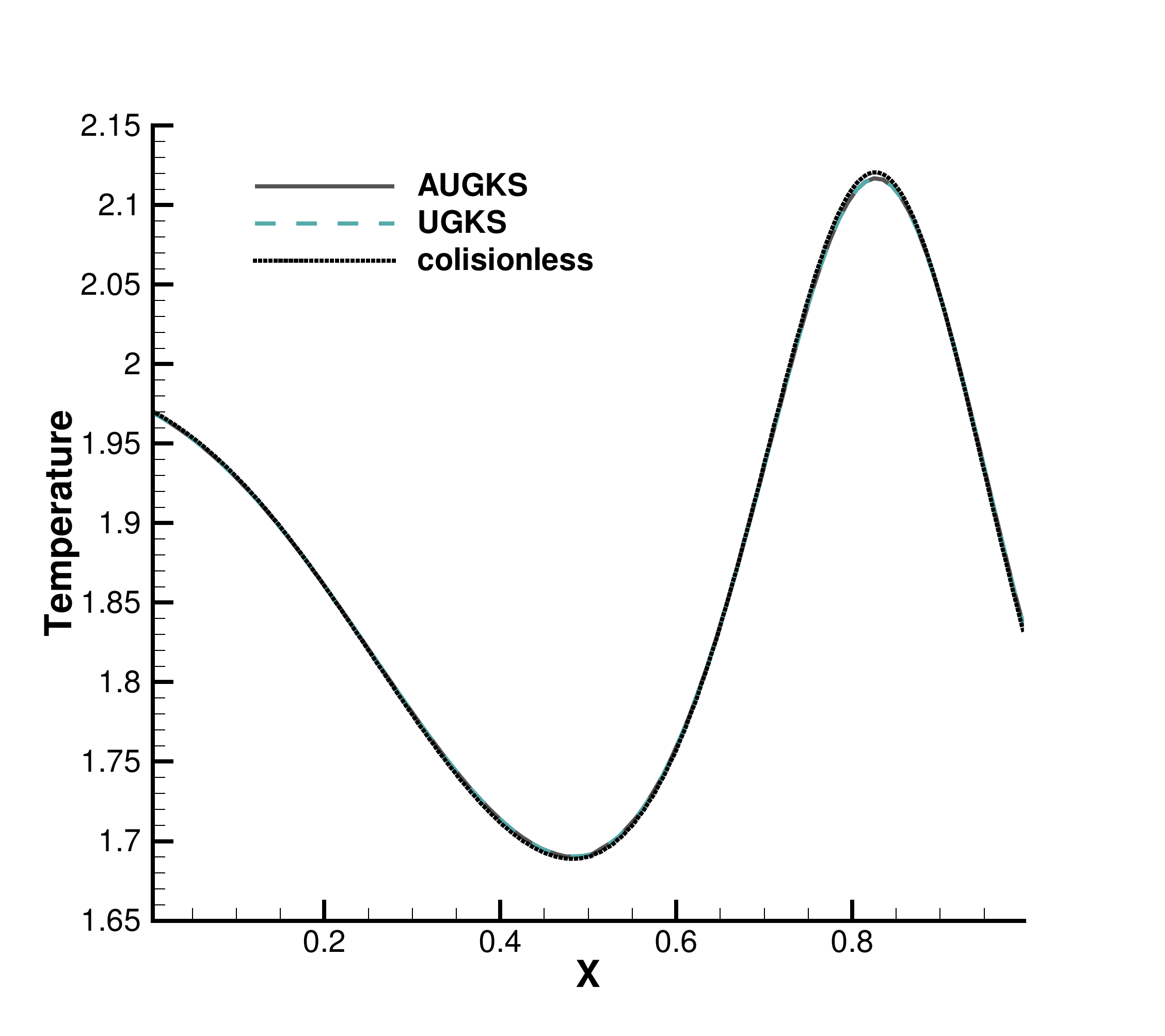}
	}
	\caption{Sod shock tube at $t=0.2$ with reference Knudsen number $Kn=1.0$.}
	\label{pic:sod 4}
\end{figure}

The first case is the Sod shock tube problem.
The flow domain $x\in [0,1]$ is divided into 100 uniform cells.
The initial condition is set as
\begin{equation*}
\begin{aligned}
&\rho=1.0,U=0.0,p=1.0,\ x\leq 0.5,\\
&\rho=0.125,U=0.0,p=0.1,\ x>0.5.
\end{aligned}
\end{equation*}
The simulations are performed with reference Knudsen numbers varying from $Kn=0.0001$ to $Kn=1.0$, corresponding to different flow regimes.
The current criterion value for velocity space transformation is set as $B=0.0001$.
The velocity space is discretized into $80$ uniform points for the update of particle distribution function.
In the AUGKS and UGKS, the full Boltzmann collision operator is solved here by the fast spectral method \cite{wu2013deterministic}.
The numerical solutions at $t=0.2$ are presented in Fig. \ref{pic:sod 1}, \ref{pic:sod 2}, \ref{pic:sod 3} and \ref{pic:sod 4}.
The reference solution of continuum flow is calculated by the continuous GKS solver with 1000 cells, and the free molecular flow solution is derived from the collisionless Boltzmann equation.

\begin{figure}[htb!]
	\centering
	\subfigure[$Kn=0.0001$]{
		\includegraphics[width=6cm]{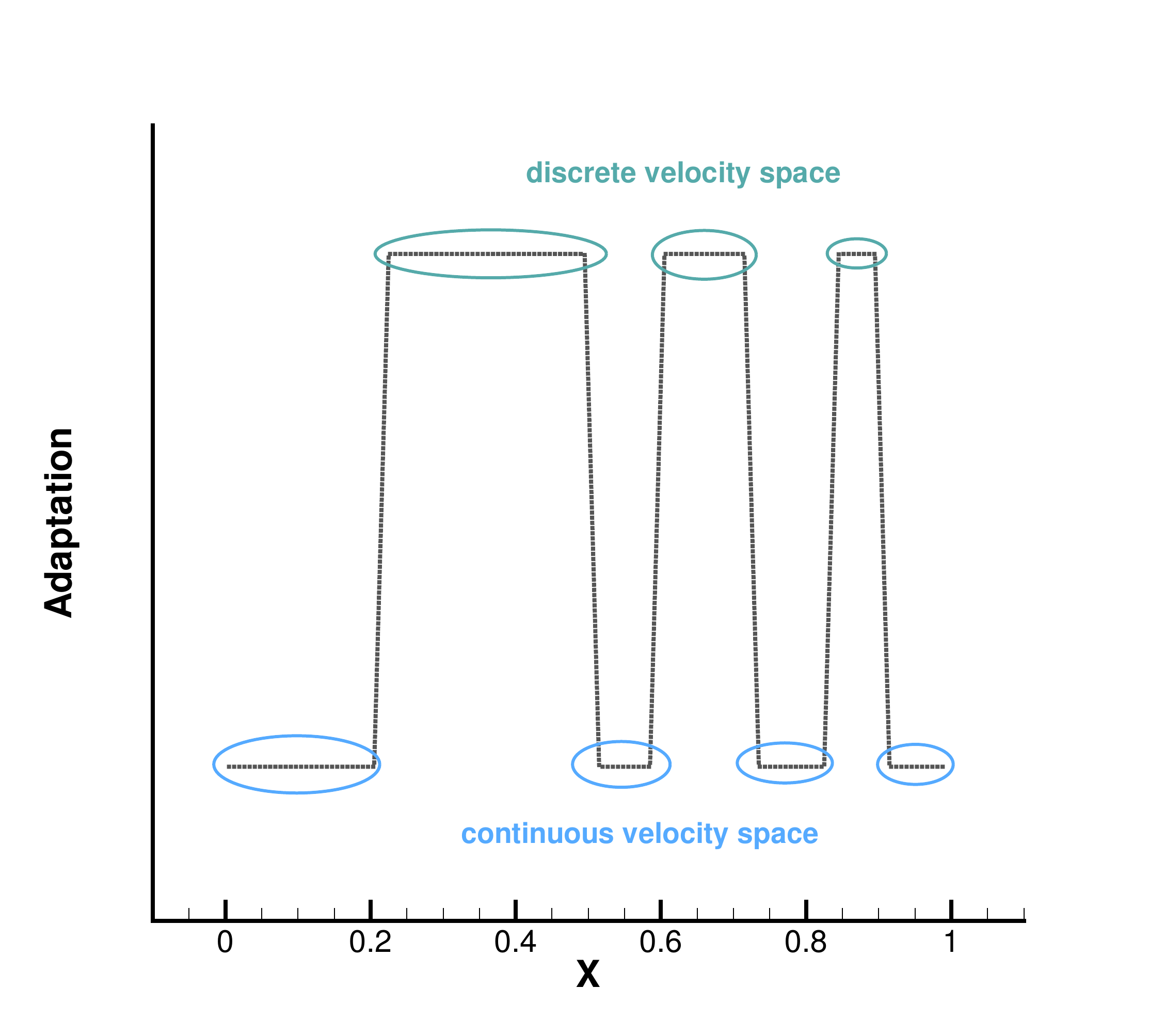}
	}
	\subfigure[$Kn=0.001$]{
		\includegraphics[width=6cm]{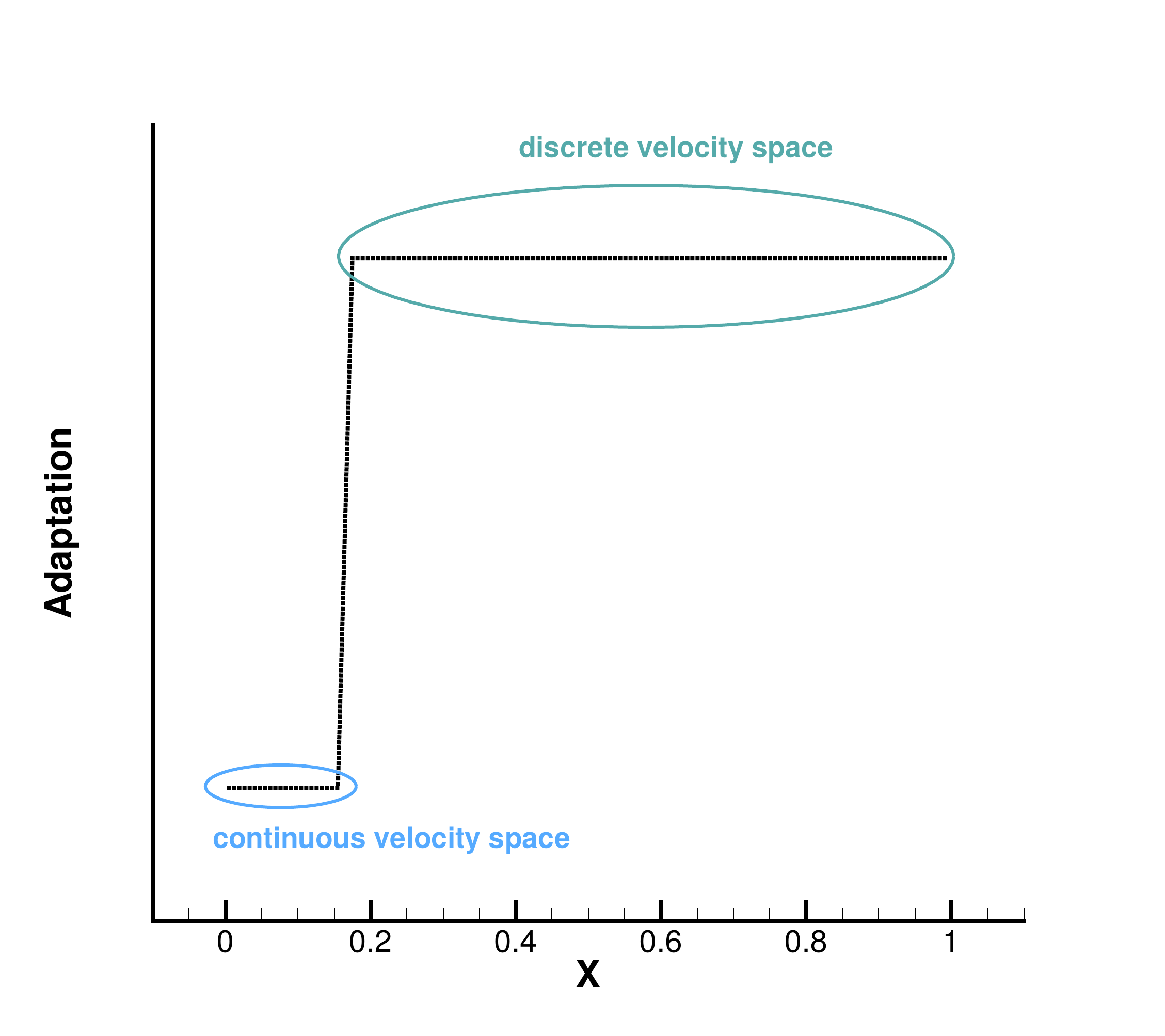}
	}
	\caption{Velocity space adaptation inside the shock tube at $t=0.2$.}
	\label{pic:sod regime}
\end{figure}

In the simulation, the region with initial homogeneous spatial distribution of flow variables is calculated with a continuous velocity space, except the central discontinuity simulated with a discretized velocity.
As time evolves, the non-equilibrium region enlarges along with the use of discrete velocity space.
As presented in Fig. \ref{pic:sod regime}a, at $Kn=0.0001$ and $t=0.2$, the flow domain is divided into some subzones, where the non-equilibrium particle distribution function inside rarefaction wave, contact discontinuity, and shock wave is fully resolved with the discretized velocity space, while in the rest near-equilibrium regions the Chapman-Enskog expansion is adopted over a continuous velocity space.
The solutions of AUGKS at $Kn=0.0001$ and $t=0.2$ are presented in Fig. \ref{pic:sod 1}a, \ref{pic:sod 1}b, \ref{pic:sod 1}c, which match the benchmark continuum and UGKS solutions accurately.
As the Knudsen number gets to $Kn=0.001$ at $t=0.2$, near-equilibrium region confines to a small part near the left tube boundary, where the distribution function has a continuous velocity space, which is shown in Fig. \ref{pic:sod regime}b.
With increasing rarefaction effect, the distributions of flow variables deviate from the NS solutions gradually and tend to collisionless Boltzmann solutions.
As the reference Knudsen number gets to $Kn=0.01$, the Navier-Stokes solutions lose its validity quickly from the initial condition, and the non-equilibrium region occupies the whole tube at $t=0.2$.
The numerical solution approaches to the collisionless Boltzmann solution at $Kn=1.0$, as shown in Fig. \ref{pic:sod 3} and \ref{pic:sod 4}.

This test case illustrates the capacity of AUGKS to simulate flow in different regimes.
The asymptotic preserving (AP) property is confirmed in the two limiting solutions.
With increasing reference Knudsen number, there is a smooth transition from the Euler solution of the Riemann problem to collisionless Boltzmann solution.
Table 1 presents the CPU time cost from the current adaptive scheme and original UGKS method.
As is shown, the AUGKS is about $3.63$ times faster than UGKS at $Kn=0.0001$.
Since the computational cost is proportional to the mesh points in the velocity space, it is expected that the computational efficiency is closely related to the size of non-equilibrium region.
When the rarefaction increases, the CPU time of AUGKS increases correspondingly, while it is still more efficient than the original UGKS.

\begin{table}  
	\label{table:sod}
	\centering
	\begin{tabular*}{7cm}{lll}  
		\hline  
		& AUGKS  & UGKS  \\
		\hline  
		Kn=0.0001  & 2042.98 & 7421.70  \\  
		Kn=0.001  & 3537.73 & 7430.82  \\  
		Kn=0.01  & 4692.52 & 7547.99  \\  
		Kn=0.1  & 5694.46 & 7275.40  \\  
		\hline 
	\end{tabular*}  
	\caption{CPU time cost in the Sod case using AUGKS and original UGKS at different Knudsen numbers.}  
\end{table}

\subsection{Rayleigh flow}

A Rayleigh flow forms over a plate which suddenly acquires a constant parallel velocity and temperature.
In this test case, we follow the setup by Sun \cite{sun2003information}.
As shown in Fig. 8, the argon gas is at rest and has a unit temperature initially.
When $t>0$, the plate suddenly moves with a constant velocity $U_w=0.0296$ and temperature $T_w=1.36$.
The momentum and energy are transported into the flow field through a shearing effect in the unsteady process.
A physical domain $y\in [0,1]$ with 100 uniform cells are set up for the simulation, and the 32 uniform points are used in the velocity space where the particle distribution function is updated directy.
In this case, the full Boltzmann collision operator in the AUGKS and UGKS is calculated by the fast spectrum method \cite{wu2013deterministic}.
The current switching criterion of velocity space is set as $B=0.0001$.

Numerical simulations are performed with a series of reference Knudsen number, and solutions at same output times are plotted in Fig. \ref{pic:rayleigh 1}, \ref{pic:rayleigh 2}, \ref{pic:rayleigh 3} and \ref{pic:rayleigh 4}.
Besides AUGKS solutions, the UGKS and DSMC results are also provided as benchmarks.
With Maxwell's fully accommodation boundary condition, Bird \cite{bird1994molecular} proposed an analytical solution from the collisionless Boltzmann equation when the time is much less than the reference mean collision time $\tau_0=\ell_0/v_0$, where $\ell_0$ is particle mean free path and $v_0$ is the mean molecular speed.
The analytical collisionless solution is also plotted in figures.

As presented in Fig. \ref{pic:rayleigh 1}, \ref{pic:rayleigh 2}, \ref{pic:rayleigh 3} and \ref{pic:rayleigh 4}, for the case at $Kn=2.66$ and $t=0.1\tau_0$, the AUGKS recovers exact collisionless Botlzmann solution. 
In the transition regime $Kn=0.266$ and $Kn=0.0266$ at $t=\tau_0$ and $t=10\tau_0$, the numerical solutions deviate from collisionless solutions gradually due to increasing intermolecular collisions.
At $t=100\tau_0$ and $Kn=0.00266$ corresponding to a near-continuum regime, the current adaptive scheme recovers the Navier-Stokes solutions with intensive intermolecular collisions.
As plotted, in all cases the AUGKS solutions match up well with the benchmark solutions from DSMC and UGKS.
It is worth mentioning that in comparison with DSMC method, the current Boltzmann-equation-based adaptive unified scheme has no statistical scattering, which is beneficial in low speed simulations.

Fig. \ref{pic:rayleigh regime} presents the velocity space adaptation inside the flow domain at the output time.
In the case with $Kn=0.00266$, in the near-wall region with large slope of macroscopic variables, the AUGKS uses a discrete velocity space, while a continuous velocity space is used in the outer domain.
As the Knudsen number increases, the enhanced dimensionless viscosity and heat conductivity lead to a large non-equilibrium region.
As a result, the non-equilibrium region enlarges faster.
For the case $Kn=0.266$, in all flow region the distribution function deviates from the Chapman-Enskog solution and its evolution must be followed with a discretized velocity space.
Table 2 presents the computational cost of the AUGKS and UGKS.
When $Kn=0.00266$, the AUGKS is $7.45$ times faster than the original UGKS.
When the Knudsen number increases, the enhanced non-equilibrium effects increase the computational cost in AUGKS.
From the current numerical experiments, it is clear that the AUGKS provides a self-adjusted algorithm from continuum to rarefied flow simulation with the consideration of both accuracy and efficiency.

\begin{figure}[htb!]
	\centering
	{
		\begin{tikzpicture}[thick]
		\node[rectangle] (bx) {};
		\draw ($(bx)+(-4,+0.5)$) -- ($(bx)+(+4,+0.5)$);
		\node at ($(bx)+(+0,+0)$) {$  T_w=1.36,U_w=0.0296$};
		\draw[line] ($(bx)+(2.5,+0.7)$) -- ($(bx)+(4,+0.7)$);
		\node at ($(bx)+(+3.25,+1)$) {$U_w$};
		\draw[dashed] ($(bx)+(-4,+4.5)$) -- ($(bx)+(+4,+4.5)$);
		\node at ($(bx)+(+0,+5)$) {$\rm Outer\  boundary$};
		\node at ($(bx)+(+0,+2.8)$) {$\rm Argon\  gas$};
		\node at ($(bx)+(+0,+2.2)$) {$T_0=1,U_0=V_0=0$};
		\draw[line] ($(bx)+(-4,+1)$) -- ($(bx)+(-4,+2)$);
		\draw[line] ($(bx)+(-4,+1)$) -- ($(bx)+(-3,+1)$);
		\node at ($(bx)+(-2.7,+1)$) {$x$};
		\node at ($(bx)+(-3.7,+1.9)$) {$y$};
		\end{tikzpicture}
	}
	\label{pic:rayleigh schematic}
	\caption{Schematic of Rayleigh problem.}
\end{figure}
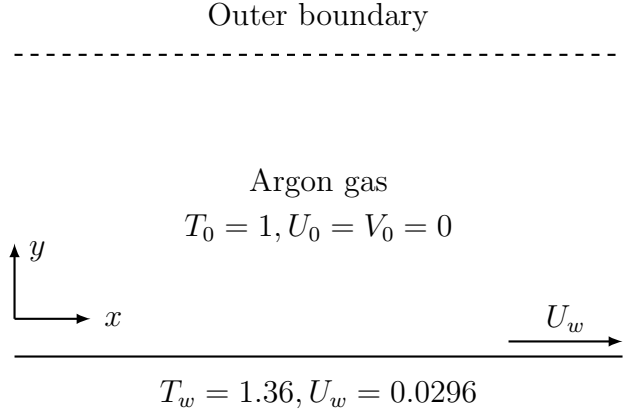

\begin{figure}[htb!]
	\centering
	\subfigure[Density and Temperature]{
		\includegraphics[width=6cm]{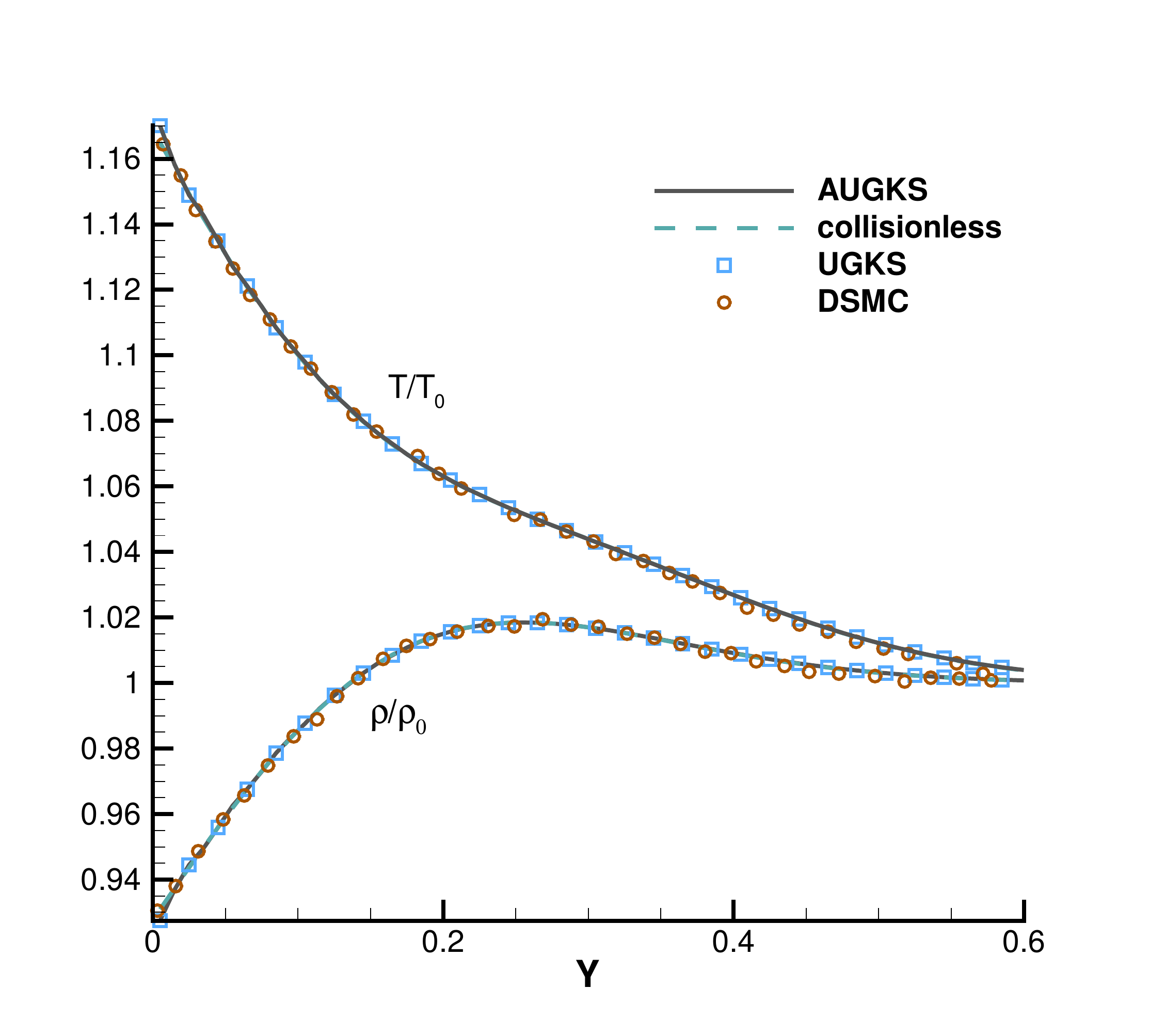}
	}
	\subfigure[Velocity]{
		\includegraphics[width=6cm]{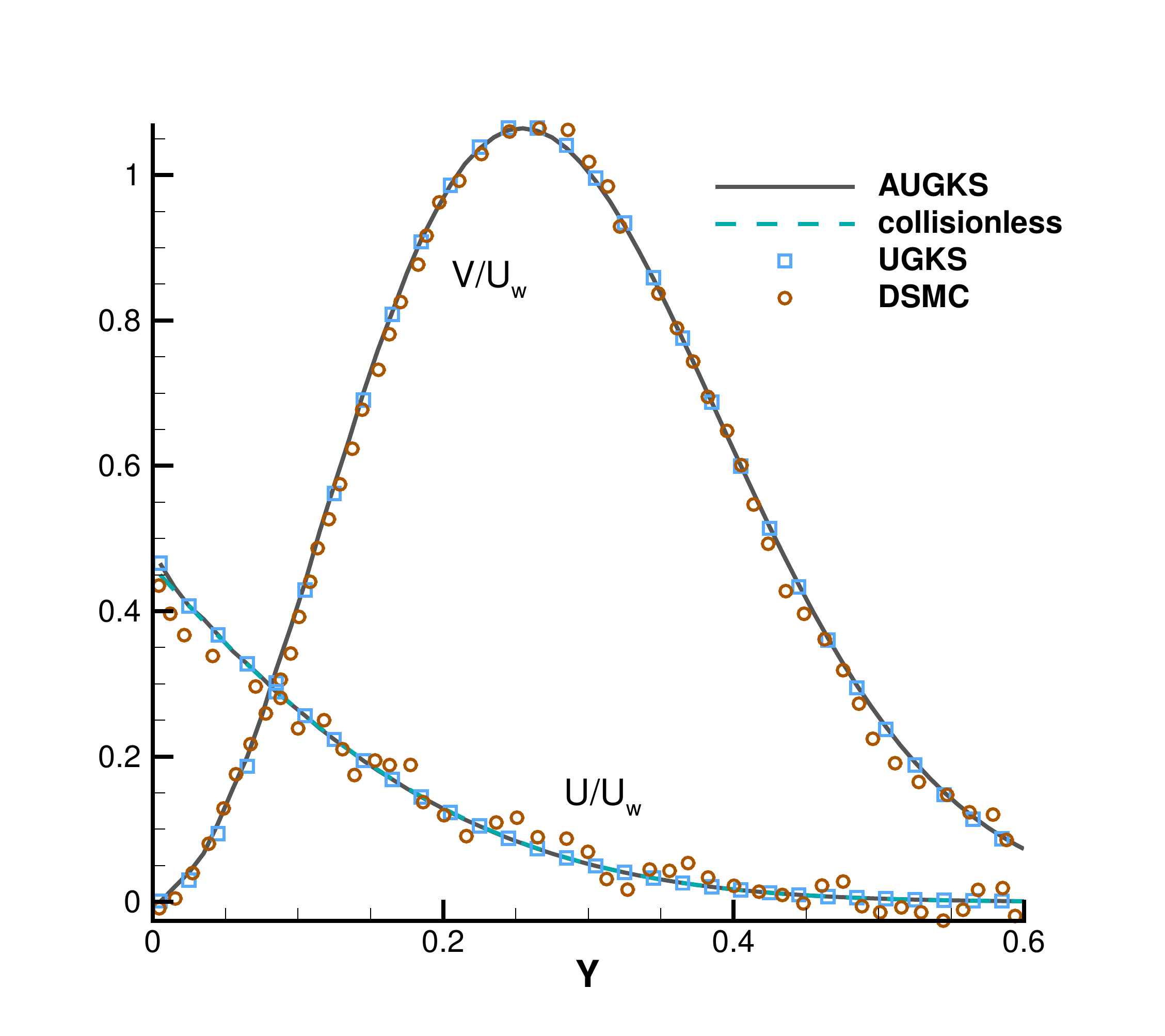}
	}
	\caption{Rayleigh flow at $t=0.1\tau_0$ with reference Knudsen number $Kn=2.66$.}
	\label{pic:rayleigh 1}
\end{figure}

\begin{figure}[htb!]
	\centering
	\subfigure[Density and Temperature]{
		\includegraphics[width=6cm]{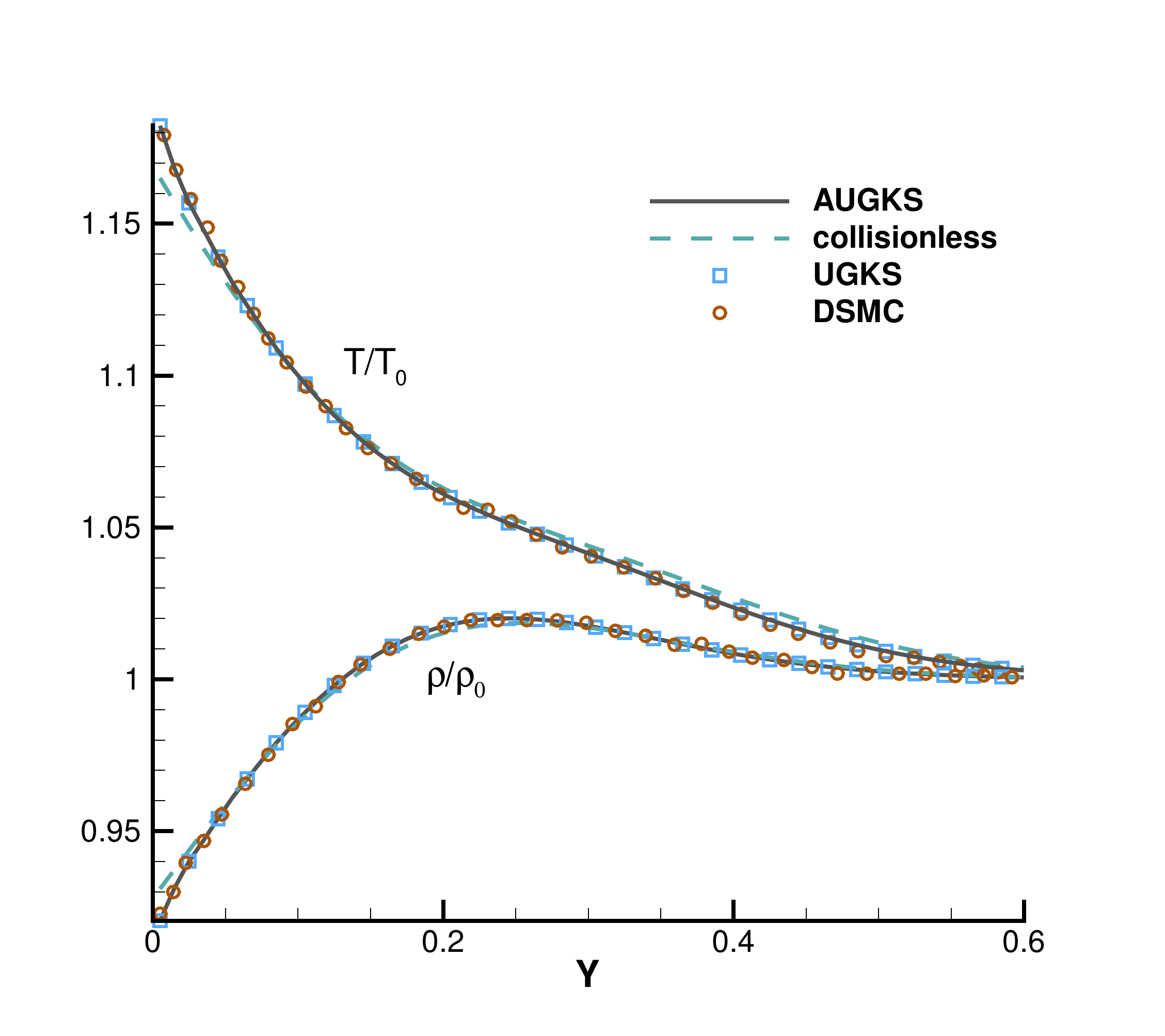}
	}
	\subfigure[Velocity]{
		\includegraphics[width=6cm]{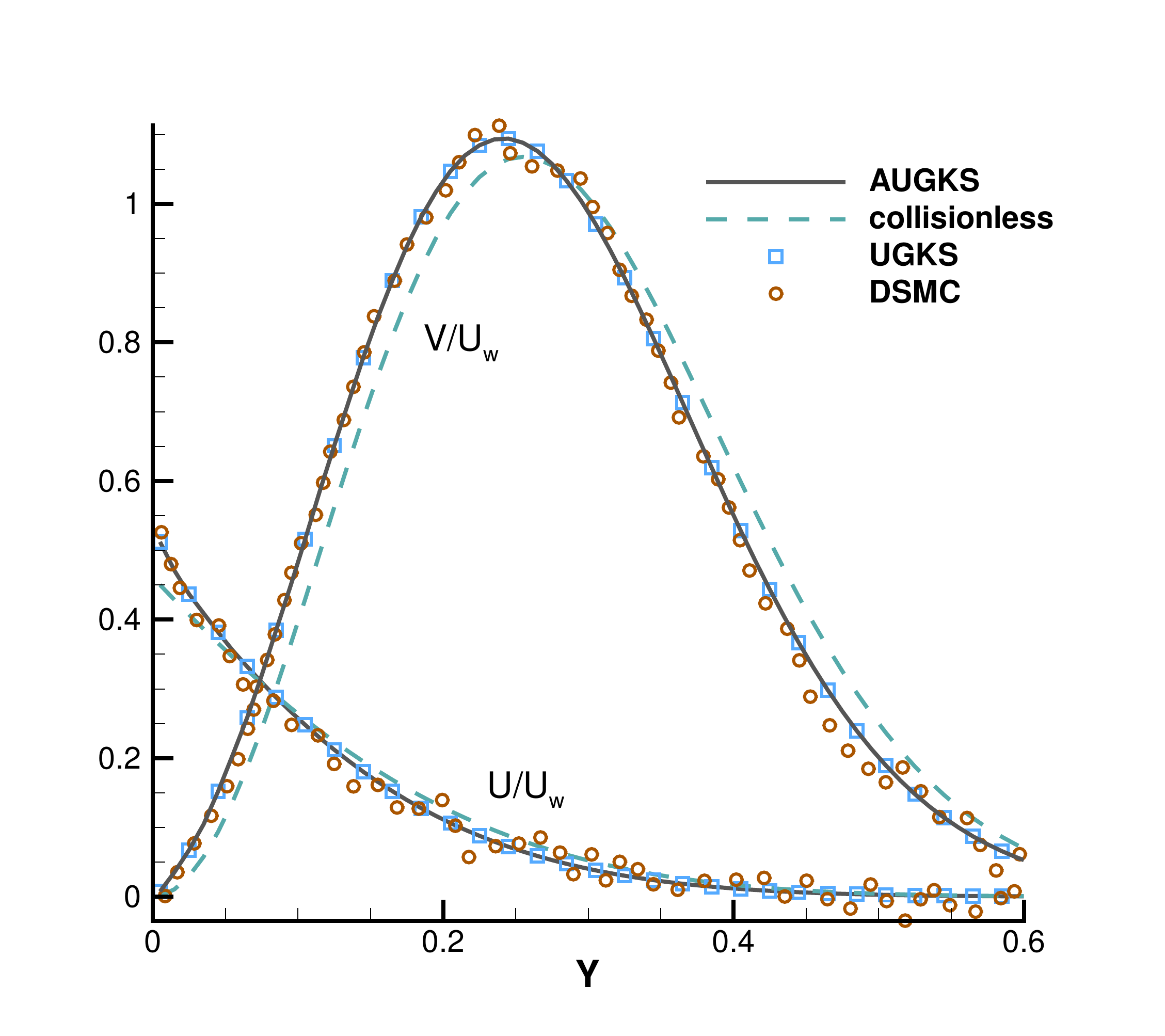}
	}
	\caption{Rayleigh flow at $t=\tau_0$ with reference Knudsen number $Kn=0.266$.}
	\label{pic:rayleigh 2}
\end{figure}

\begin{figure}[htb!]
	\centering
	\subfigure[Density and Temperature]{
		\includegraphics[width=6cm]{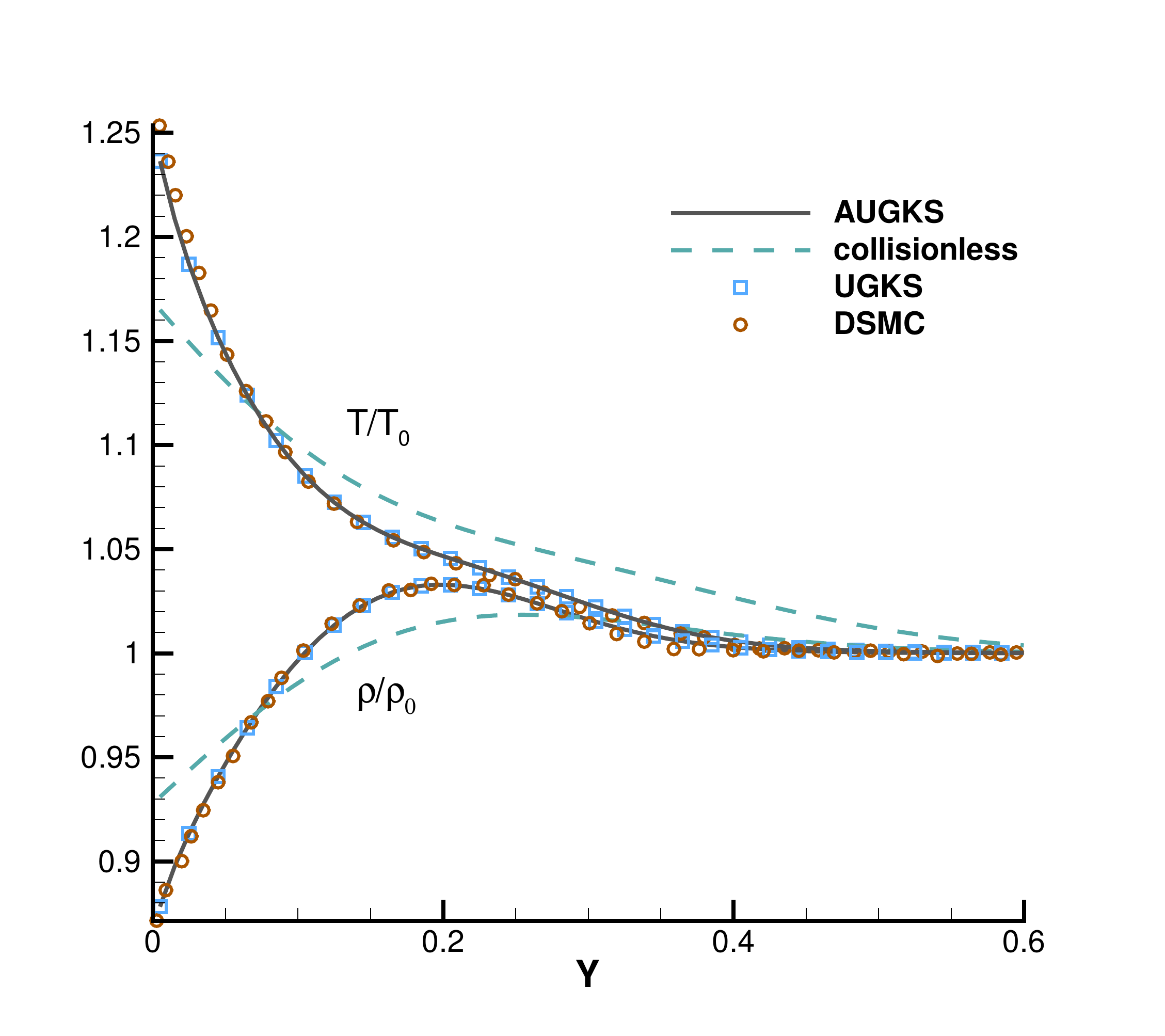}
	}
	\subfigure[Velocity]{
		\includegraphics[width=6cm]{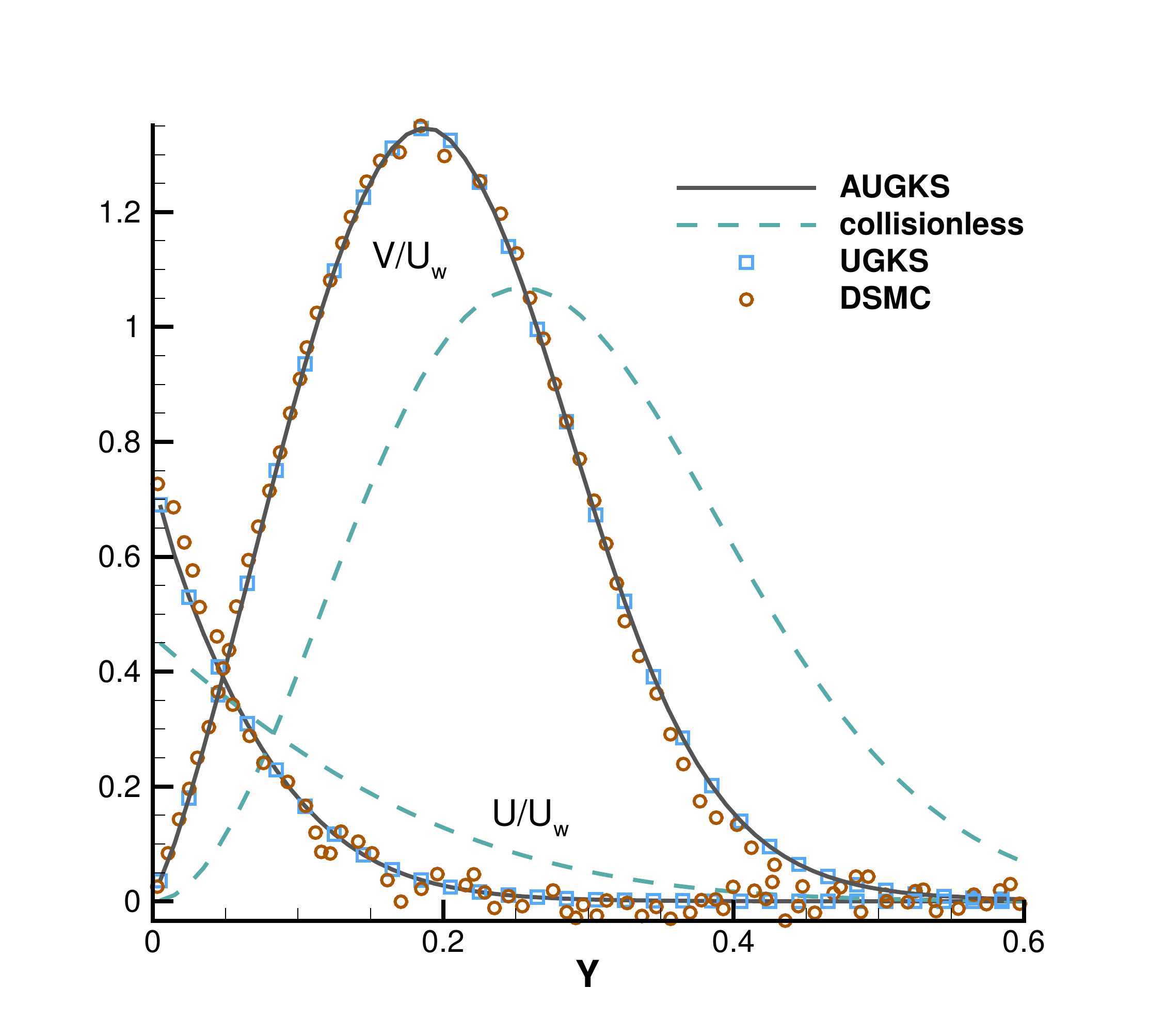}
	}
	\caption{Rayleigh flow at $t=10\tau_0$ with reference Knudsen number $Kn=0.0266$.}
	\label{pic:rayleigh 3}
\end{figure}

\begin{figure}[htb!]
	\centering
	\subfigure[Density and Temperature]{
		\includegraphics[width=6cm]{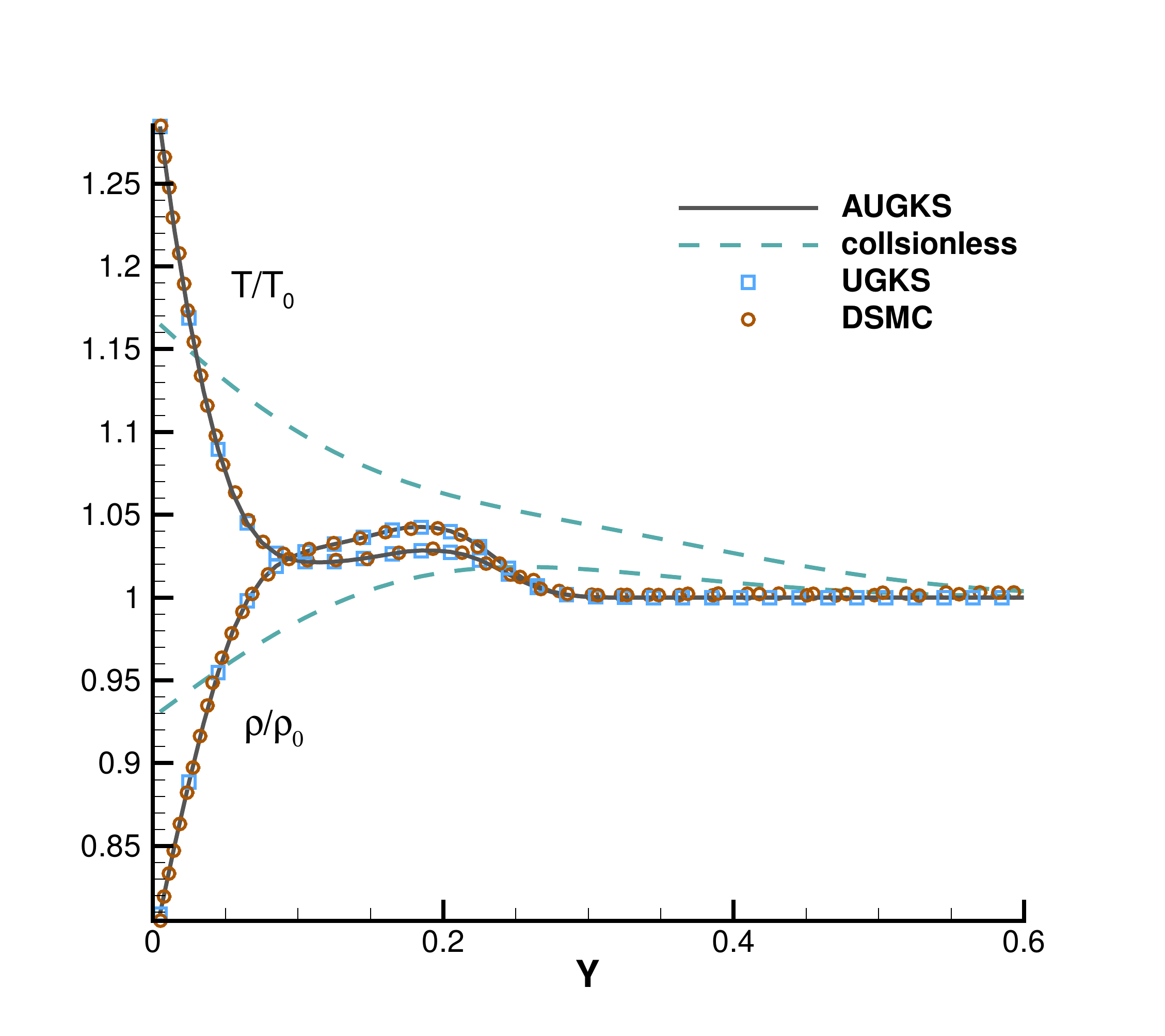}
	}
	\subfigure[Velocity]{
		\includegraphics[width=6cm]{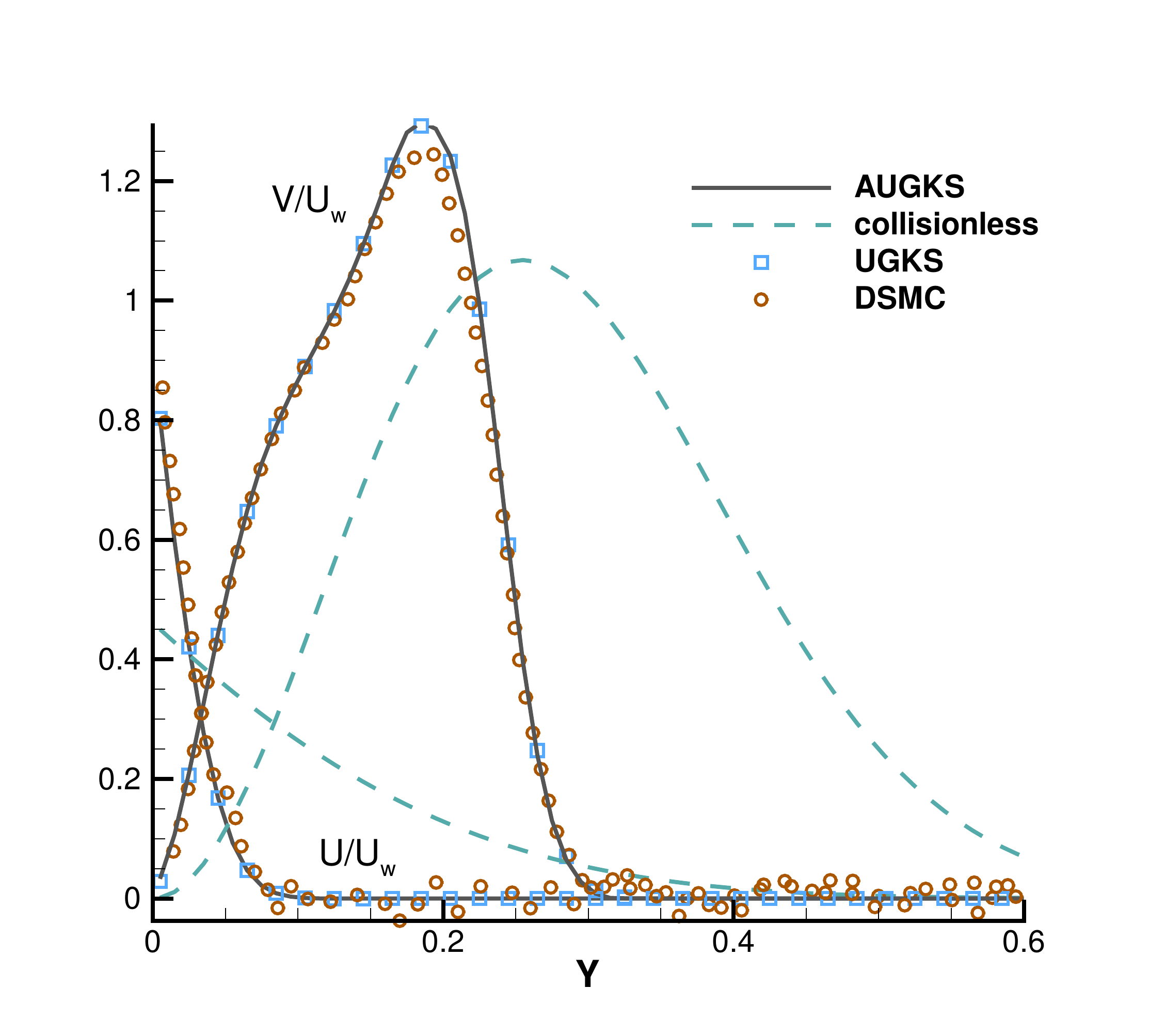}
	}
	\caption{Rayleigh flow at $t=100\tau_0$ with reference Knudsen number $Kn=0.00266$.}
	\label{pic:rayleigh 4}
\end{figure}

\begin{figure}[htb!]
	\centering
	\includegraphics[width=8cm]{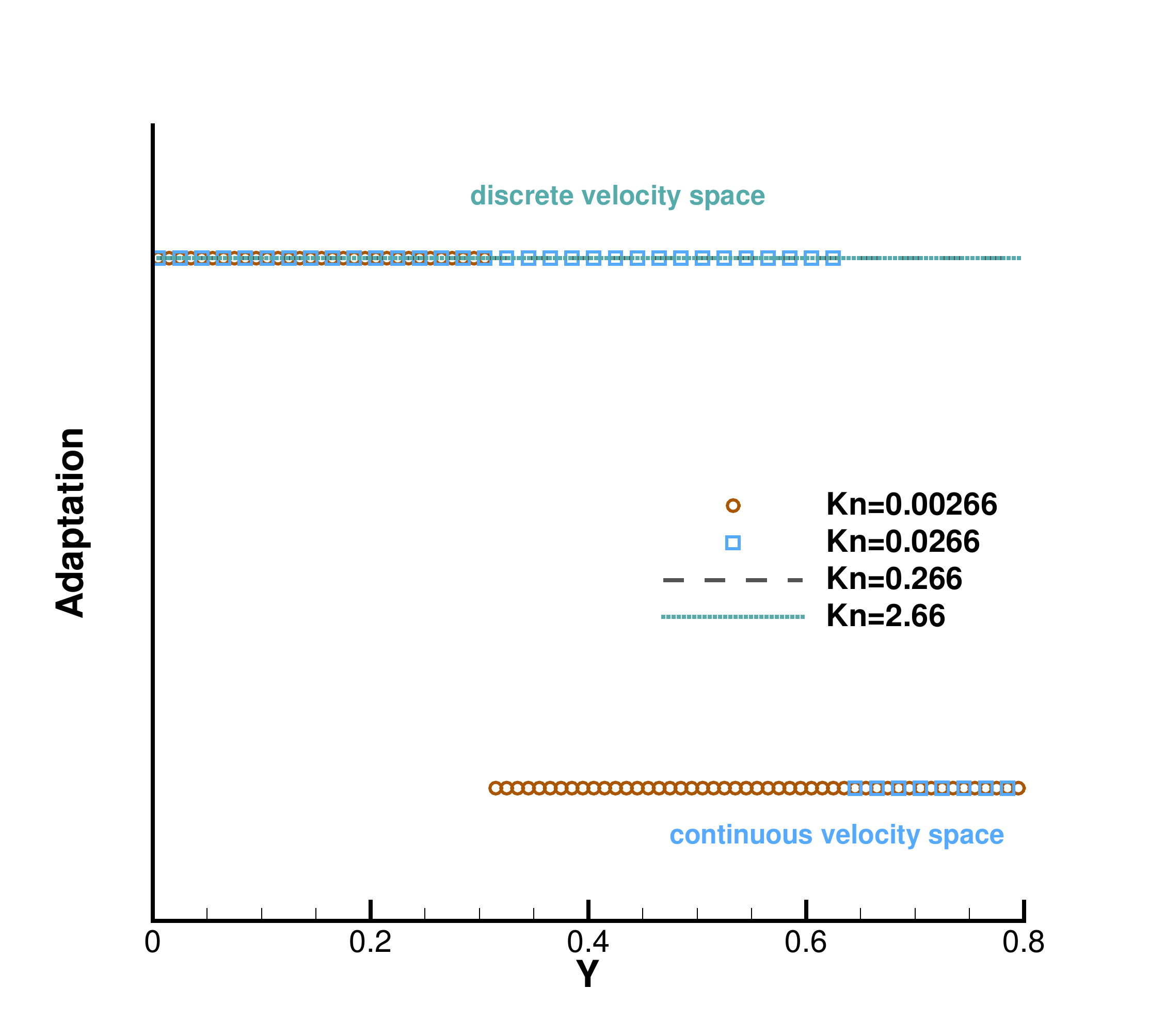}
	\caption{Velocity space adaptation in the Rayleigh flow at the output instant with different reference Knudsen numbers.}
	\label{pic:rayleigh regime}
\end{figure}

\begin{table}  
	\label{table:rayleigh}
	\centering
	\begin{tabular*}{7cm}{lll}  
		\hline  
		& AUGKS  & UGKS  \\
		\hline  
		Kn=0.00266  & 1003.59 & 7477.10  \\  
		Kn=0.0266  & 2223.62 & 7460.21  \\  
		Kn=0.266  & 3751.75. & 7438.06  \\  
		Kn=2.66  & 4637.19 & 7457.87  \\  
		\hline 
	\end{tabular*}  
	\caption{CPU time cost in the Rayleigh problem using the AUGKS and UGKS at different Knudsen numbers.}  
\end{table}

\subsection{Nozzle flow}

The nozzle flow connecting different flow regimes is an ideal case to test the capacity of AUGKS in capturing multiple scale flow dynamics.
The schematic of the nozzle problem is presented in Fig. 14.
The argon gas is enclosed in a rectangular box $x\in [0,2.2],y\in[-0.5,0.5]$.
The velocity space is discretized into $28\times 28$ uniform points for the update of particle distribution function.
The switching criterion of velocity space in this case is set as $B=0.0005$.
In this case, the collision term in AUGKS and UGKS is the Shakhov model.
The computational domain is divided into two parts which are connected through a nozzle.
The gas density in the left is $100$ times higher than taht in the right part, and the initial Knudsen numbers are $Kn_L=0.0001$ and $Kn_R=0.01$ respectively.
Except that, the initial gas is at rest and has the same temperature inside two subdomains, same as the cavity wall.
All the walls are set up as Maxwell's diffusive boundary.
The nozzle has two entrances with cross sections $\Delta y_L=0.13$ and $\Delta y_R=0.33$ along a length $\Delta x=0.14$, from which a jet flow will be formed.
The simulation is performed till $t=50\tau_0$, where $\tau_0=\ell_0/v_0$ is the mean collision time of initial argon gas in the right domain.

Fig. \ref{pic:nozzle contour1} and \ref{pic:nozzle contour2} presents the solution contours of $U-$velocity and temperature at two times $t=5\tau_0$ and $t=20\tau_0$.
The upper part of color contours are the results calculated by the AUGKS (flood) and UGKS (lines), while the lower part is the Navier-Stokes solutions provided by the GKS with a continuous velocity space.
As is shown, the bow shock and expansion cooling region behind shock are captured by the two methods.
However, it is clear that at $Kn=0.01$, the Navier-Stokes equations lose validity to quantitatively describe the flow evolution in the right domain, and it is necessary to use kinetic method to get accurate solutions here.
Fig. \ref{pic:nozzle line1} and \ref{pic:nozzle line2} presents the solutions along the horizontal center line of the box at times $t=5\tau_0$ and $t=20\tau_0$.
It is clear that the AUGKS provide equivalent solutions with the NS ones in the near-equilibrium left region, and with the Boltzmann solutions in the non-equilibrium right region.
This test demonstrates the multiscale capability of the adaptive method to get the physical solutions in the corresponding flow regimes.

Fig. \ref{pic:nozzle adaptation1}, \ref{pic:nozzle adaptation2} and \ref{pic:nozzle adaptation3} present the components of spatial slope $a$ used in the velocity space switching criterion, the mean collision time and the corresponding velocity space adaptation of the flow domain at three times $t=\tau_0,\ 5\tau_0,\ 20\tau_0$.
As shown, the shock and expansion waves are the major sources for large flow gradients inside the domain.
Accompanying with the high-density jet into the right domain, the mean collision time decreases in the jet region.
With time increasing, the local flow structure becomes more complicated, leading to a large non-equilibrium region.
As a result, the Chapman-Enskog expansion fails in the places where the strong non-equilibrium effects appear, and the discretized velocity space has to be used in AUGKS.
Table 3 presents the computational cost of AUGKS, UGKS and GKS in this case.
With the current setup of physical mesh and velocity space, the continuous GKS solver is about 25 times faster than the UGKS with discretized velocity space.
The AUGKS is $1.7$ times faster than the original UGKS.
At the initial stage in the simulation, the memory cost in the AUGKS is on the same order as the GKS, which is about $1/30$ of the UGKS.
As flow evolves, the number of cells associated with discretized velocity space increases, and the corresponding memory cost becomes larger.
At the final time $t=50\tau_0$, there are about 429 out of 914 total cells using continuous velocity space, and the corresponding memory burden is $53\%$ of the original UGKS.

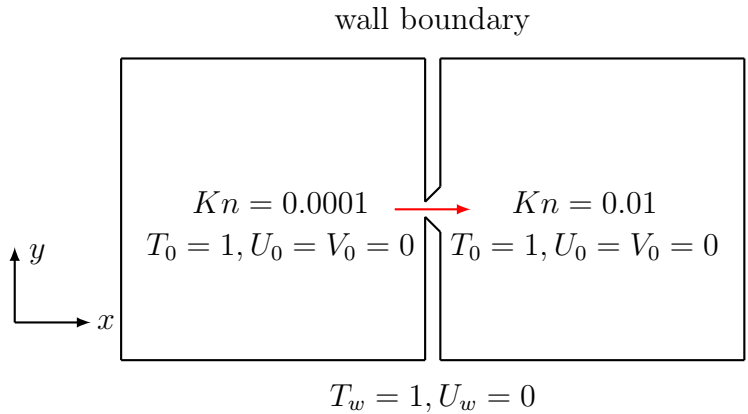
\begin{figure}[htb!]
	\centering
	{
		\begin{tikzpicture}[thick]
		\node[rectangle] (bx) {};
		\draw ($(bx)+(-4.1,+0.5)$) -- ($(bx)+(-0.1,+0.5)$);
		\draw ($(bx)+(+0.1,+0.5)$) -- ($(bx)+(+4.1,+0.5)$);
		\draw ($(bx)+(-4.1,+4.5)$) -- ($(bx)+(-0.1,+4.5)$);
		\draw ($(bx)+(+0.1,+4.5)$) -- ($(bx)+(+4.1,+4.5)$);
		\draw ($(bx)+(+4.1,+0.5)$) -- ($(bx)+(+4.1,+4.5)$);
		\draw ($(bx)+(-4.1,+0.5)$) -- ($(bx)+(-4.1,+4.5)$);
		\draw ($(bx)+(-0.1,+0.5)$) -- ($(bx)+(-0.1,+2.4)$);
		\draw ($(bx)+(-0.1,+2.6)$) -- ($(bx)+(-0.1,+4.5)$);
		\draw ($(bx)+(0.1,+2.8)$) -- ($(bx)+(0.1,+4.5)$);
		\draw ($(bx)+(0.1,+0.5)$) -- ($(bx)+(0.1,+2.2)$);
		\draw ($(bx)+(-0.1,+2.6)$) -- ($(bx)+(0.1,+2.8)$);
		\draw ($(bx)+(-0.1,+2.4)$) -- ($(bx)+(0.1,+2.2)$);
		\node at ($(bx)+(+0,+0)$) {$ T_w=1,U_w=0$};
		\node at ($(bx)+(+0,+5)$) {$\rm wall\  boundary$};
		\draw[line][red] ($(bx)+(-0.5,+2.5)$) -- ($(bx)+(+0.5,+2.5)$);
		\node at ($(bx)+(-2,+2.6)$) {$Kn=0.0001$};
		\node at ($(bx)+(+2,+2.6)$) {$Kn=0.01$};
		\node at ($(bx)+(-2,+2)$) {$T_0=1,U_0=V_0=0$};
		\node at ($(bx)+(+2,+2)$) {$T_0=1,U_0=V_0=0$};
		\draw[line] ($(bx)+(-5.5,+1)$) -- ($(bx)+(-5.5,+2)$);
		\draw[line] ($(bx)+(-5.5,+1)$) -- ($(bx)+(-4.5,+1)$);
		\node at ($(bx)+(-4.3,+1)$) {$x$};
		\node at ($(bx)+(-5.2,+1.9)$) {$y$};
		\end{tikzpicture}
	}
	\label{pic:nozzle schematic}
	\caption{Schematic of Nozzle jet problem.}
\end{figure}

\begin{figure}[htb!]
	\centering
	\subfigure[U-velocity]{
		\includegraphics[width=10cm]{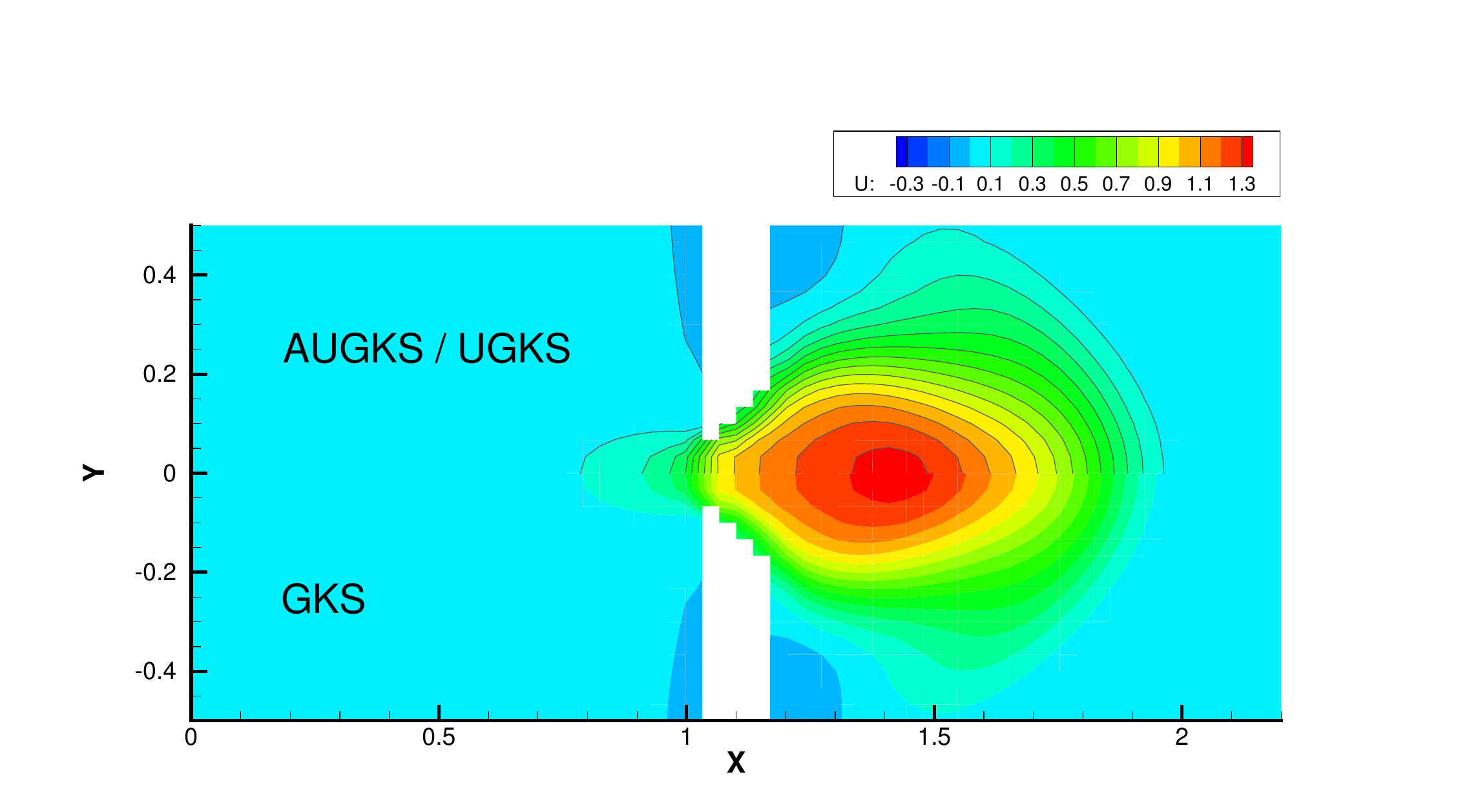}
	}
	\subfigure[Temperature]{
		\includegraphics[width=10cm]{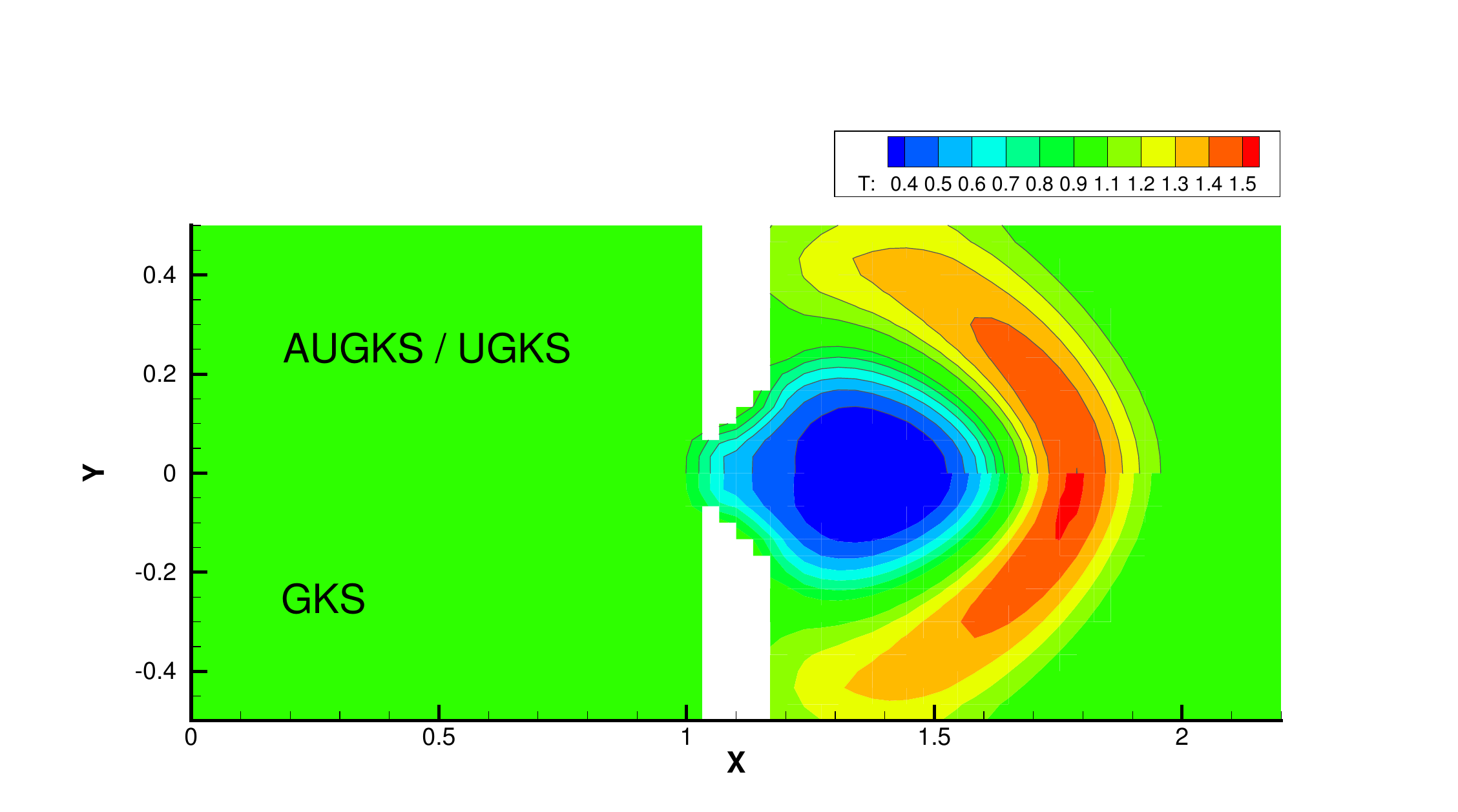}
	}
	\caption{Density and Temperature contours at $t=5\tau_0$ (upper flood: AUGKS, upper lines: UGKS, lower: GKS).}
	\label{pic:nozzle contour1}
\end{figure}

\begin{figure}[htb!]
	\centering
	\subfigure[U-velocity]{
		\includegraphics[width=10cm]{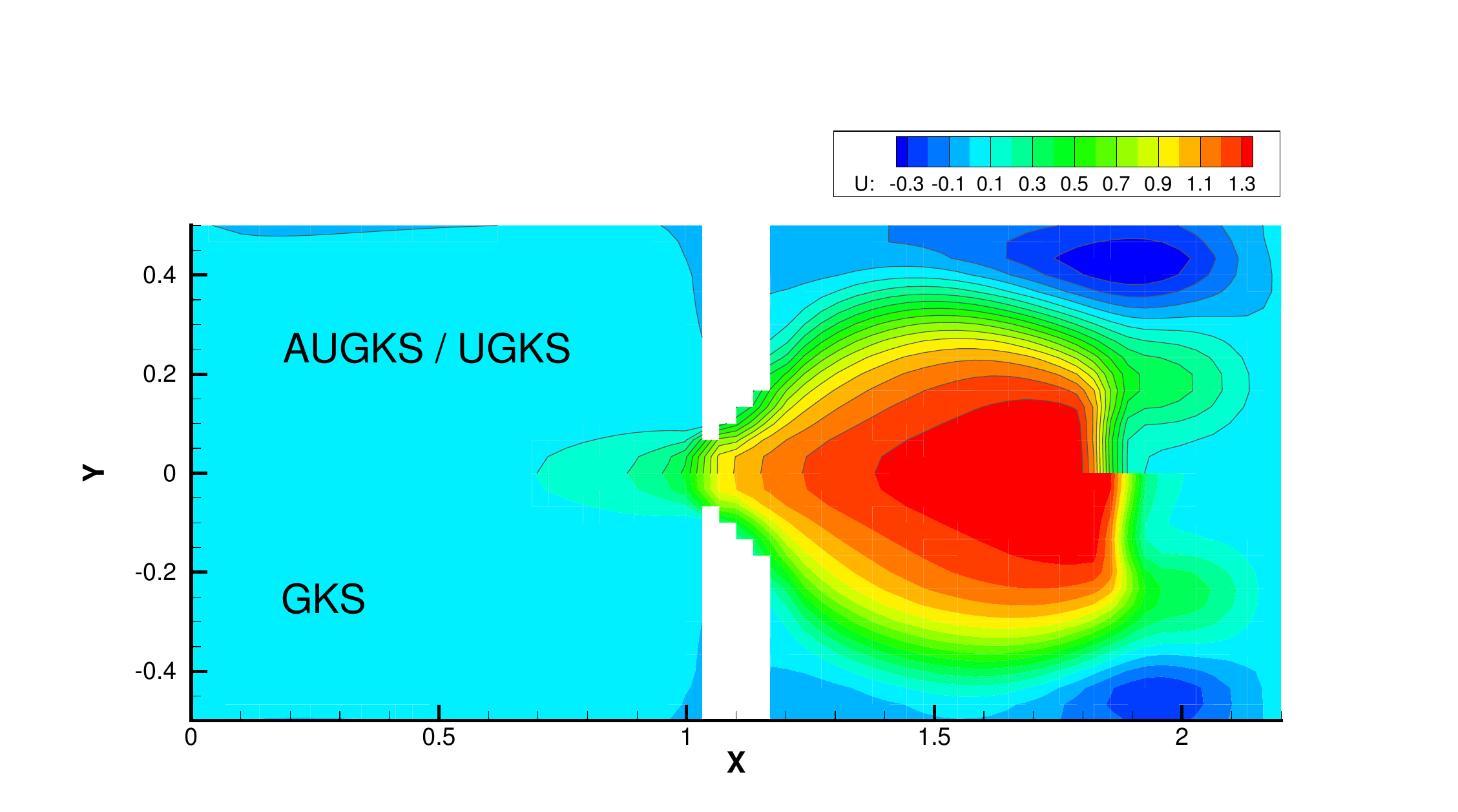}
	}
	\subfigure[Temperature]{
		\includegraphics[width=10cm]{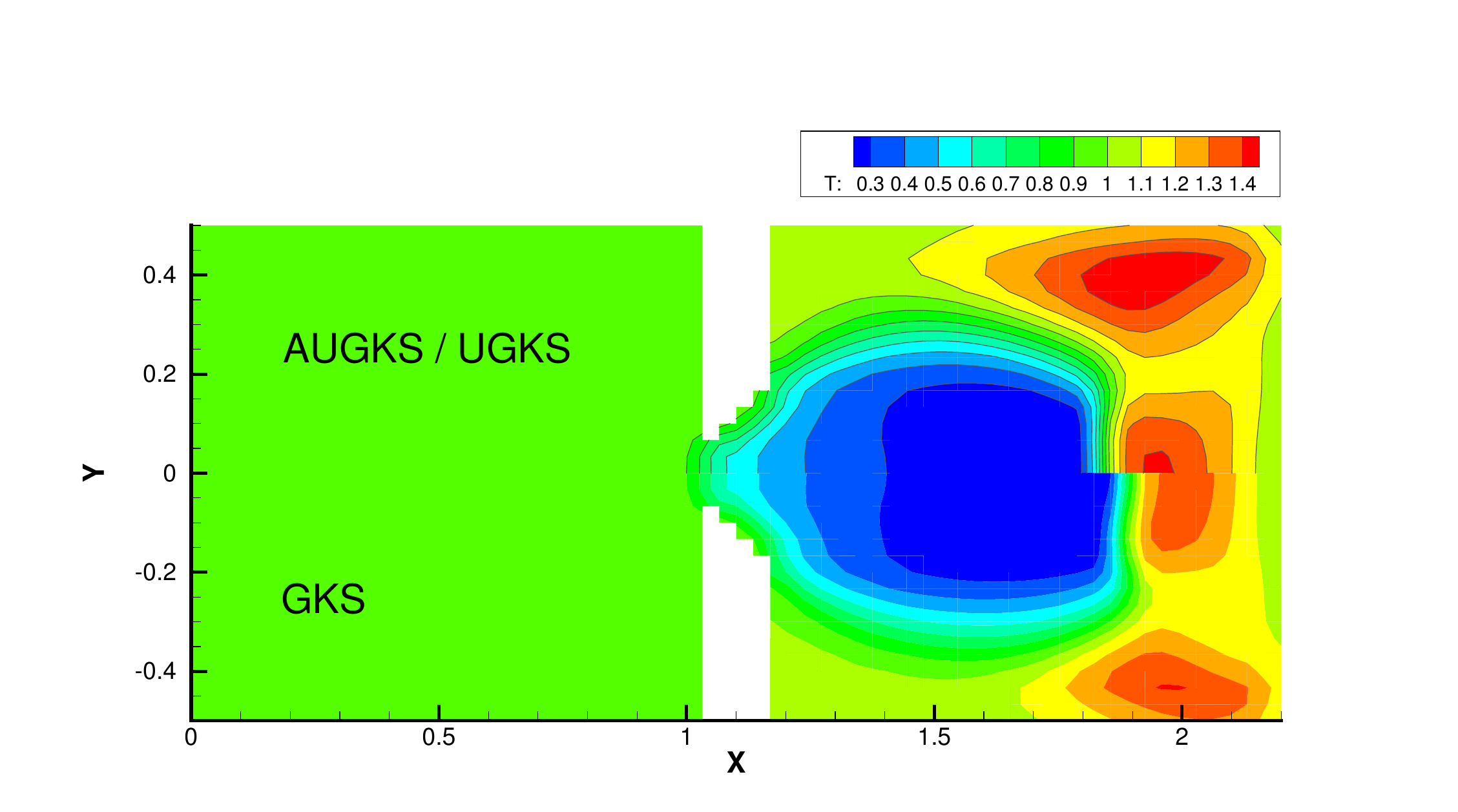}
	}
	\caption{Density and Temperature contours at $t=20\tau_0$ (upper flood: AUGKS, upper lines: UGKS, lower: GKS).}
	\label{pic:nozzle contour2}
\end{figure}

\begin{figure}[htb!]
	\centering
	\subfigure[Density]{
		\includegraphics[width=5cm]{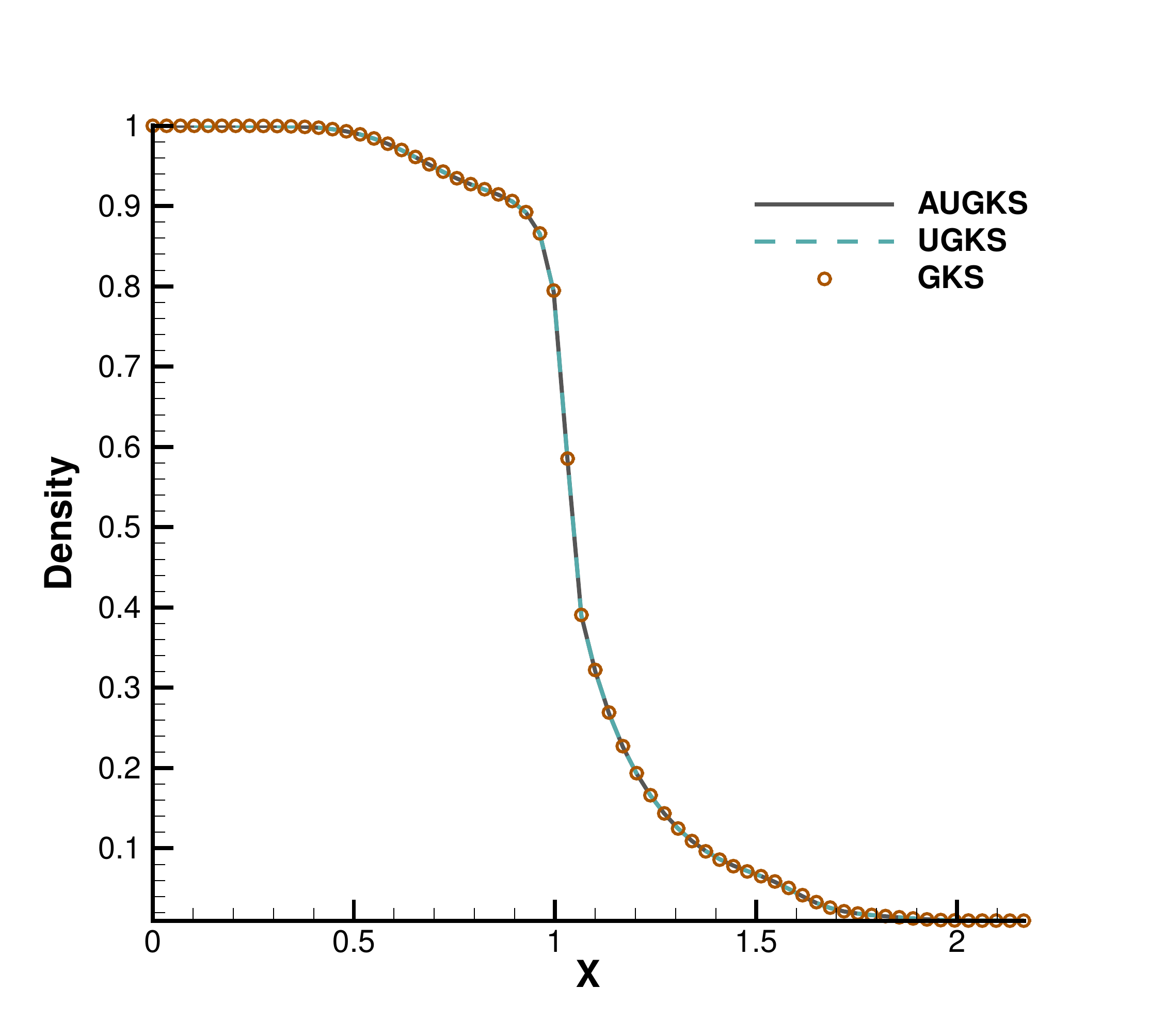}
	}
	\subfigure[U-velocity]{
		\includegraphics[width=5cm]{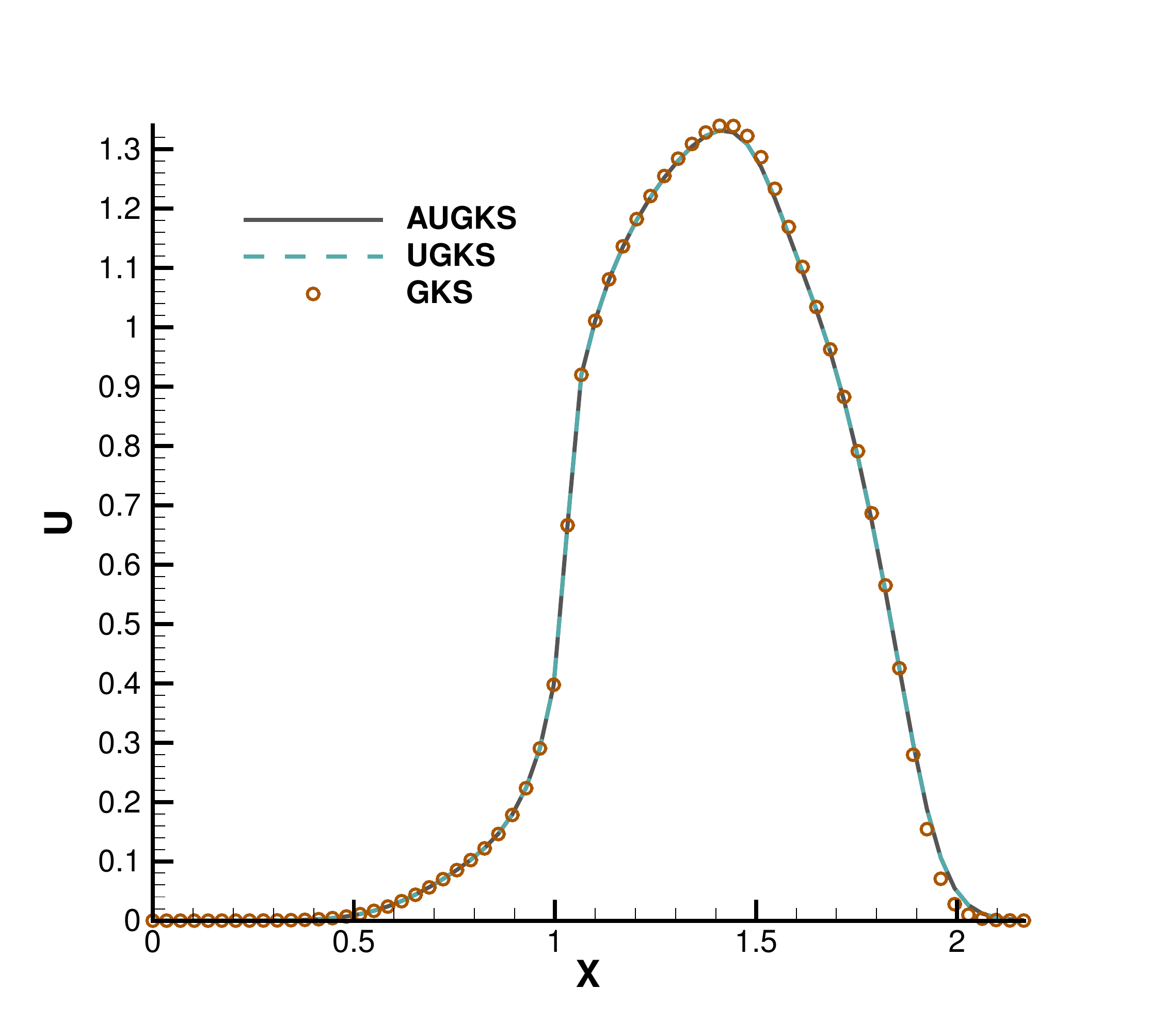}
	}
	\subfigure[Temperature]{
		\includegraphics[width=5cm]{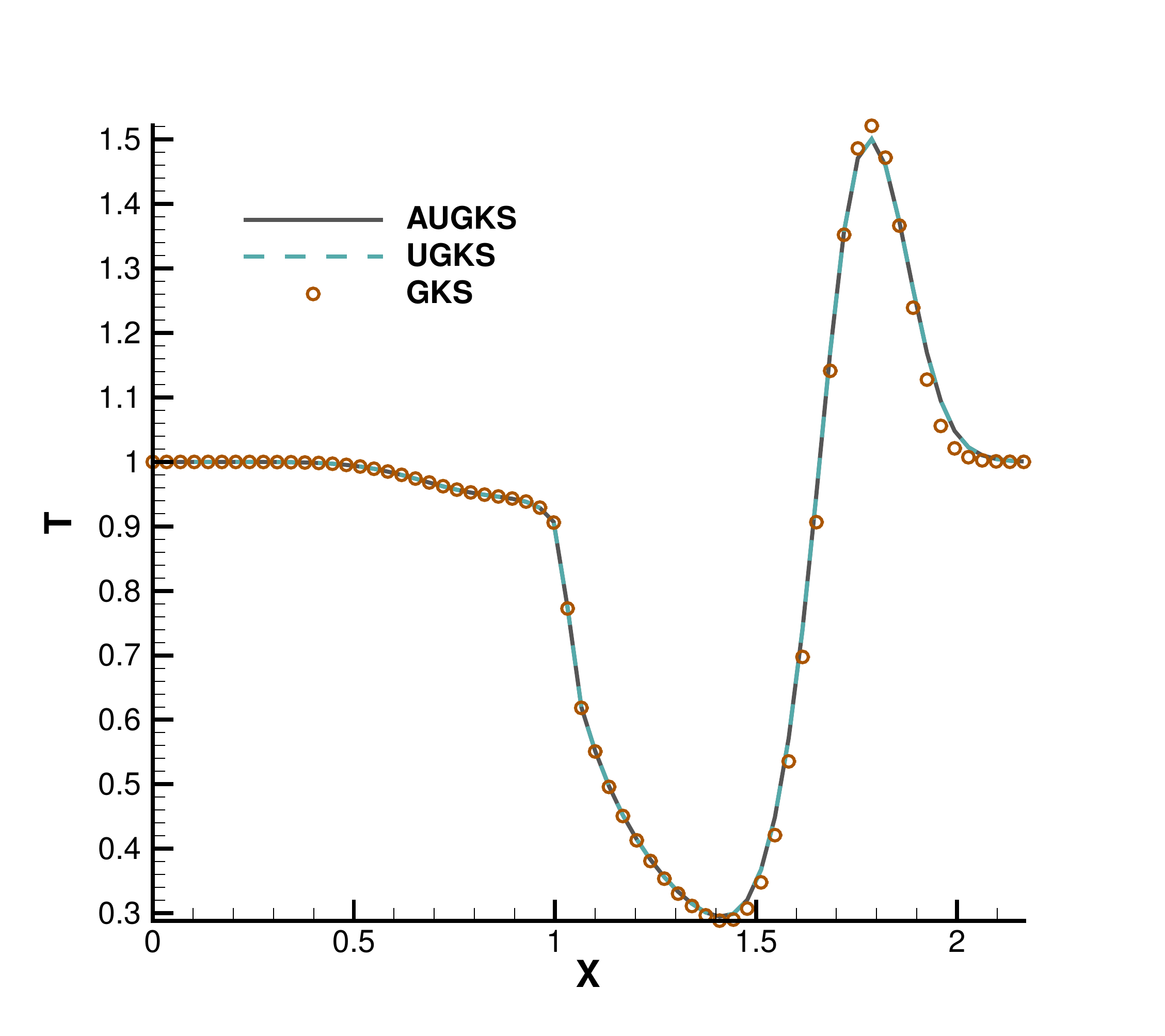}
	}
	\caption{Solutions along the horizontal central line at $t=5\tau_0$.}
	\label{pic:nozzle line1}
\end{figure}

\begin{figure}[htb!]
	\centering
	\subfigure[Density]{
		\includegraphics[width=5cm]{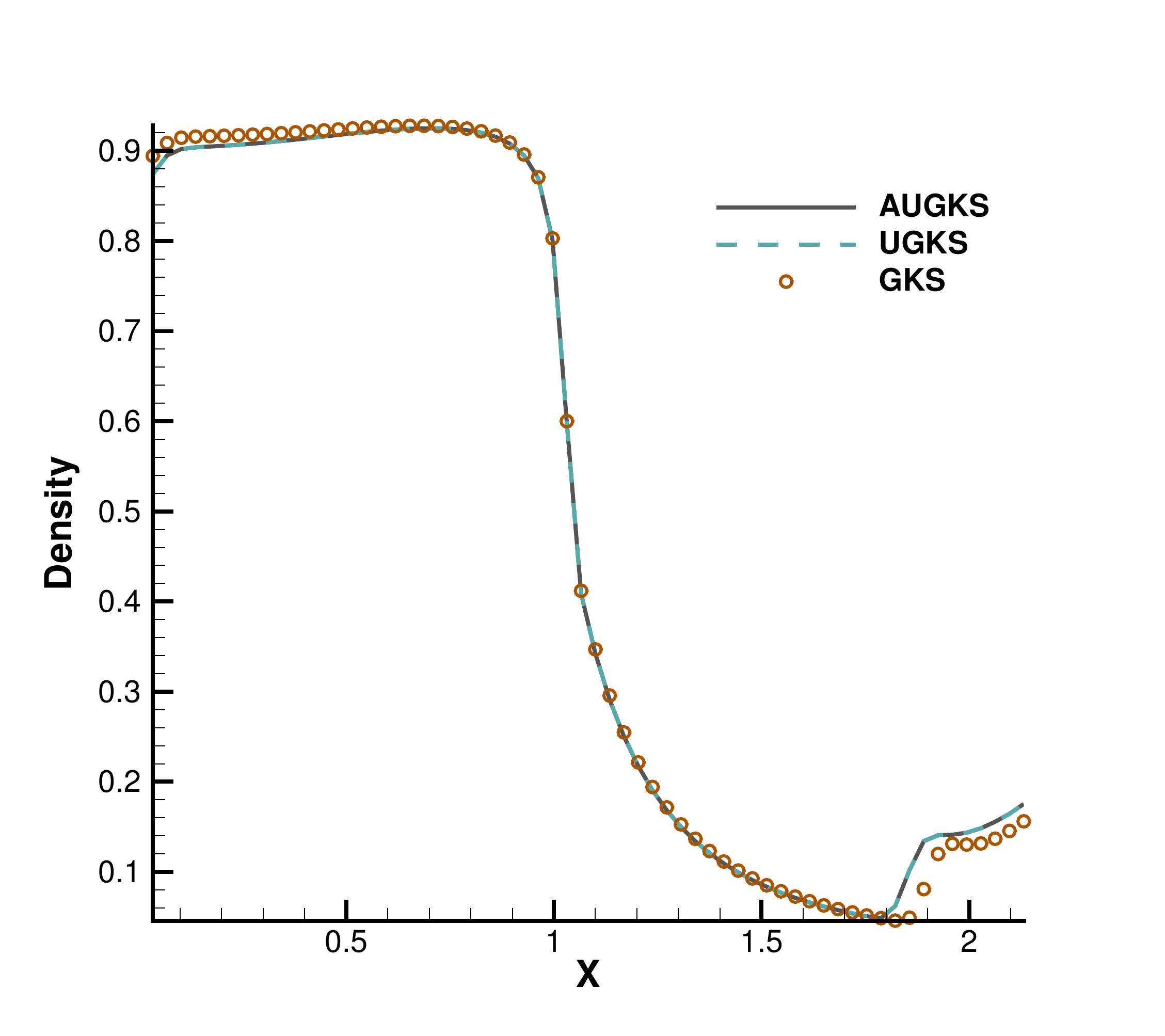}
	}
	\subfigure[U-velocity]{
		\includegraphics[width=5cm]{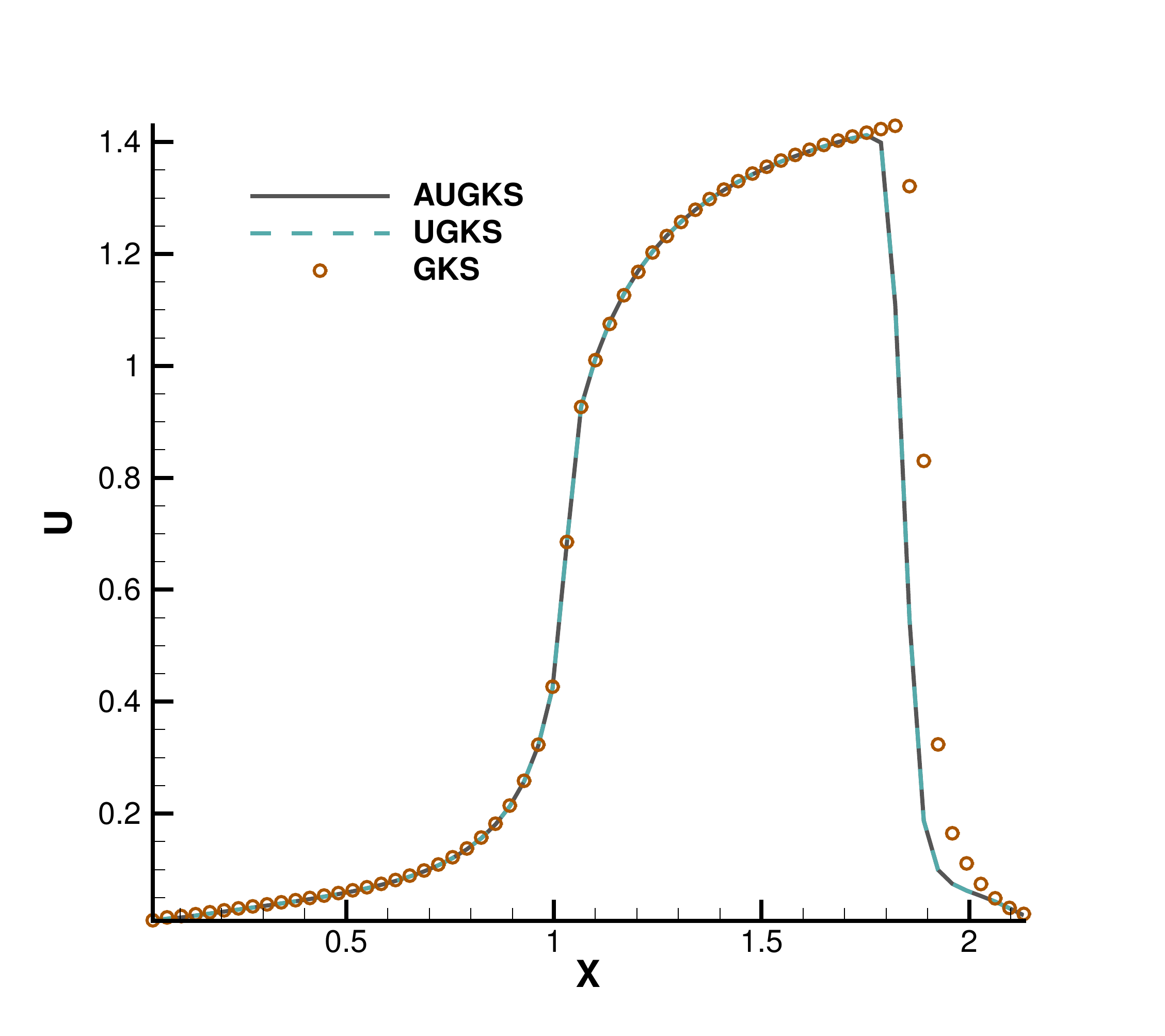}
	}
	\subfigure[Temperature]{
		\includegraphics[width=5cm]{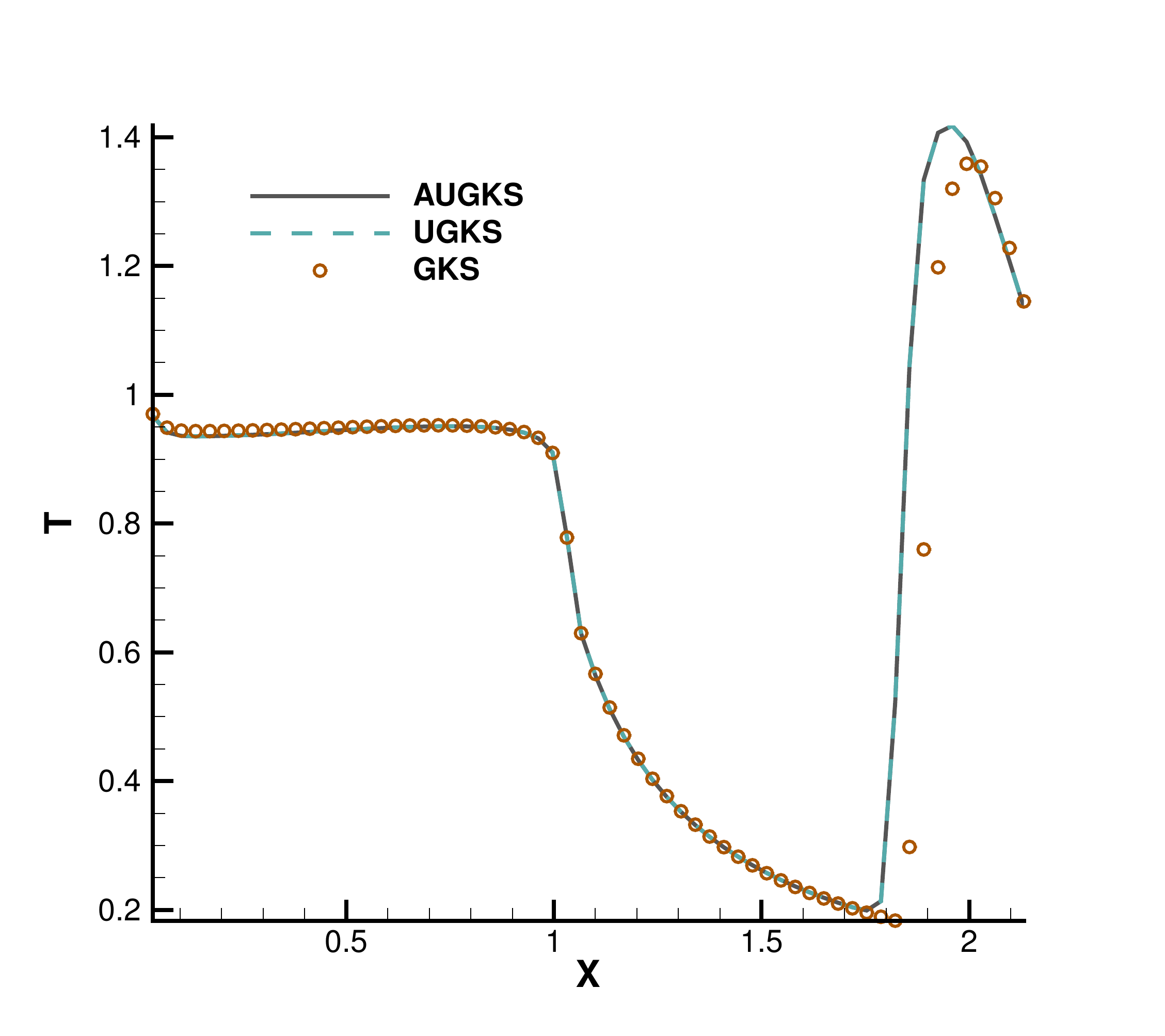}
	}
	\caption{Solutions along the horizontal central line at $t=20\tau_0$.}
	\label{pic:nozzle line2}
\end{figure}

\begin{table}  
	\label{table:nozzle}
	\centering
	\begin{tabular*}{10cm}{llll}  
		\hline  
		& AUGKS  & UGKS & GKS \\
		\hline  
		CPU time (s) &526.25  & 898.69 & 35.25 \\  
		Memory (KB) $(t=0)$  &3614  & 74828 & 2636 \\  
		Memory (KB) $(t=50\tau_0)$  &35125  & 74836 & 2640 \\  
		\hline 
	\end{tabular*}  
	\caption{CPU time and memory cost in the nozzle flow at $t=50\tau_0$.}  
\end{table}  

\begin{figure}[htb!]
	\centering
	\subfigure[$a_1$]{
		\includegraphics[width=12cm]{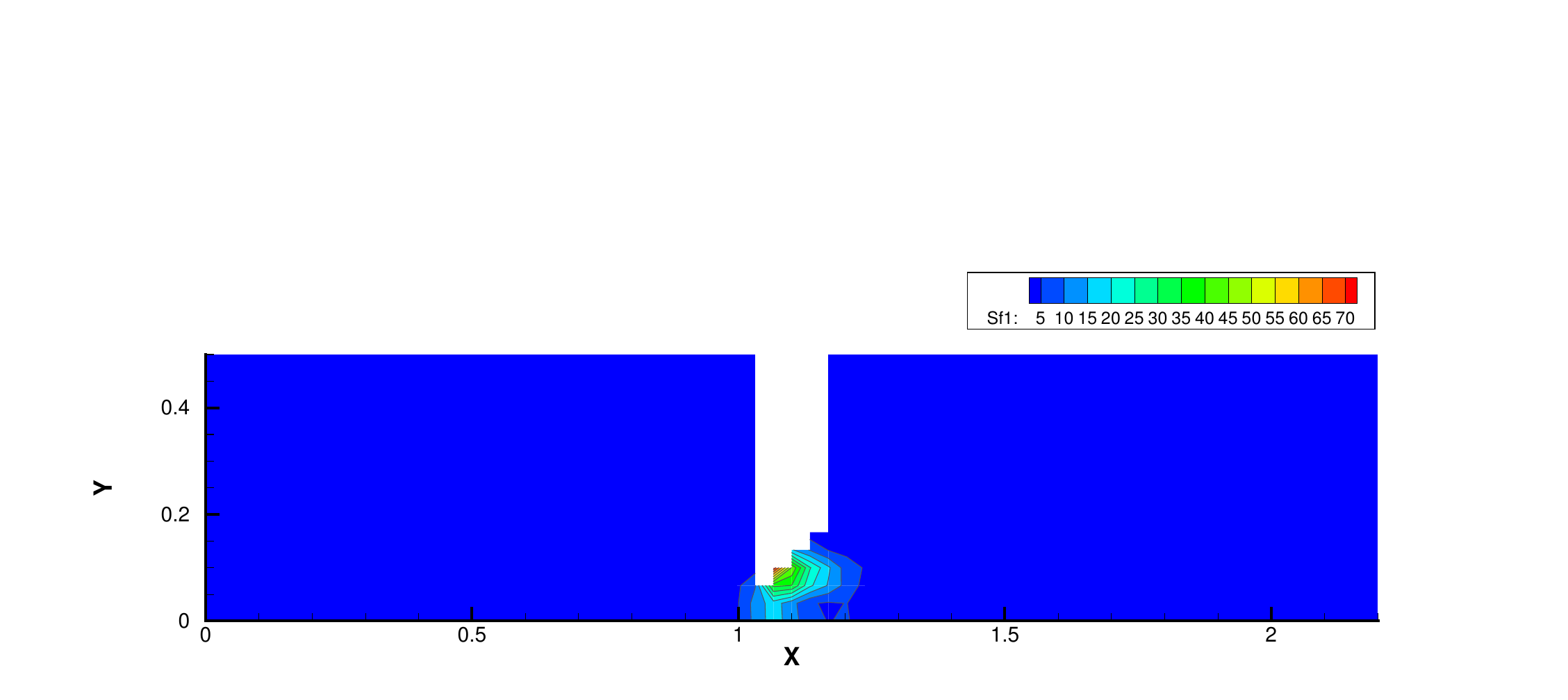}
	}
	\subfigure[Colliison time]{
		\includegraphics[width=12cm]{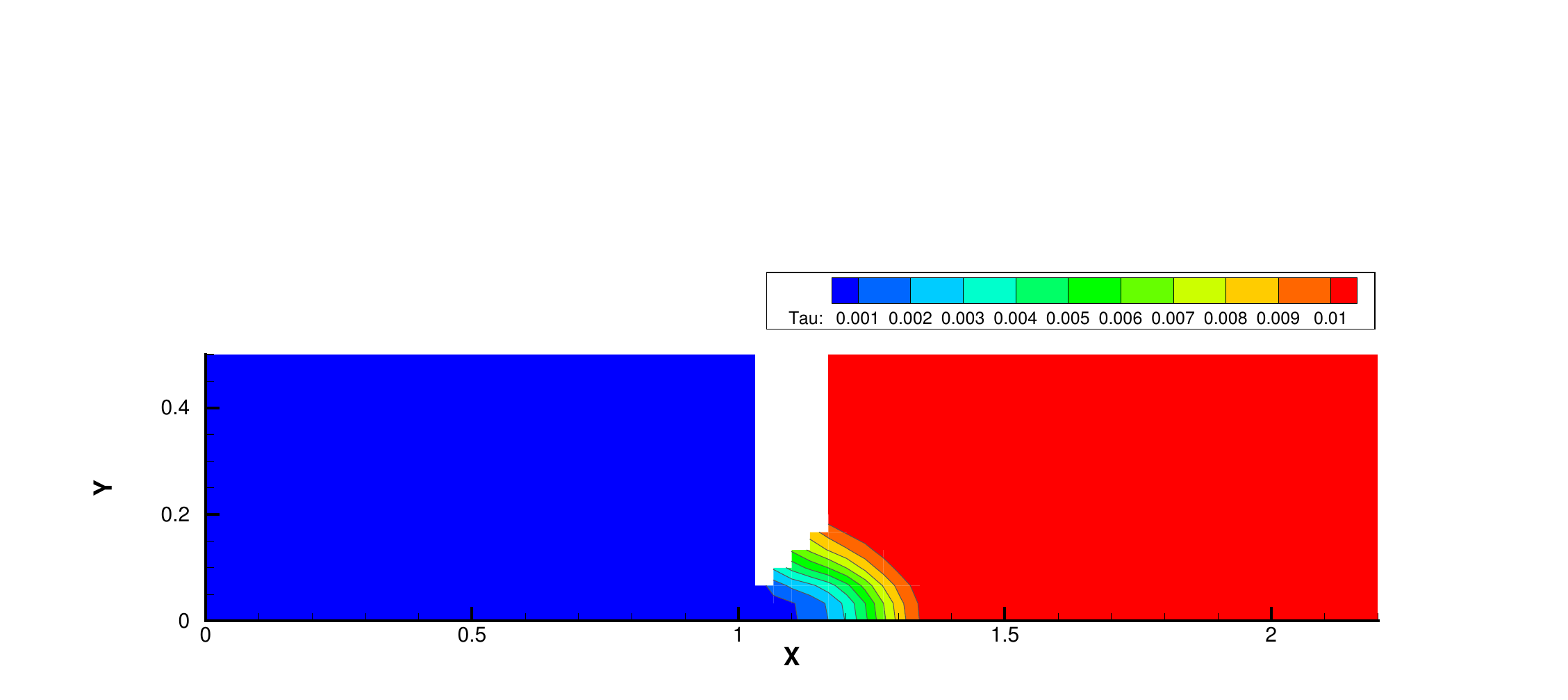}
	}
	\subfigure[Velocity adaptation]{
		\includegraphics[width=12cm]{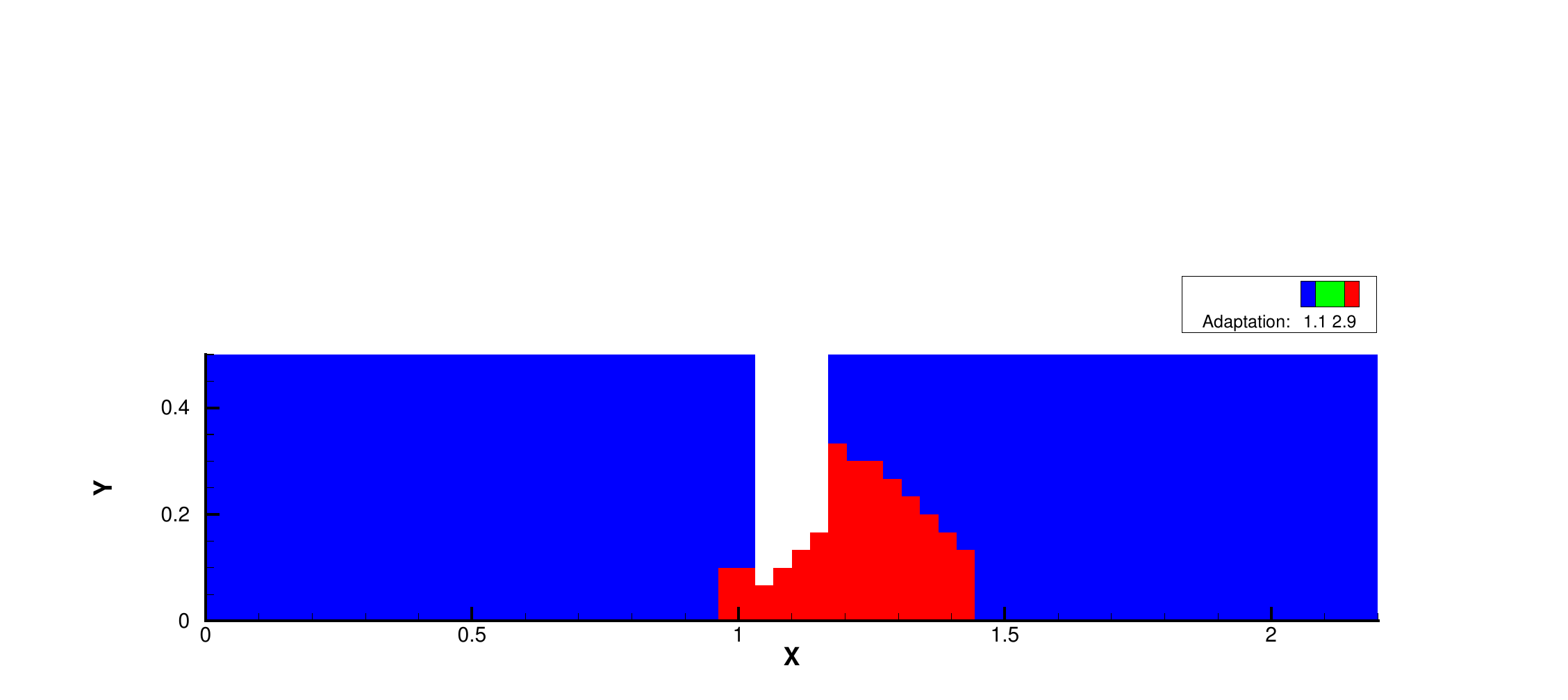}
	}
	\caption{Velocity space adaptation in the nozzle flow at $t=\tau_0$.}
	\label{pic:nozzle adaptation1}
\end{figure}
\begin{figure}[htb!]
	\centering
	\subfigure[$a_1$]{
		\includegraphics[width=12cm]{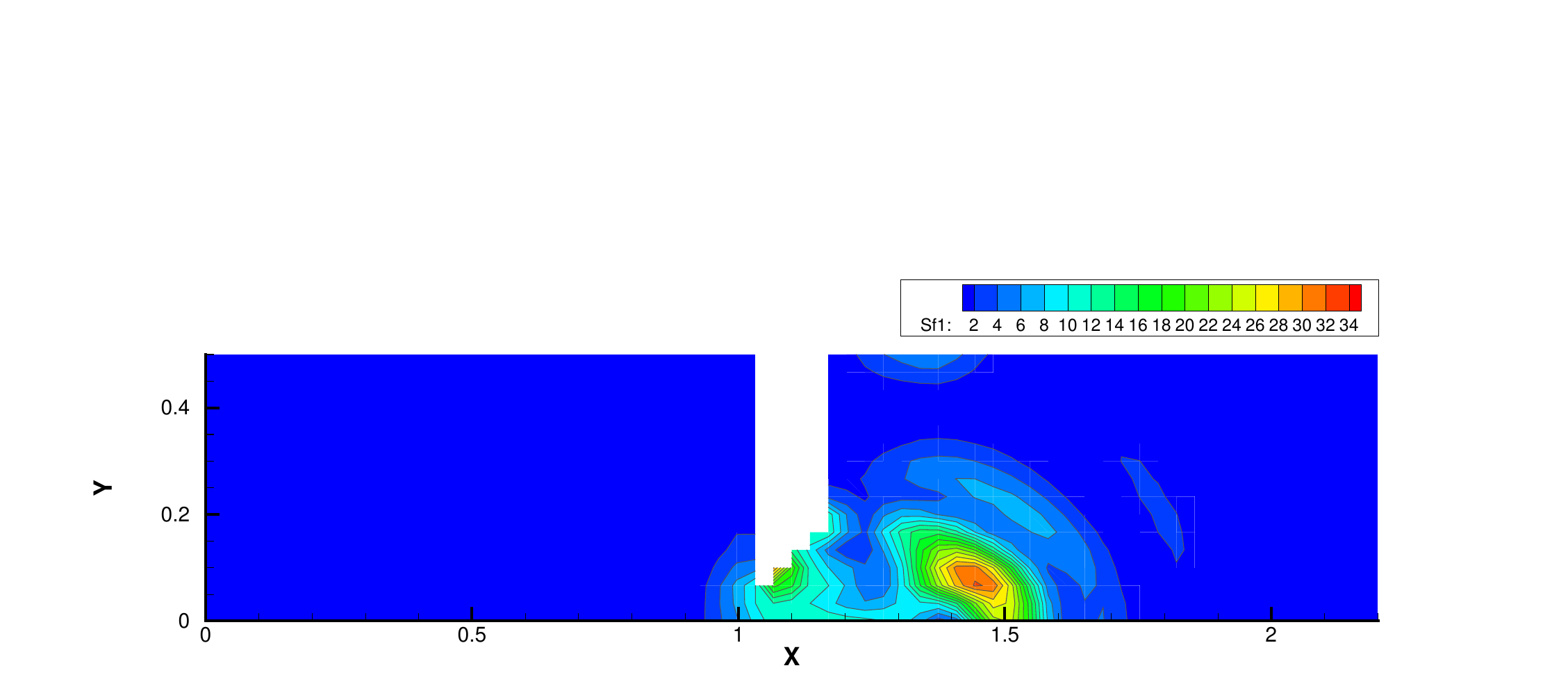}
	}
	\subfigure[Colliison time]{
		\includegraphics[width=12cm]{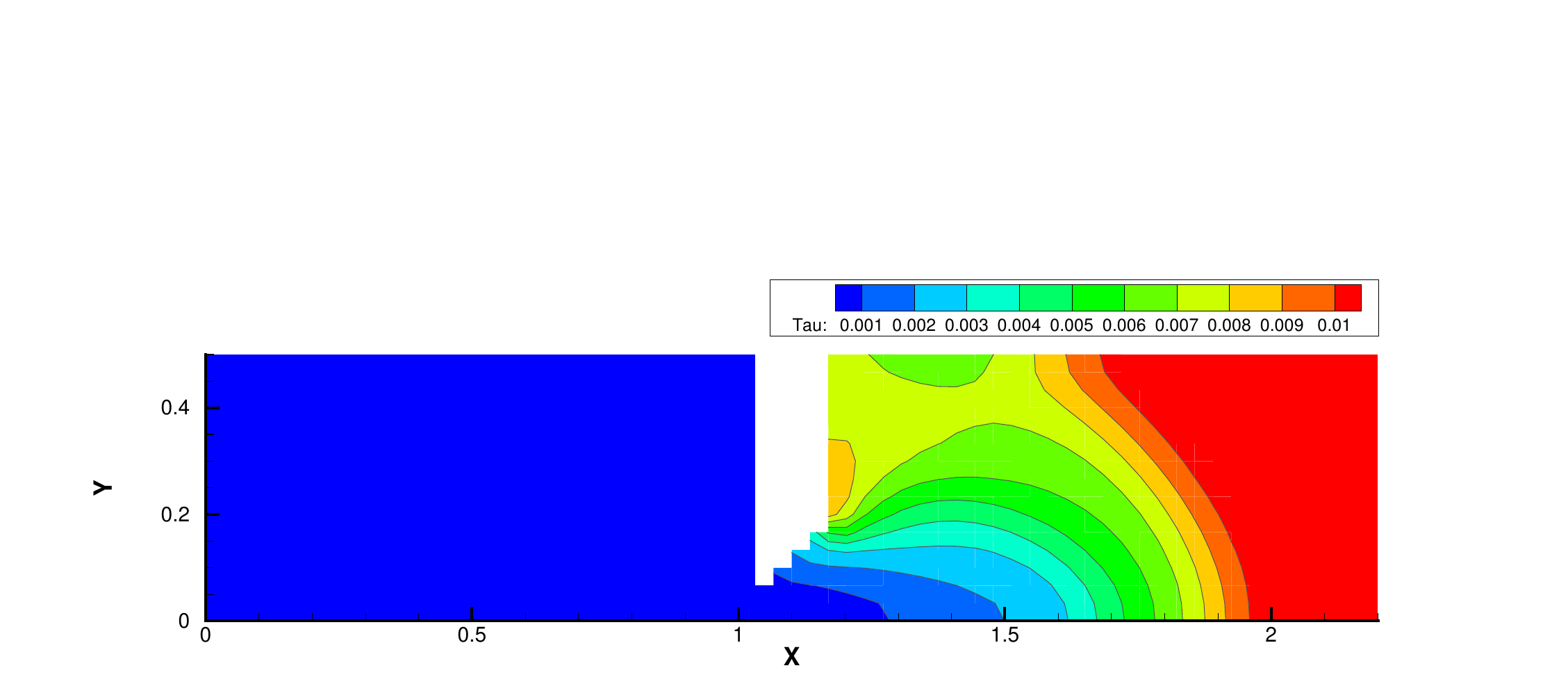}
	}
	\subfigure[Velocity adaptation]{
		\includegraphics[width=12cm]{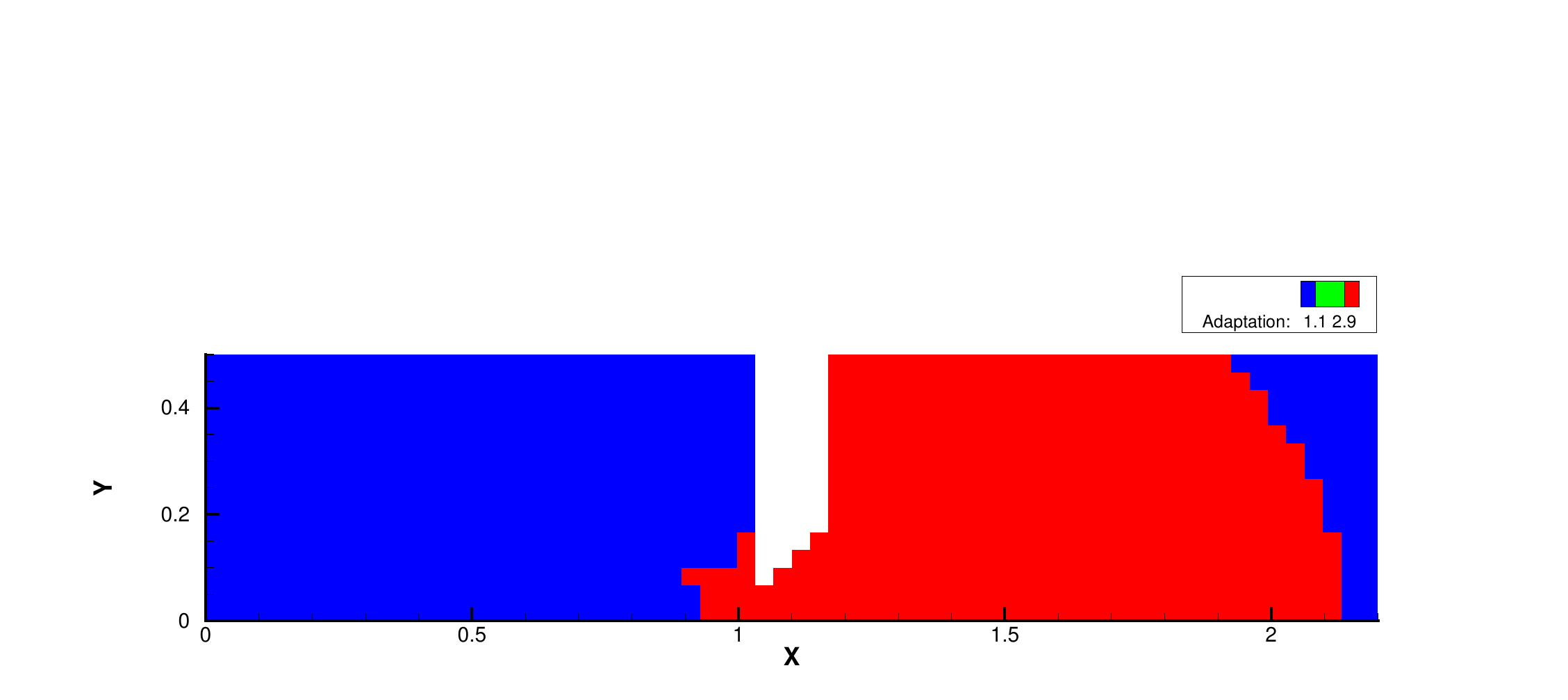}
	}
	\caption{Velocity space adaptation in the nozzle flow at $t=5\tau_0$.}
	\label{pic:nozzle adaptation2}
\end{figure}

\begin{figure}[htb!]
	\centering
	\subfigure[$a_1$]{
		\includegraphics[width=12cm]{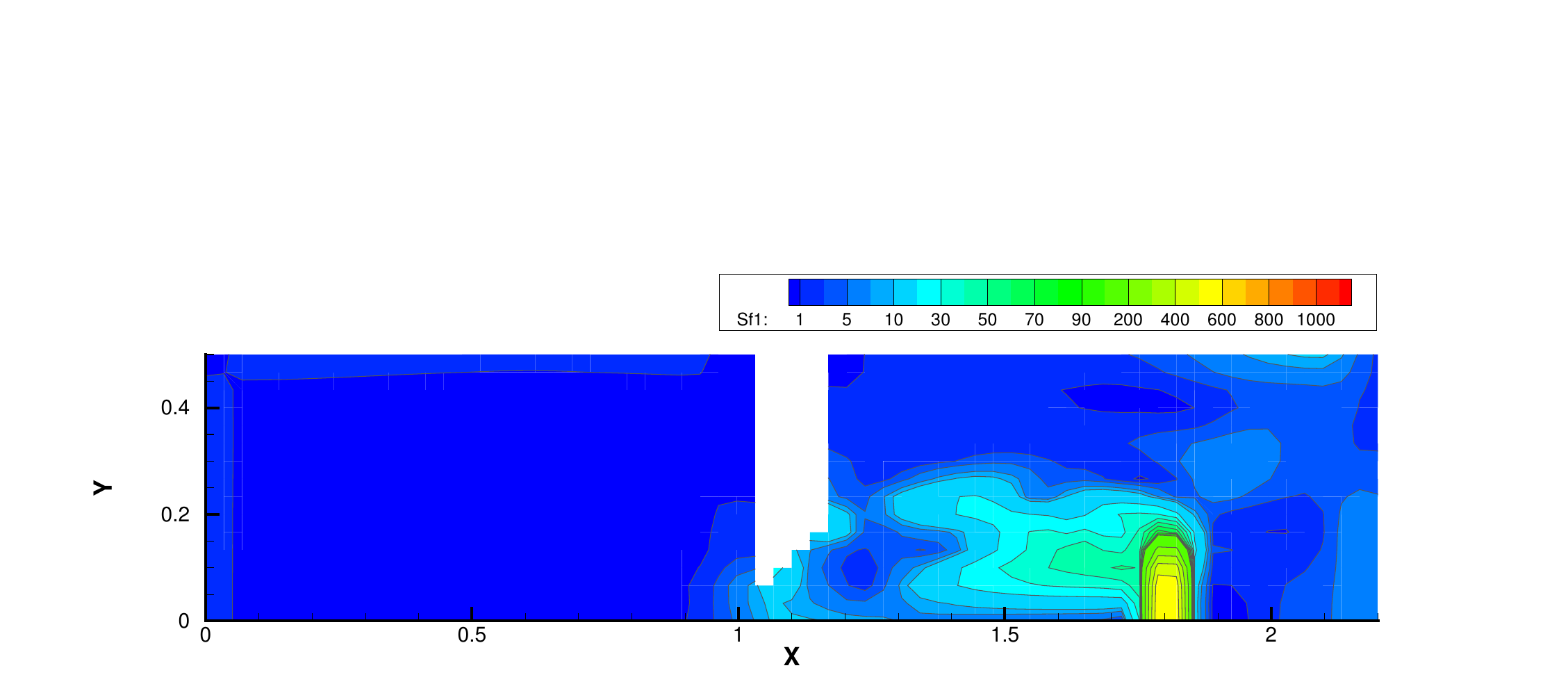}
	}
	\subfigure[Colliison time]{
		\includegraphics[width=12cm]{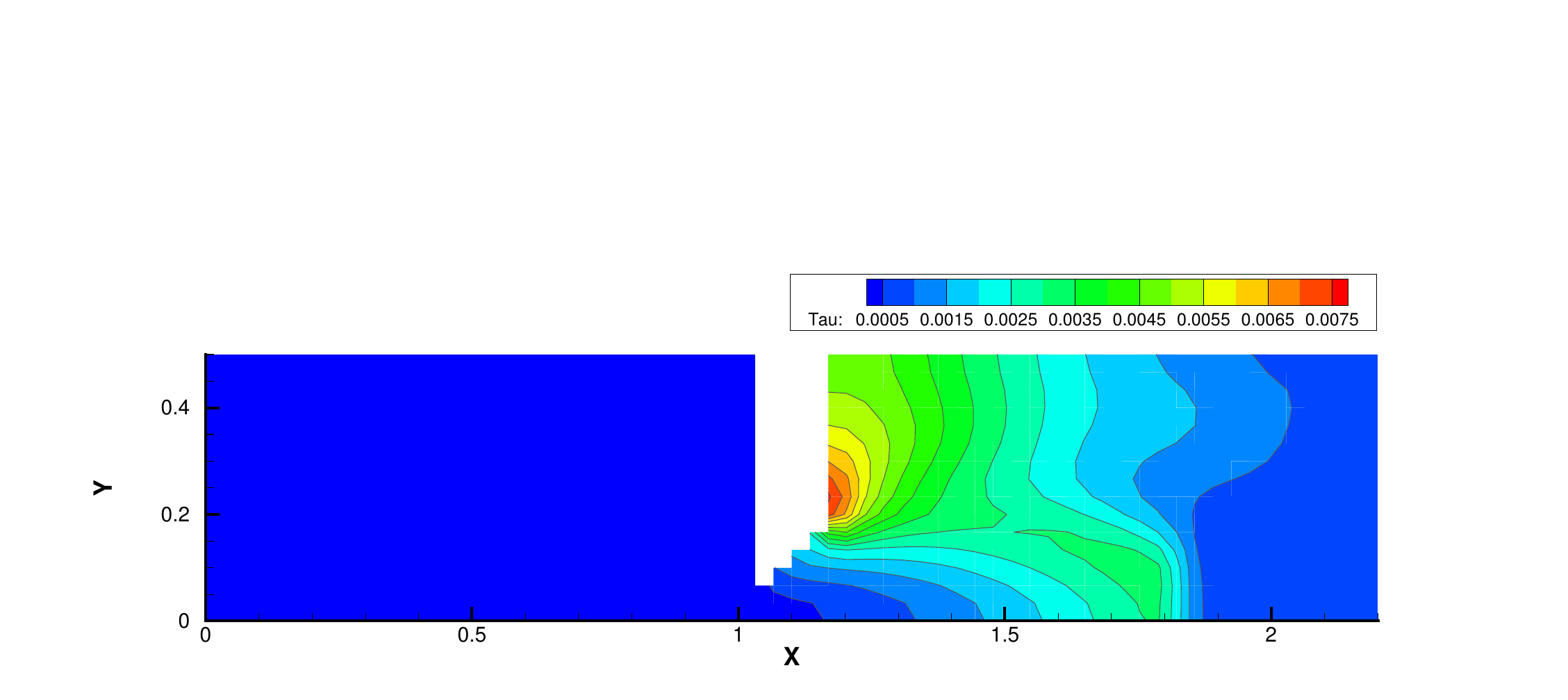}
	}
	\subfigure[Velocity adaptation]{
		\includegraphics[width=12cm]{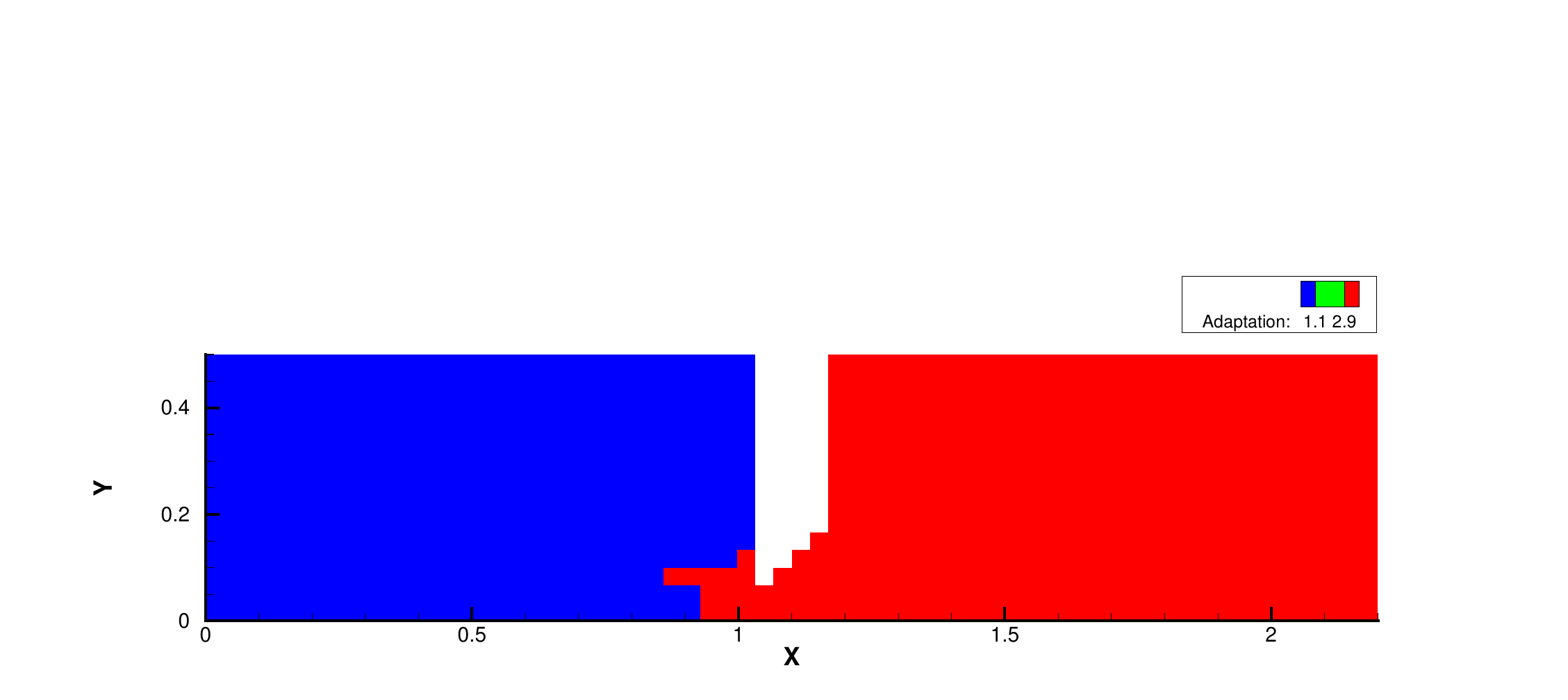}
	}
	\caption{Velocity space adaptation in the nozzle flow at $t=20\tau_0$.}
	\label{pic:nozzle adaptation3}
\end{figure}

\subsection{Flow around circular cylinder}

In the previous studies, the AUGKS in the simulation of unsteady flows is well validated.
In this case, the flow passing through a circular cylinder is used to test the performance of current adaptive scheme for steady flow.
The incoming gas has a uniform velocity with Mach number $Ma=5$ and the same temperature as the cylinder wall.
The reference Knudsen number is set up as $Kn = 0.001$ and $Kn = 0.01$ relative to cylinder radius, and the corresponding dynamic viscosity is $\mu_{ref}=7.313\times 10^{-4}$ and $\mu_{ref}=7.313\times 10^{-4}$.
In the calculation, $60$ cells in radial direction and $100$ cells in circumferential direction are used in physical domain.
The velocity space is discretized into $41\times 41$ velocity points for the update of particle distribution function.
In this case, the collision term for particle distribution function in AUGKS and UGKS is modeled as the Shakhov model.
The cylinder's wall is set up as Maxwell's diffusive boundary.
The switching criterion of particle velocity space is set as $B=0.0005$.

For steady problem, the computational time can be further reduced with the help of the GKS with a continuous velocity space.
A convergent coarse flow field can be first obtained by the GKS, and then used as the initial state in the subsquent adaptive method.
The method for the computation of steady flow is the following.
\begin{enumerate}
	\item {From initial setup, use the GKS solver in the entire domain and obtain a convergent flow field.}
	\item {Use the calculated macroscopic flow variables as the initial flow condition to get the particle distribution function with the discretized Chapman-Enskog form by Eq.(\ref{eqn:reconstruct distribution}).}
	\item {Adapt the velocity space based on the current switching criterion in Eq.(\ref{eqn:velocity adaptation}).}
	\item {Use continuous velocity space in near-equilibrium flow region and discretized velocity space in non-equilibrium one, and redo the velocity adaptation after some iterations until a convergent flow field is obtained.}
\end{enumerate}

\begin{figure}[htb!]
	\centering
	\subfigure[U-velocity]{
		\includegraphics[width=7.5cm]{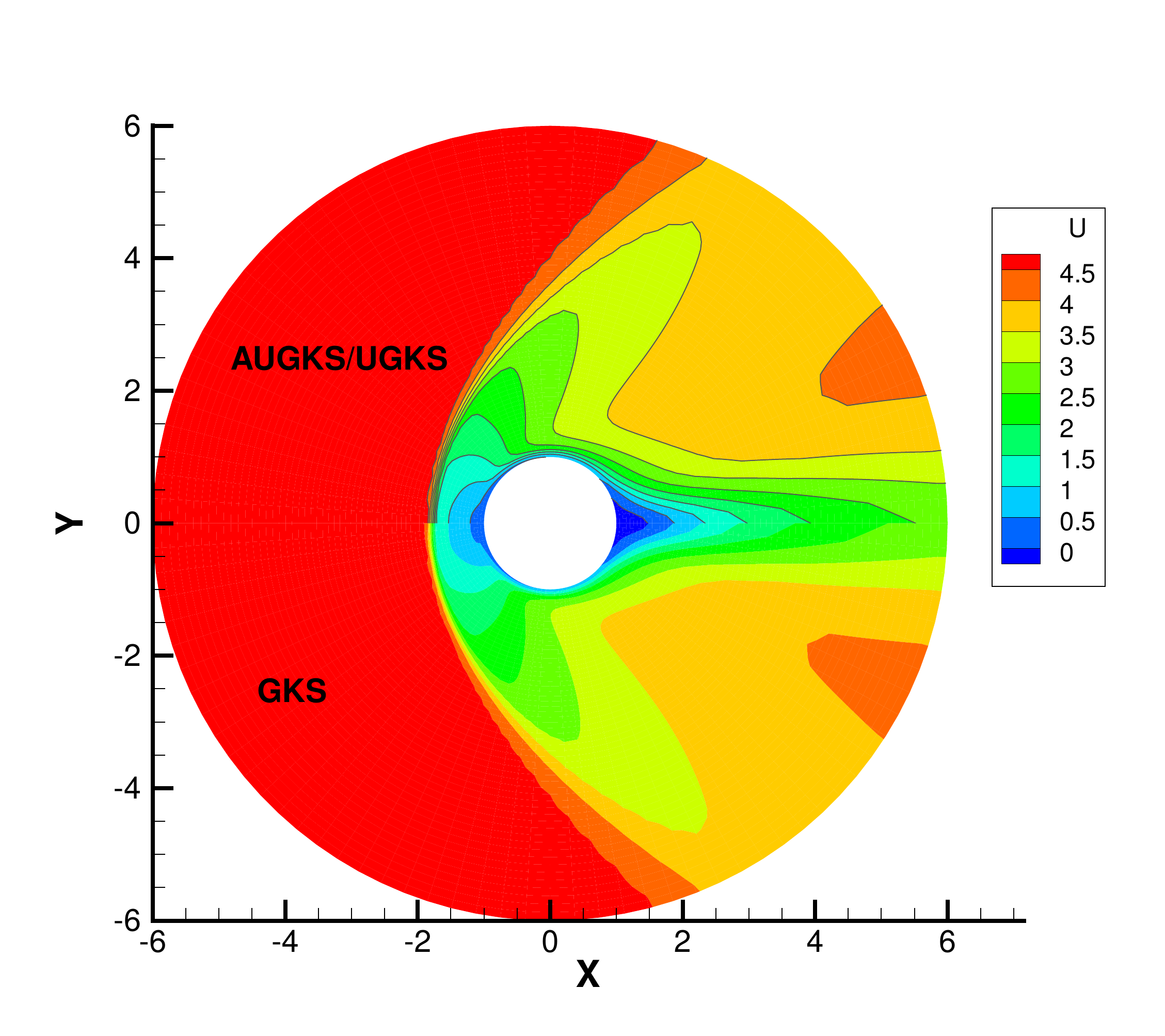}
	}
	\subfigure[Temperature]{
		\includegraphics[width=7.5cm]{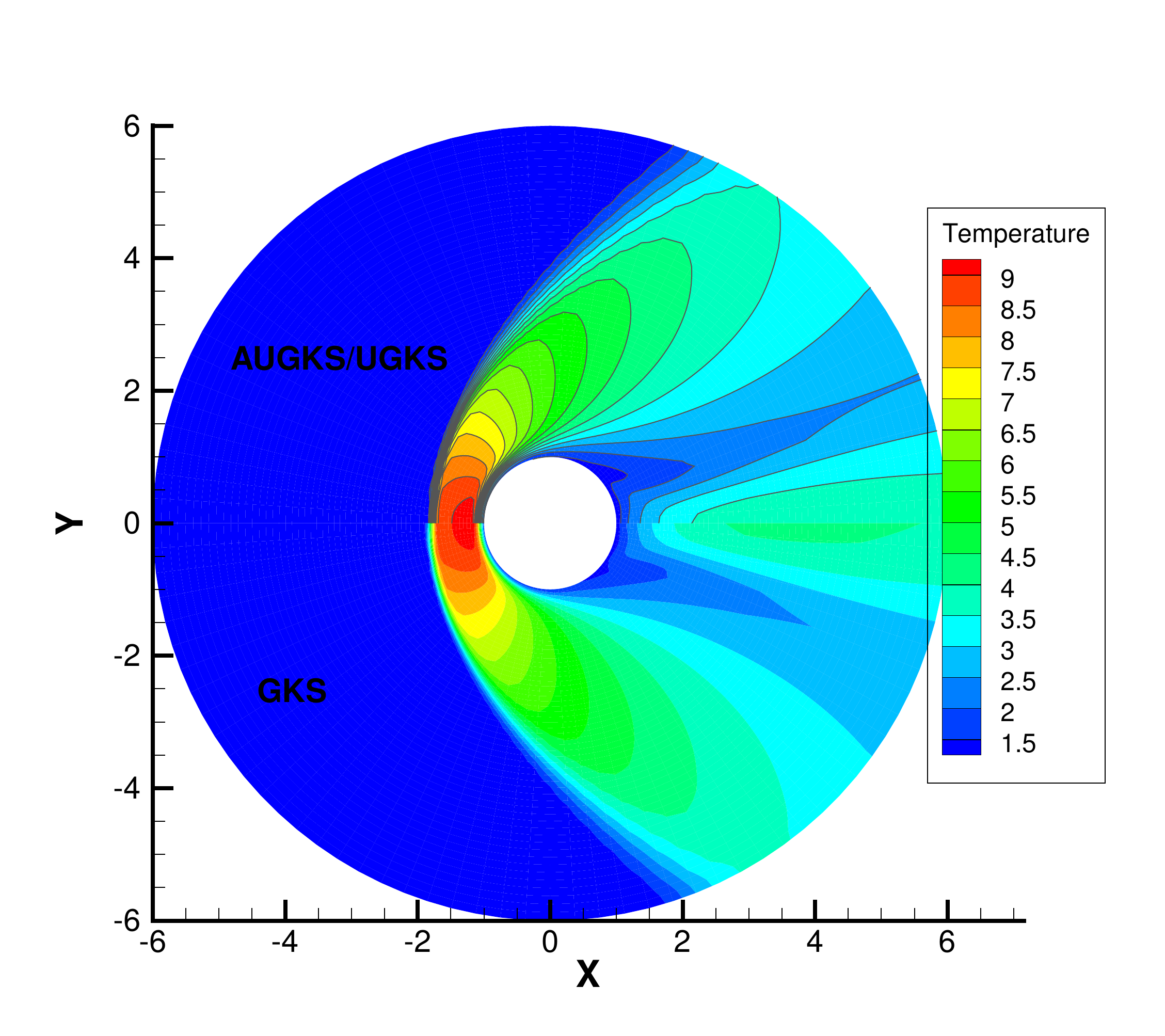}
	}
	\caption{Density and Temperature contours in the flow passing through cylinder at $Kn=0.001$ (upper flood: AUGKS, upper lines: UGKS, lower flood: GKS).}
	\label{pic:cylinder contour1}
\end{figure}

\begin{figure}[htb!]
	\centering
	\subfigure[U-velocity]{
		\includegraphics[width=7.5cm]{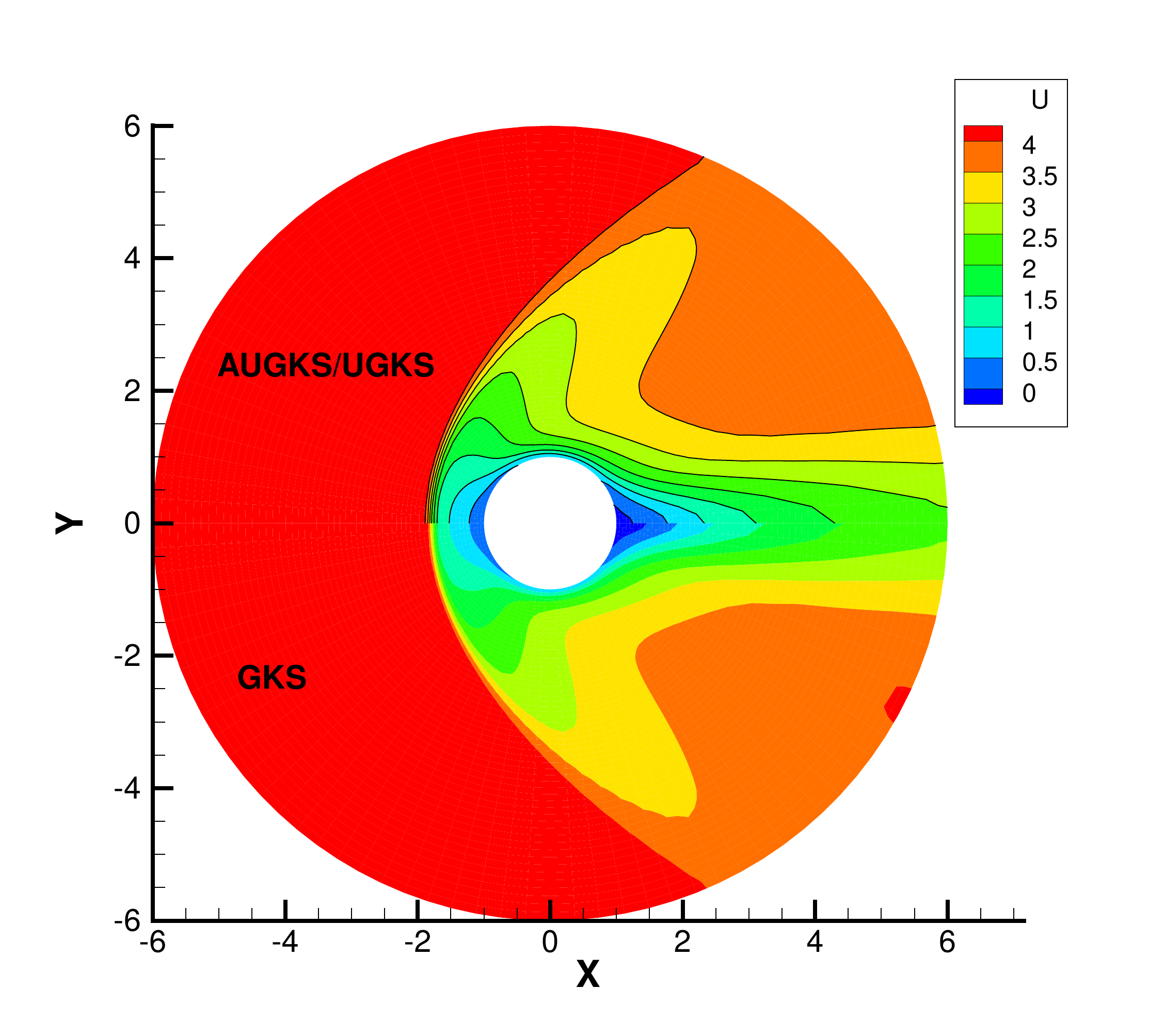}
	}
	\subfigure[Temperature]{
		\includegraphics[width=7.5cm]{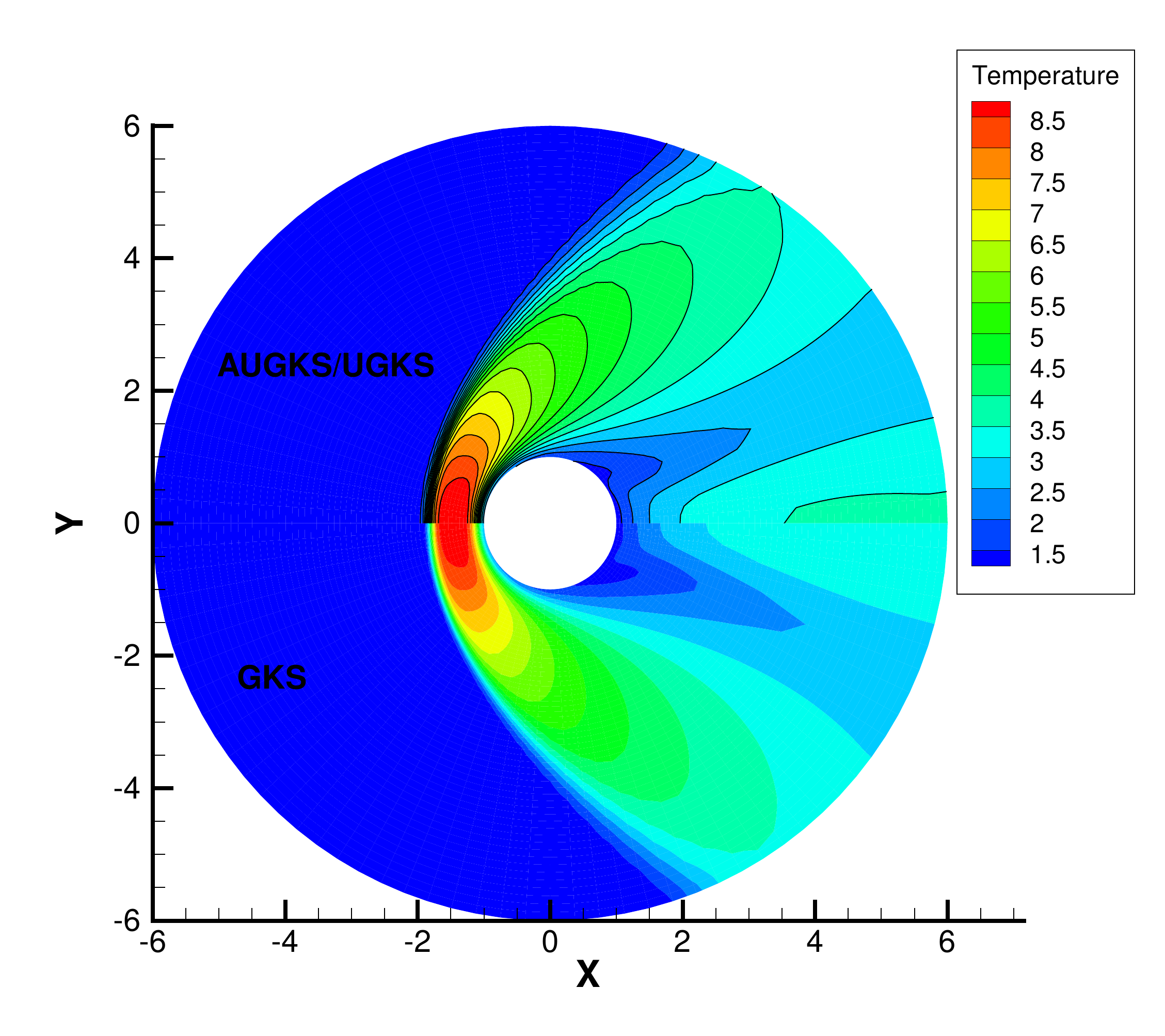}
	}
	\caption{Density and Temperature contours in the flow passing through cylinder at $Kn=0.01$ (upper flood: AUGKS, upper lines: UGKS, lower flood: GKS).}
	\label{pic:cylinder contour2}
\end{figure}

Fig. \ref{pic:cylinder contour1} and Fig. \ref{pic:cylinder contour2} presents the solution contours of $U-$velocity and temperature calculated by the AUGKS, UGKS and GKS methods respectively around the cylinder.
The upper part of contours is the results of AUGKS (flood) and UGKS (solid line) solutions, and the lower part is the GKS solutions.
As shown, the bow shock and expansion cooling region behind shock are well captured by all methods.

\begin{figure}[htb!]
	\centering
	\subfigure[Density]{
		\includegraphics[width=5cm]{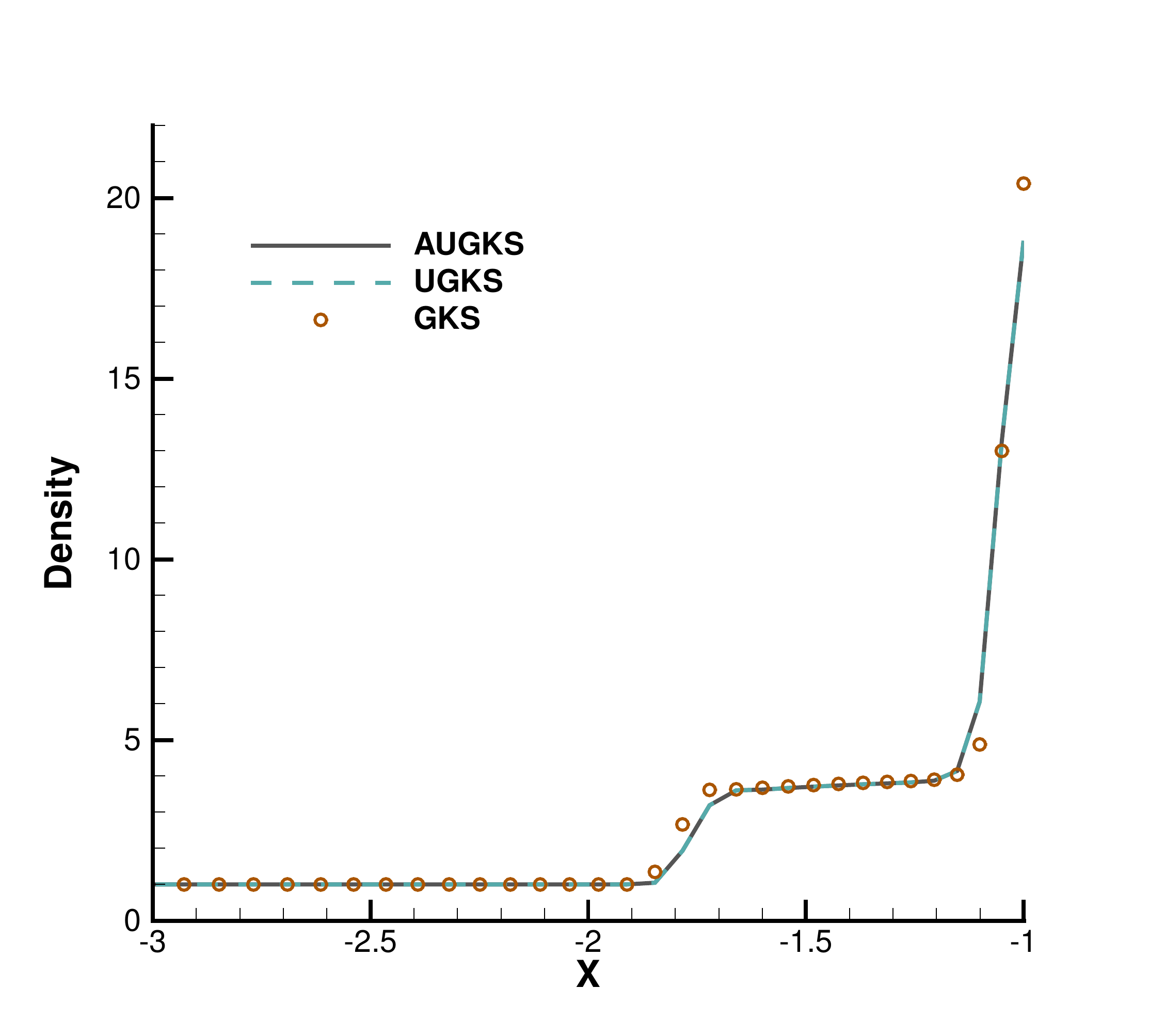}
	}
	\subfigure[U-velocity]{
		\includegraphics[width=5cm]{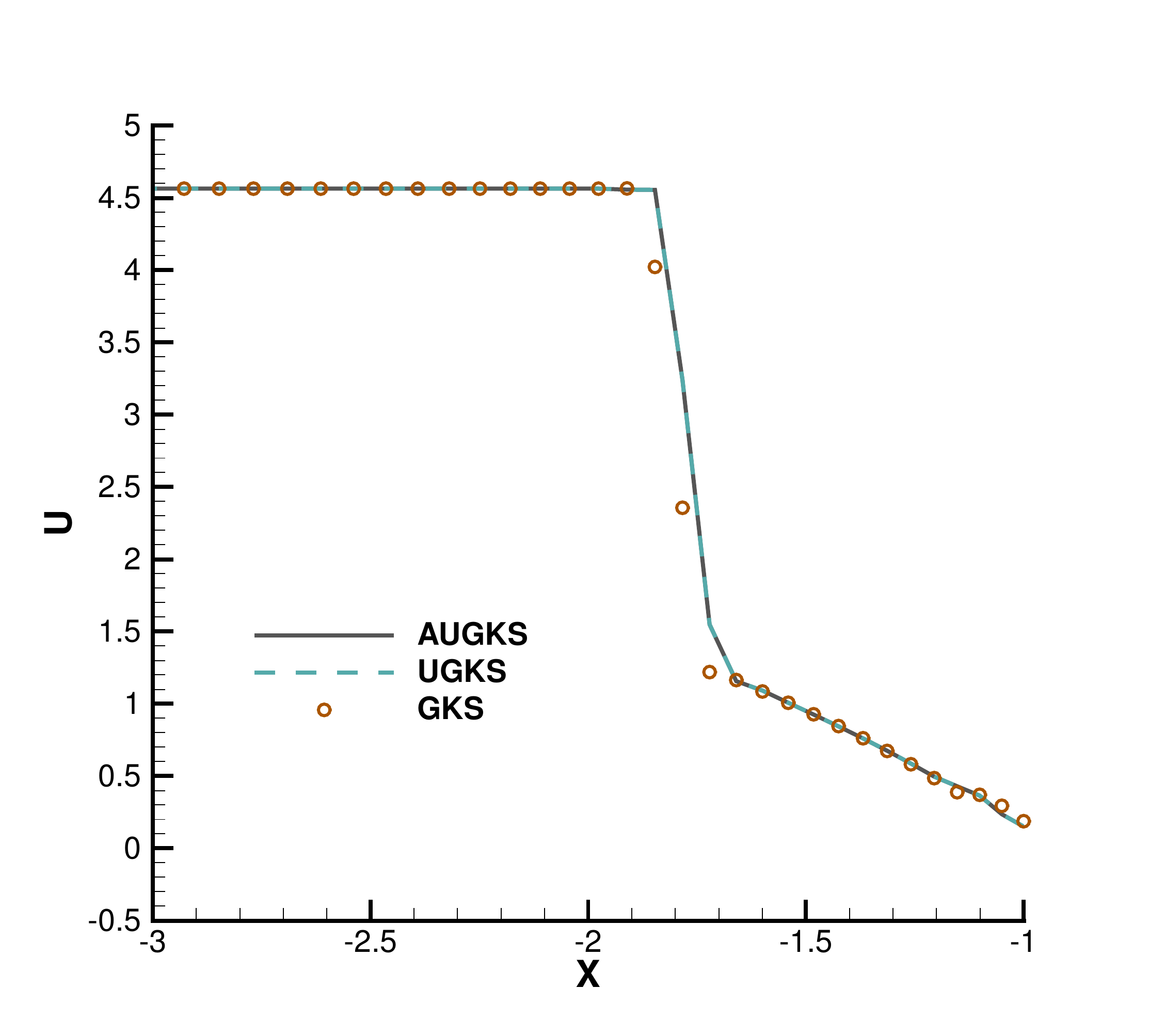}
	}
	\subfigure[Temperature]{
		\includegraphics[width=5cm]{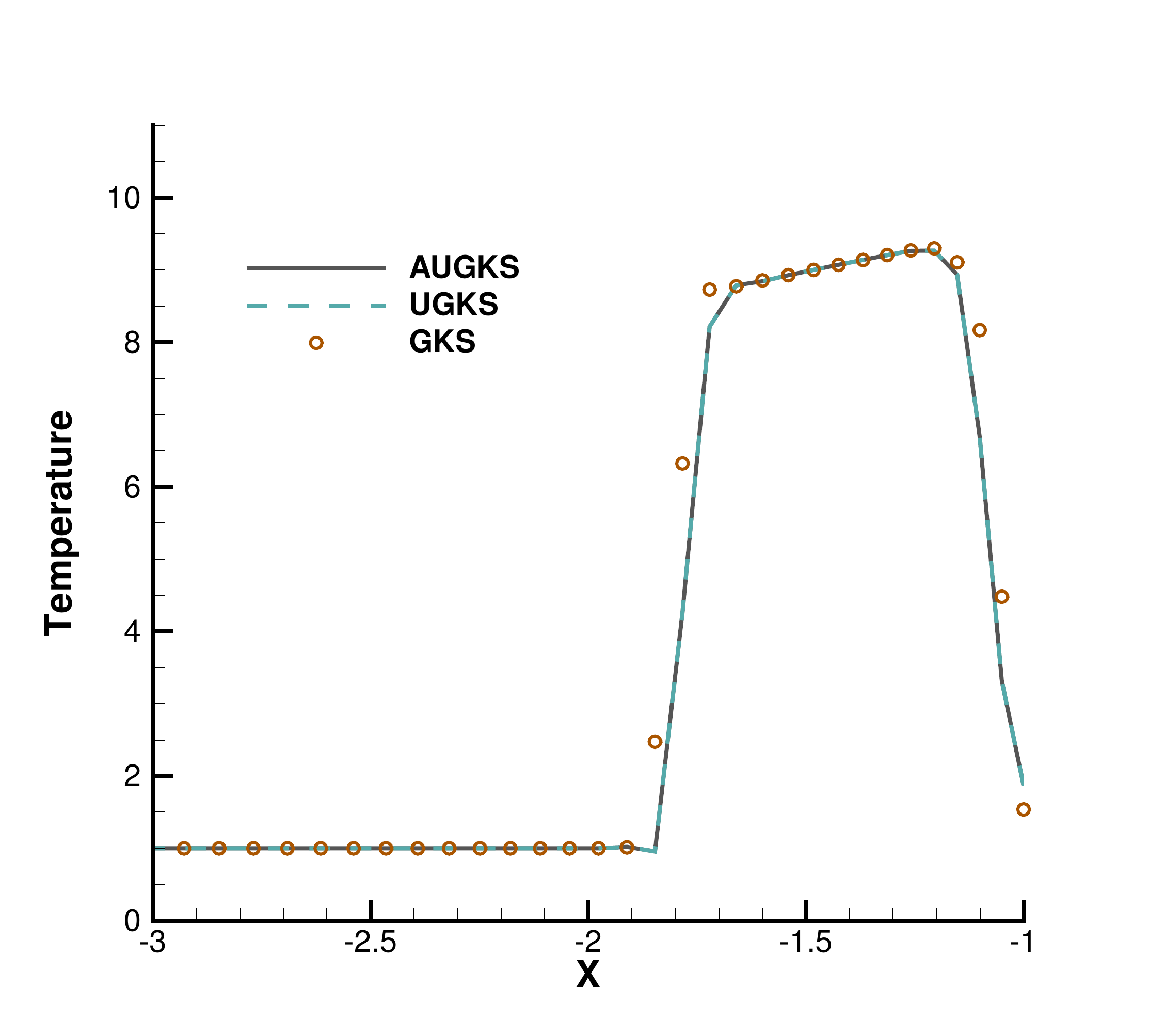}
	}
	\caption{Solutions along the horizontal central line in front of cylinder at $Kn=0.001$.}
	\label{pic:cylinder frontline1}
\end{figure}

\begin{figure}[htb!]
	\centering
	\subfigure[Density]{
		\includegraphics[width=5cm]{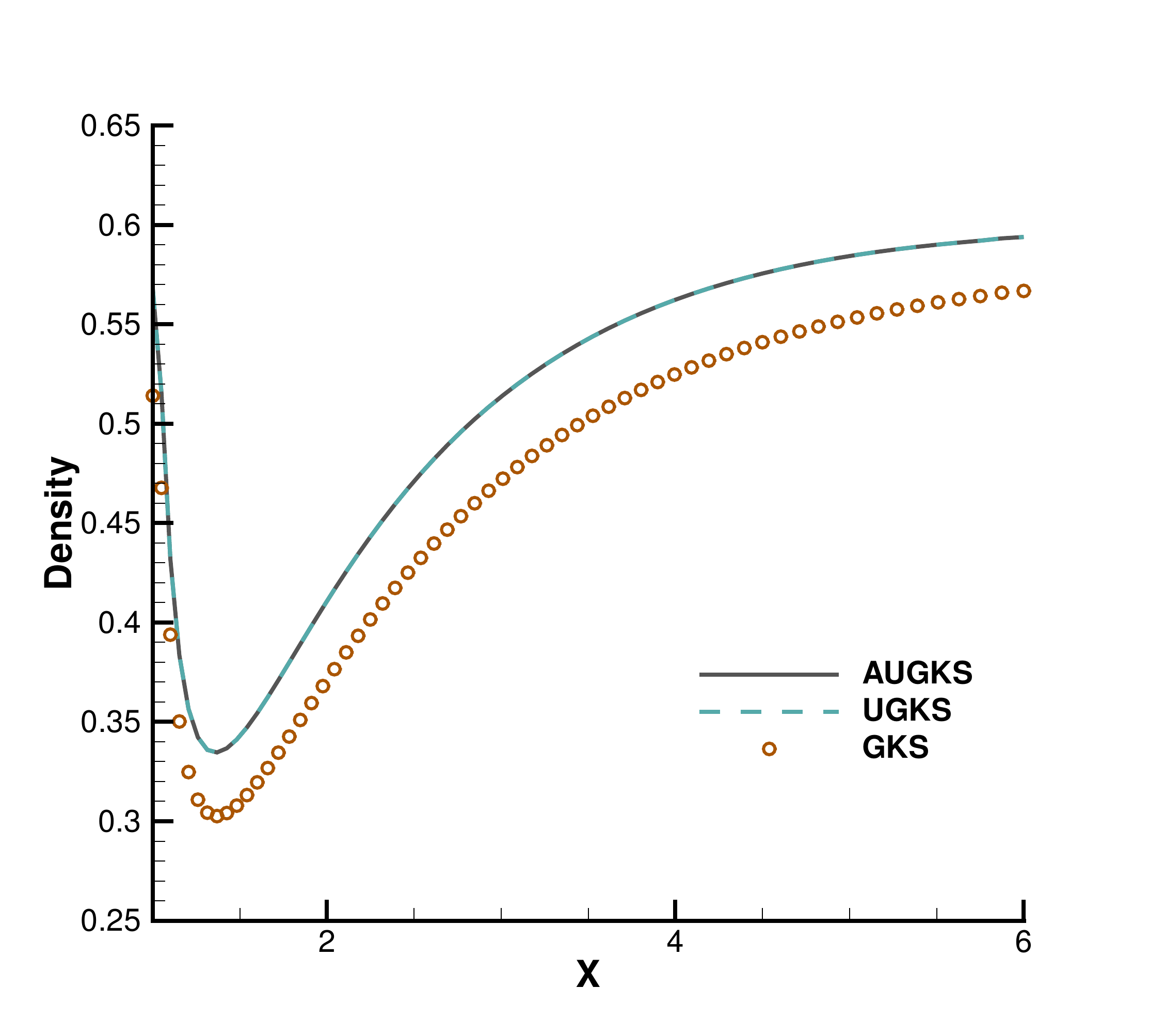}
	}
	\subfigure[U-velocity]{
		\includegraphics[width=5cm]{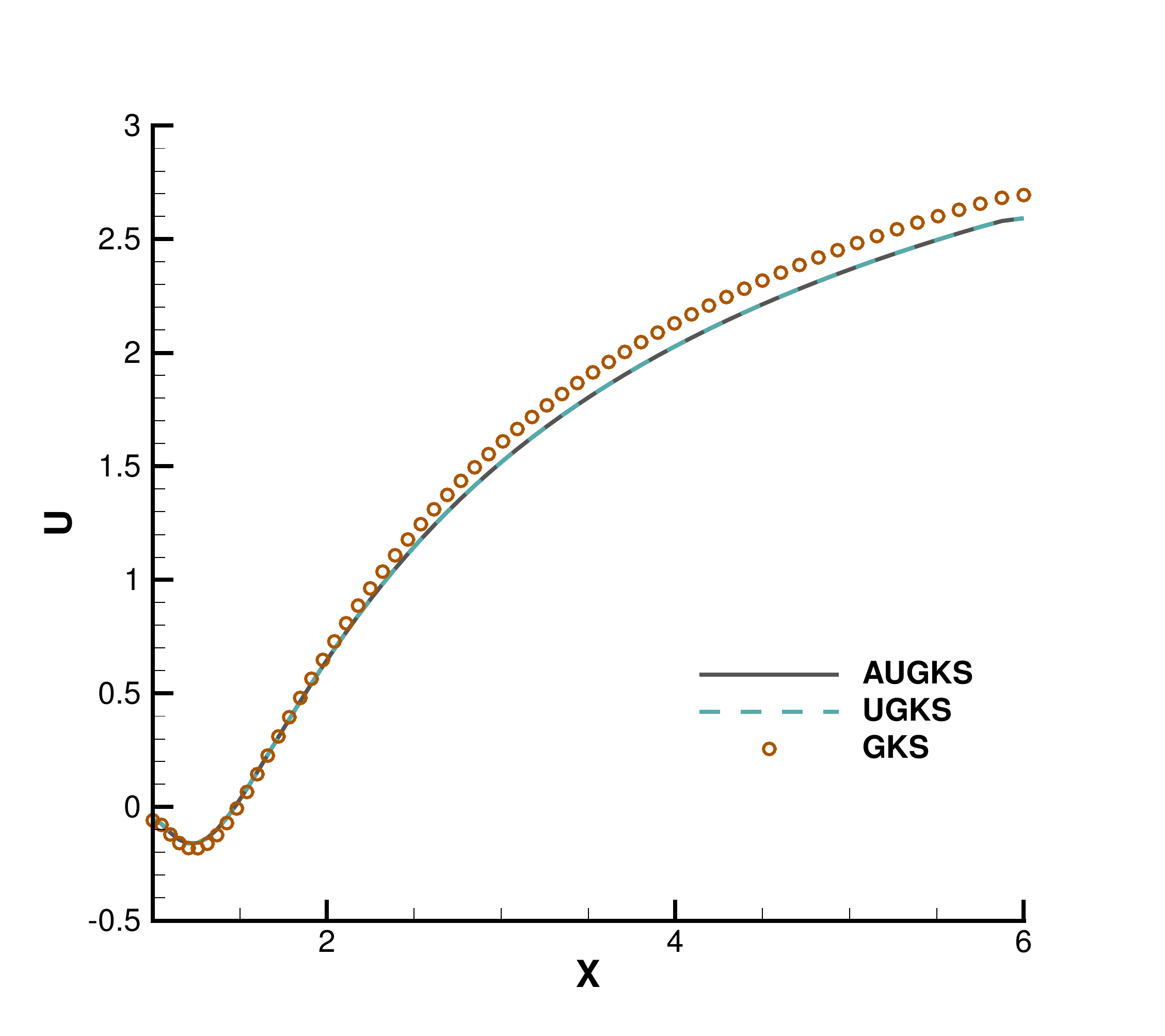}
	}
	\subfigure[Temperature]{
		\includegraphics[width=5cm]{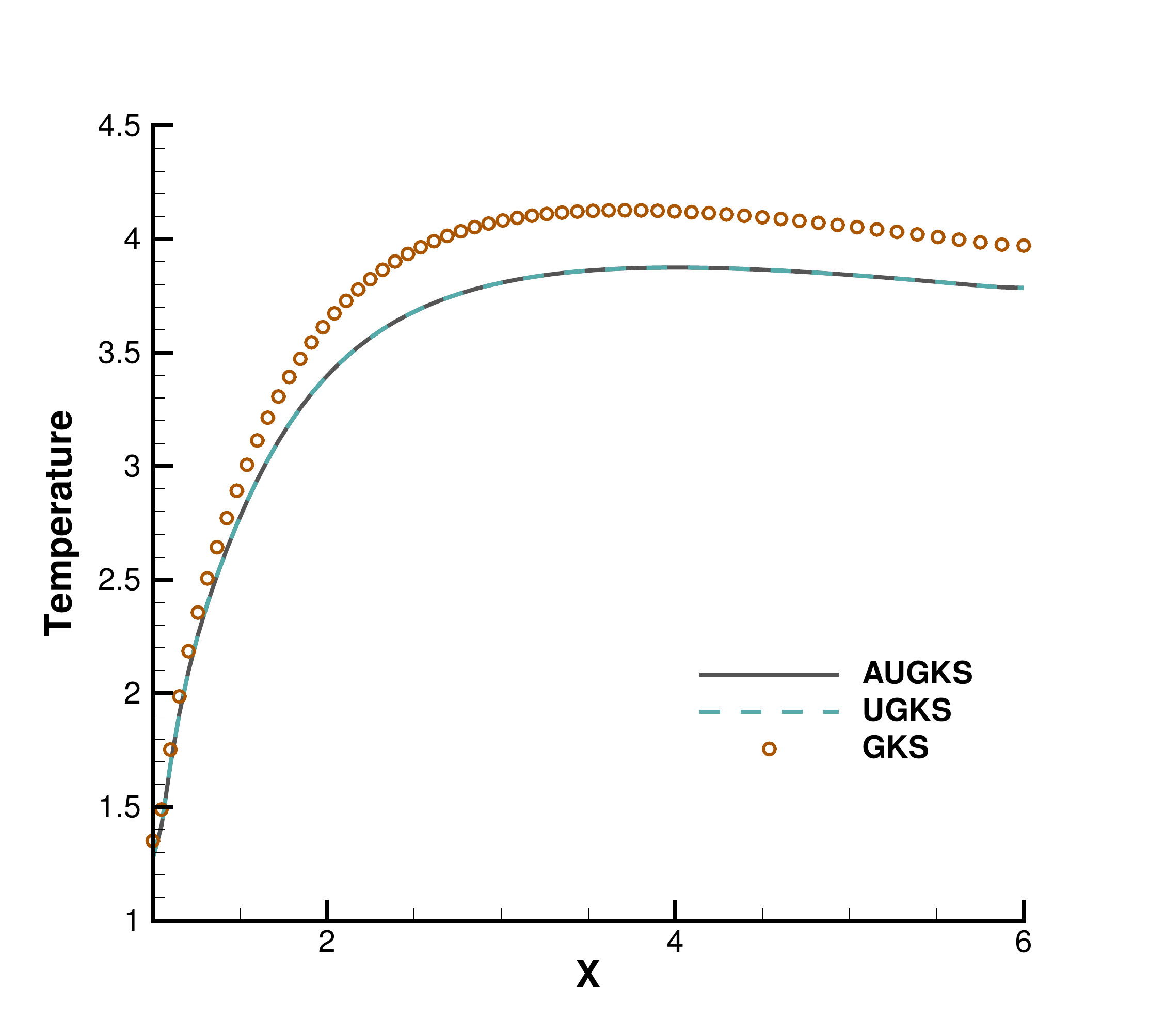}
	}
	\caption{Solutions along the horizontal central line behind cylinder at $Kn=0.001$.}
	\label{pic:cylinder behindline1}
\end{figure}

\begin{figure}[htb!]
	\centering
	\subfigure[Density]{
		\includegraphics[width=5cm]{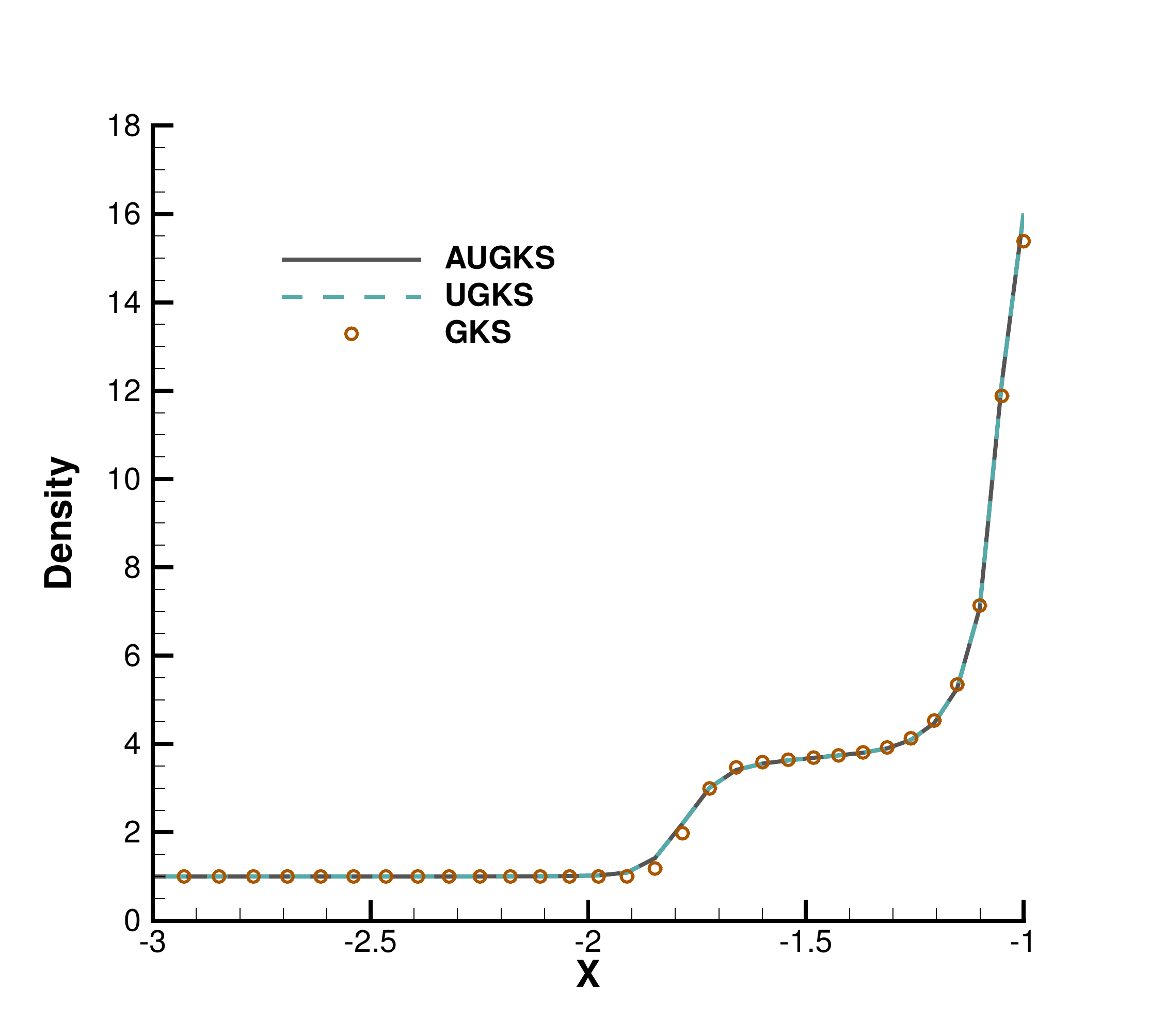}
	}
	\subfigure[U-velocity]{
		\includegraphics[width=5cm]{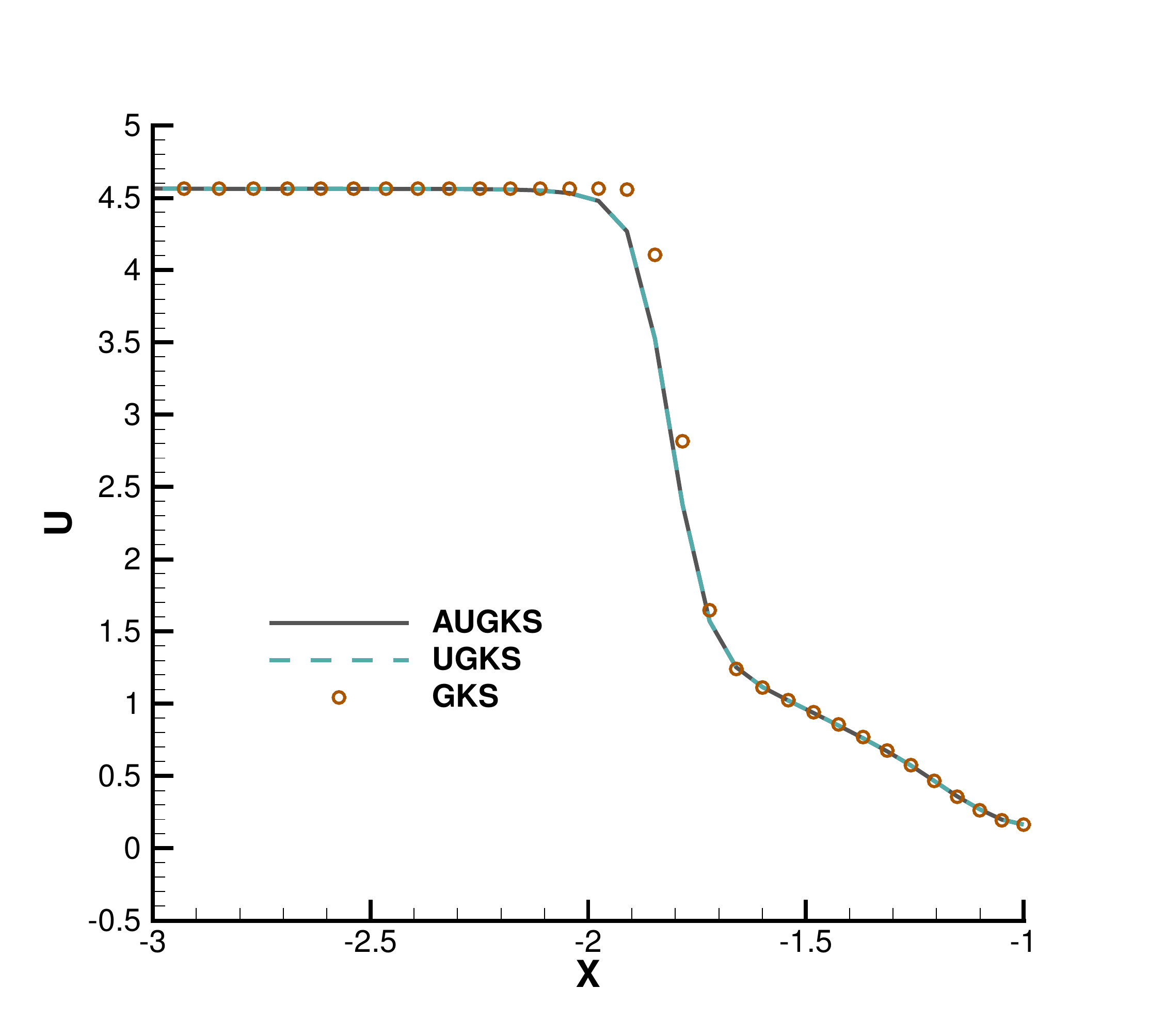}
	}
	\subfigure[Temperature]{
		\includegraphics[width=5cm]{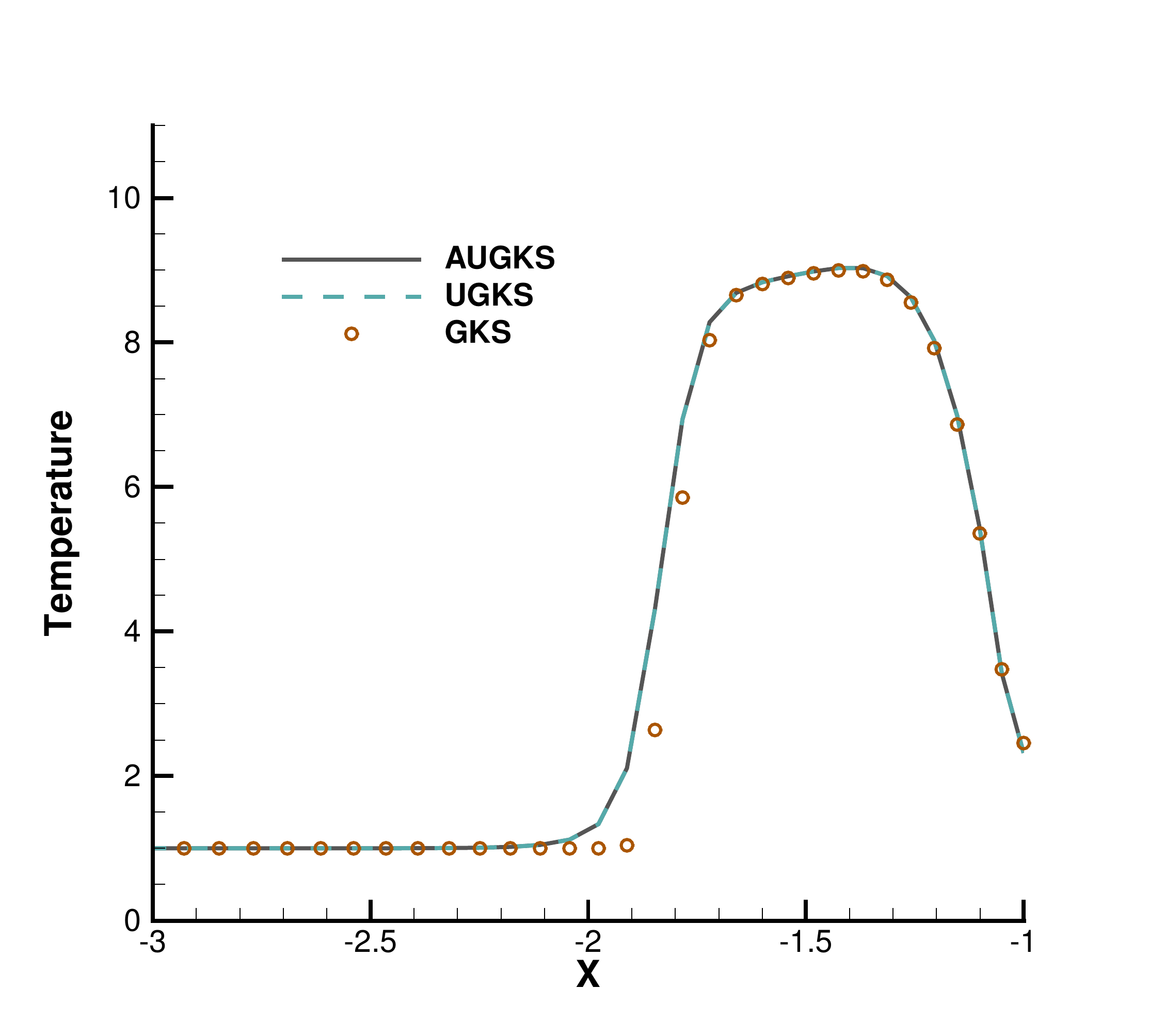}
	}
	\caption{Solutions along the horizontal central line in front of cylinder at $Kn=0.01$.}
	\label{pic:cylinder frontline2}
\end{figure}

\begin{figure}[htb!]
	\centering
	\subfigure[Density]{
		\includegraphics[width=5cm]{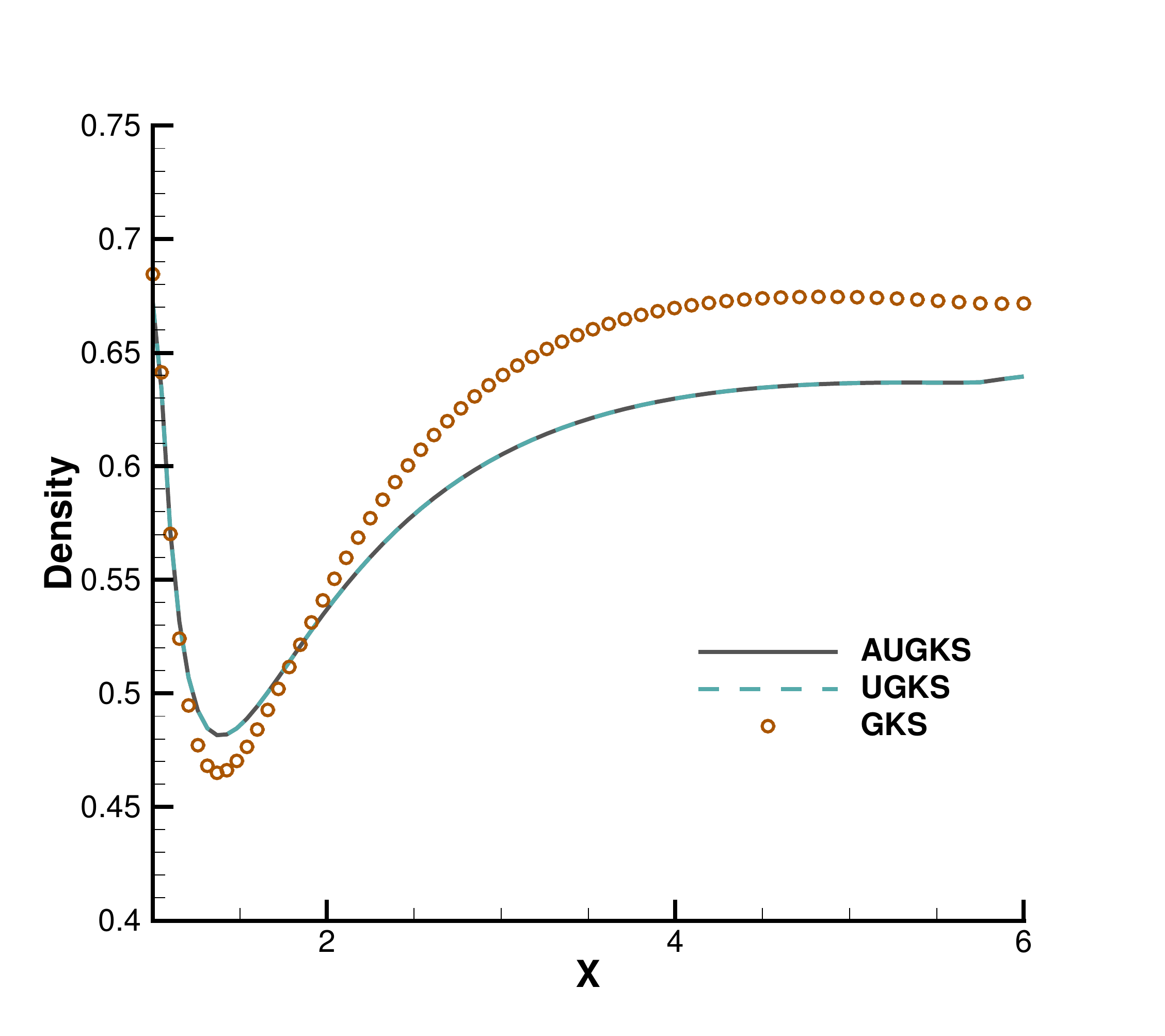}
	}
	\subfigure[U-velocity]{
		\includegraphics[width=5cm]{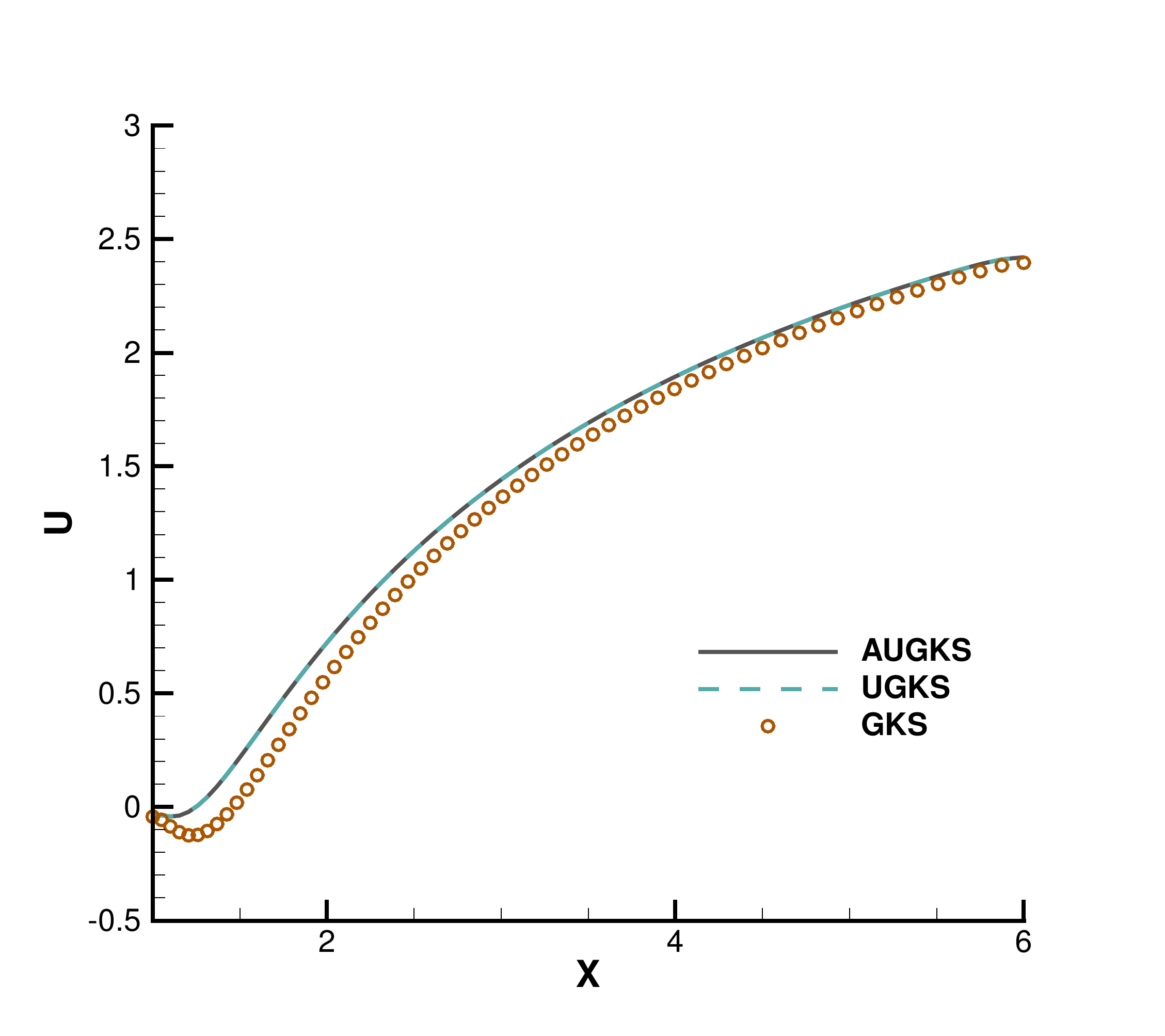}
	}
	\subfigure[Temperature]{
		\includegraphics[width=5cm]{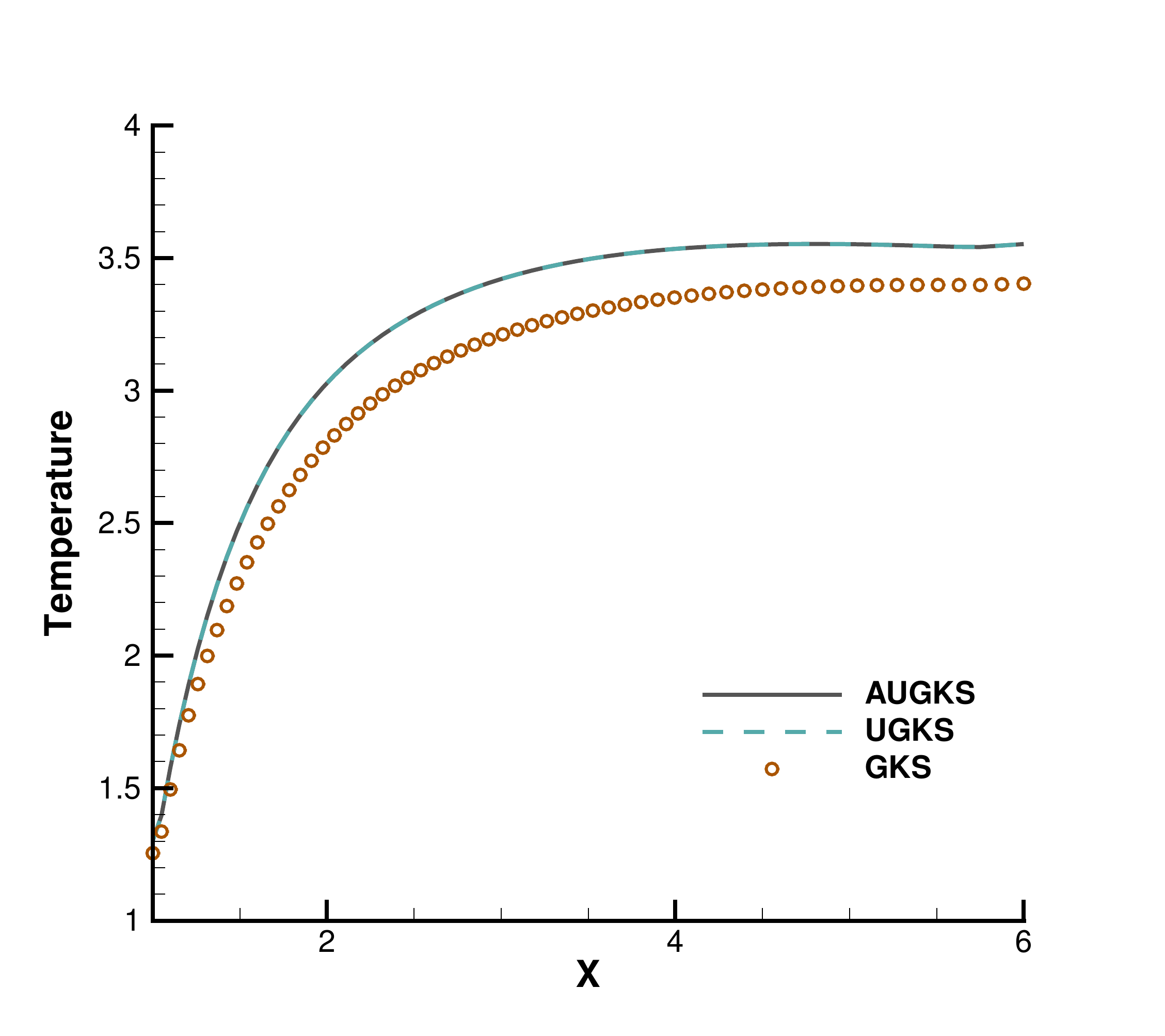}
	}
	\caption{Solutions along the horizontal central line behind cylinder at $Kn=0.01$.}
	\label{pic:cylinder behindline2}
\end{figure}

Fig. \ref{pic:cylinder frontline1}, \ref{pic:cylinder behindline1}, \ref{pic:cylinder frontline2}, \ref{pic:cylinder behindline2} present the solutions along the horizontal center line in front of and behind the cylinder.
At $Kn=0.001$, the cell size and time step in the computation are much larger than particle mean free path and collision time.
Due to the limited time-space resolution, all the three methods become shock-capturing schemes, and a sharp shock profile is obtained in front of the cylinder in Fig. \ref{pic:cylinder frontline1}.
Near the cylinder wall, due to the non-equilibrium gas dynamics in gas-surface interaction, there is a slight difference in the solutions provided by UGKS and GKS.
At the same time, the gas density reduces much in the wake region behind cylinder with emerging rarefied effects, and there is a significant difference between UGKS and GKS solutions in Fig. \ref{pic:cylinder behindline1}.

When the reference Knudsen number gets to $Kn=0.01$, the increasing particle mean free path collision leads to a wider shock structure.
This non-equilibrium evolution is provided in the scale-dependent interface solution used in AUGKS and UGKS.
However, the Chapman-Enskog expansion can only provide incomplete information about this process in continuous GKS solver.
As a result, the GKS presents a narrower shock profile than that in AUGKS and UGKS in Fig. \ref{pic:cylinder frontline2}.
In the wake region, due to the increasing collision time at $Kn=0.01$, the results provided by GKS differ significantly from AUGKS and UGKS solutions in Fig. \ref{pic:cylinder behindline2}.
The GKS with a continuous velocity space fails to predict physical solutions in these regions, and particle distribution function is updated explicitly in the AUGKS with a discretized particle velocity space.
It is clear that the current velocity adaptive unified scheme captures the equivalent physical solutions as the NS in the near-equilibrium region and the UGKS ones in the non-equilibrium region.

\begin{figure}[htb!]
	\centering
	\subfigure[$a_1$]{
		\includegraphics[width=7.5cm]{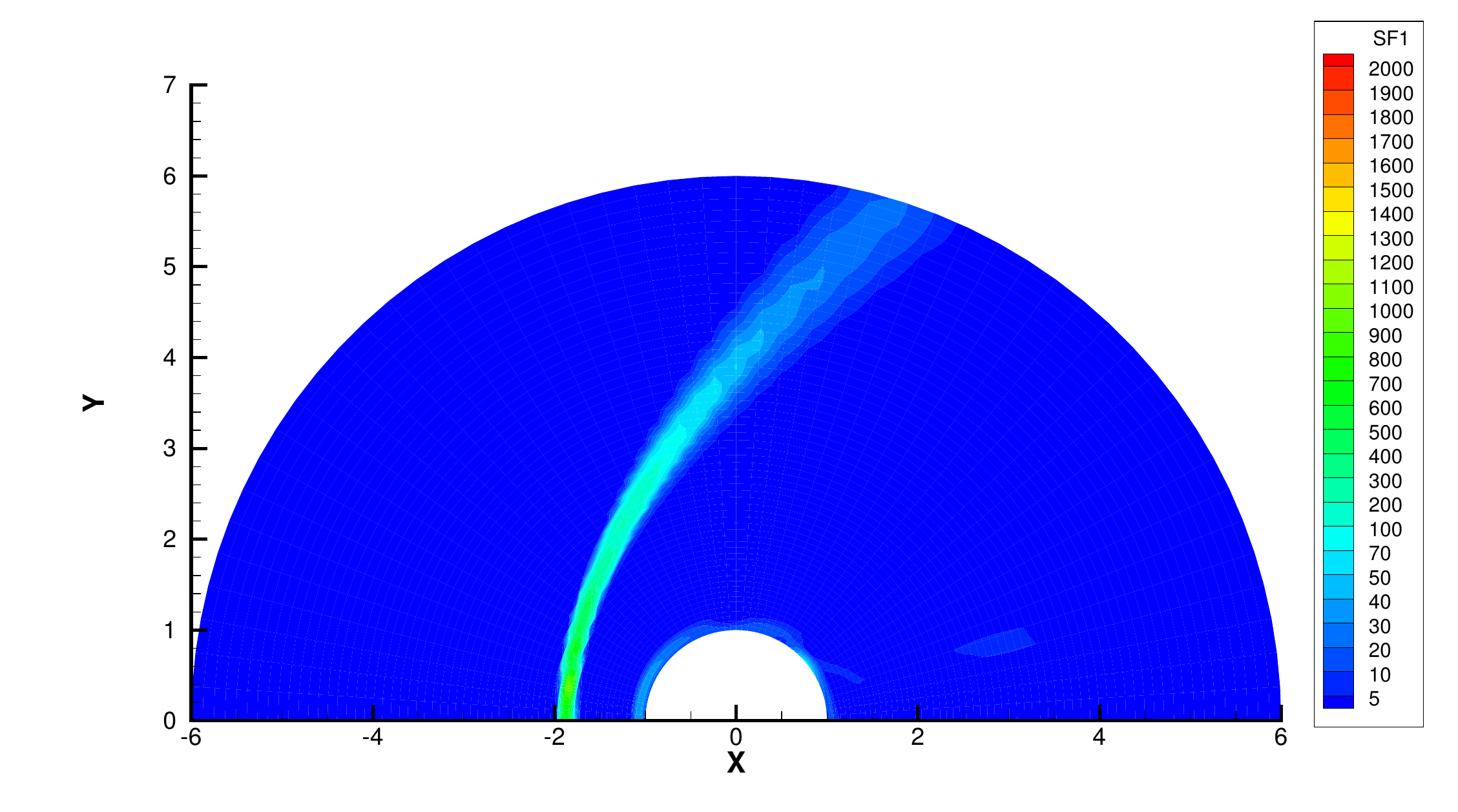}
	}
	\subfigure[$a_3$]{
		\includegraphics[width=7.5cm]{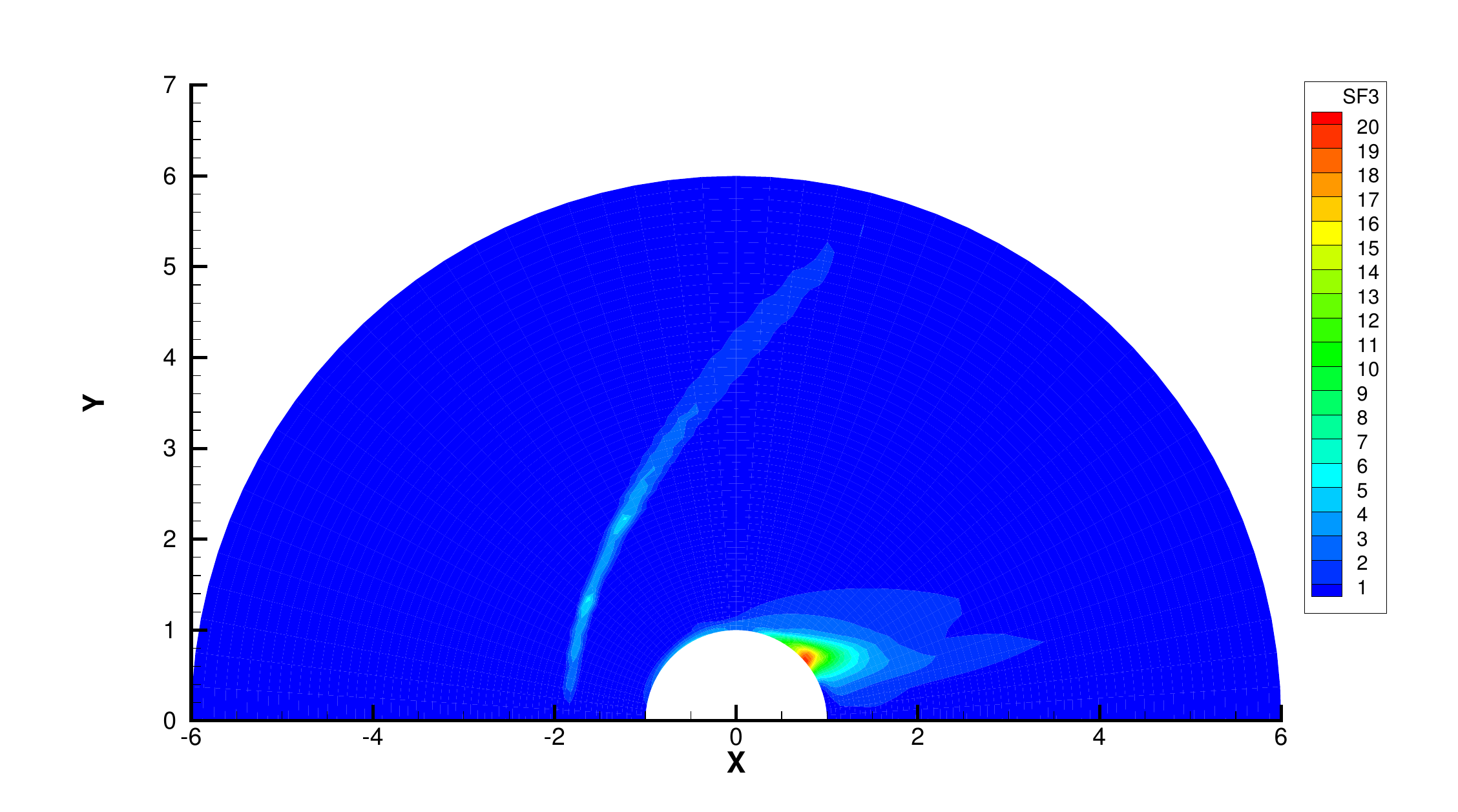}
	}
	\subfigure[Collision time]{
		\includegraphics[width=7.5cm]{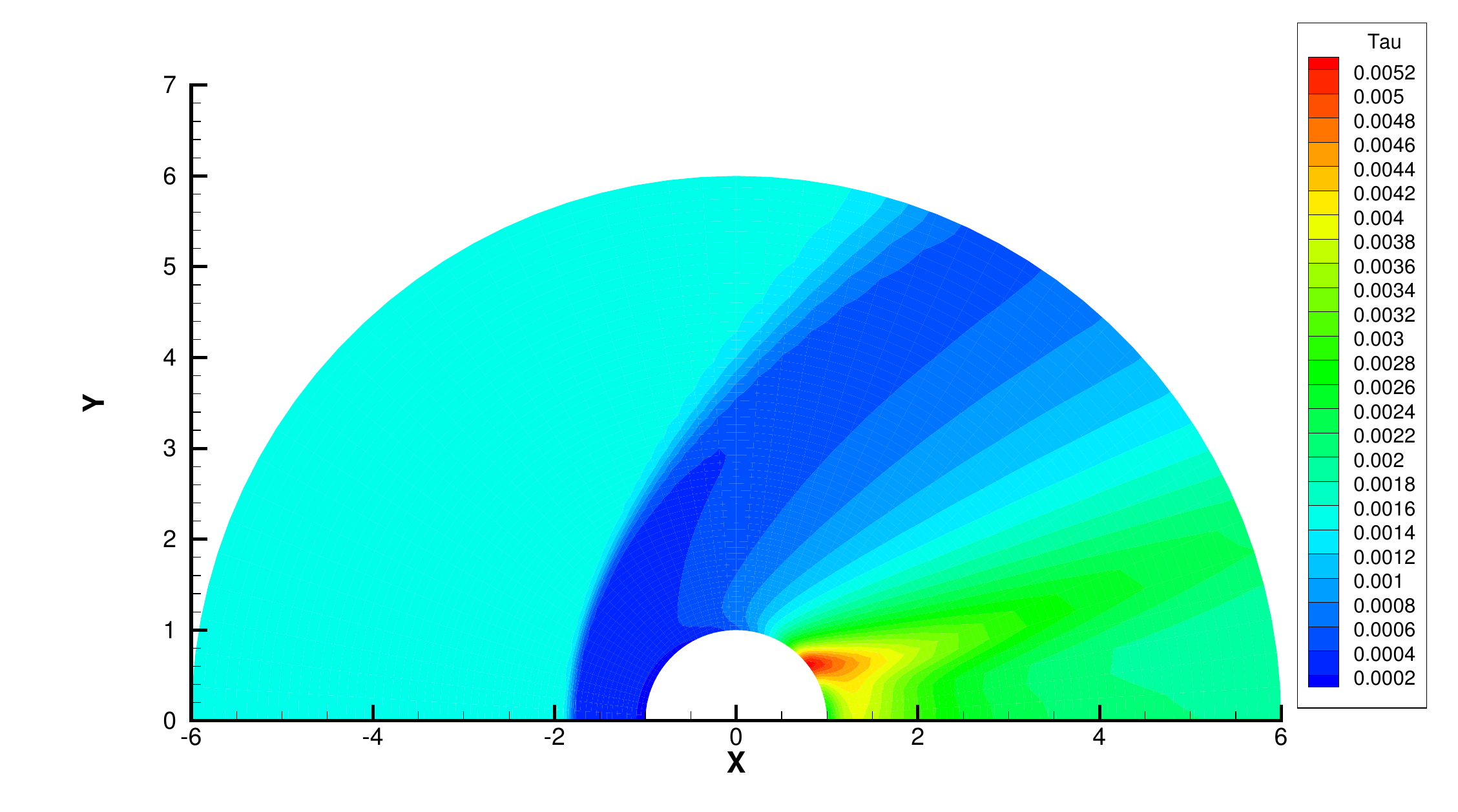}
	}
	\subfigure[Velocity adaptation]{
		\includegraphics[width=7.5cm]{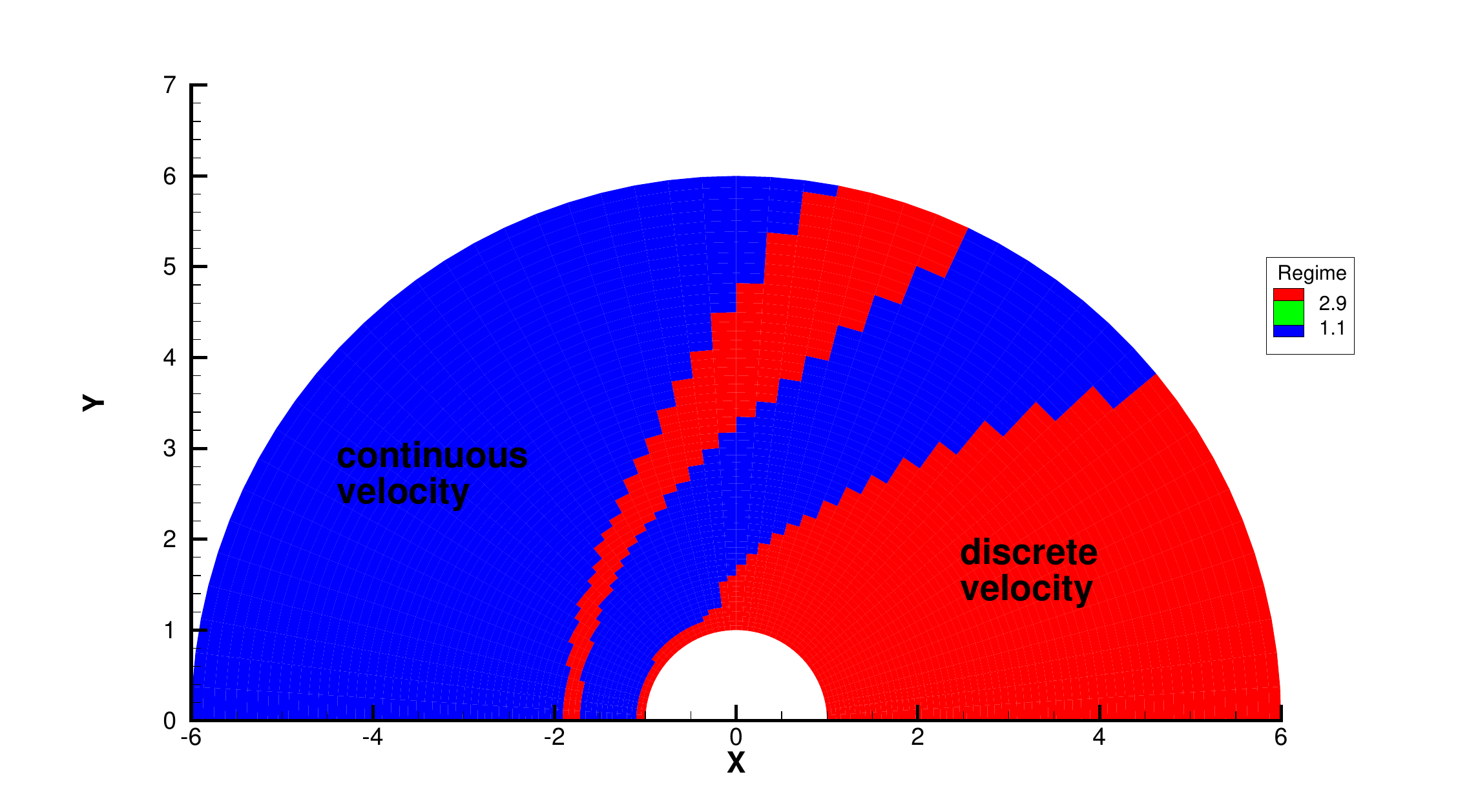}
	}
	\caption{Velocity space adaptation in the flow domain at $Kn=0.001$.}
	\label{pic:cylinder criterion1}
\end{figure}

\begin{figure}[htb!]
	\centering
	\subfigure[$a_1$]{
		\includegraphics[width=7.5cm]{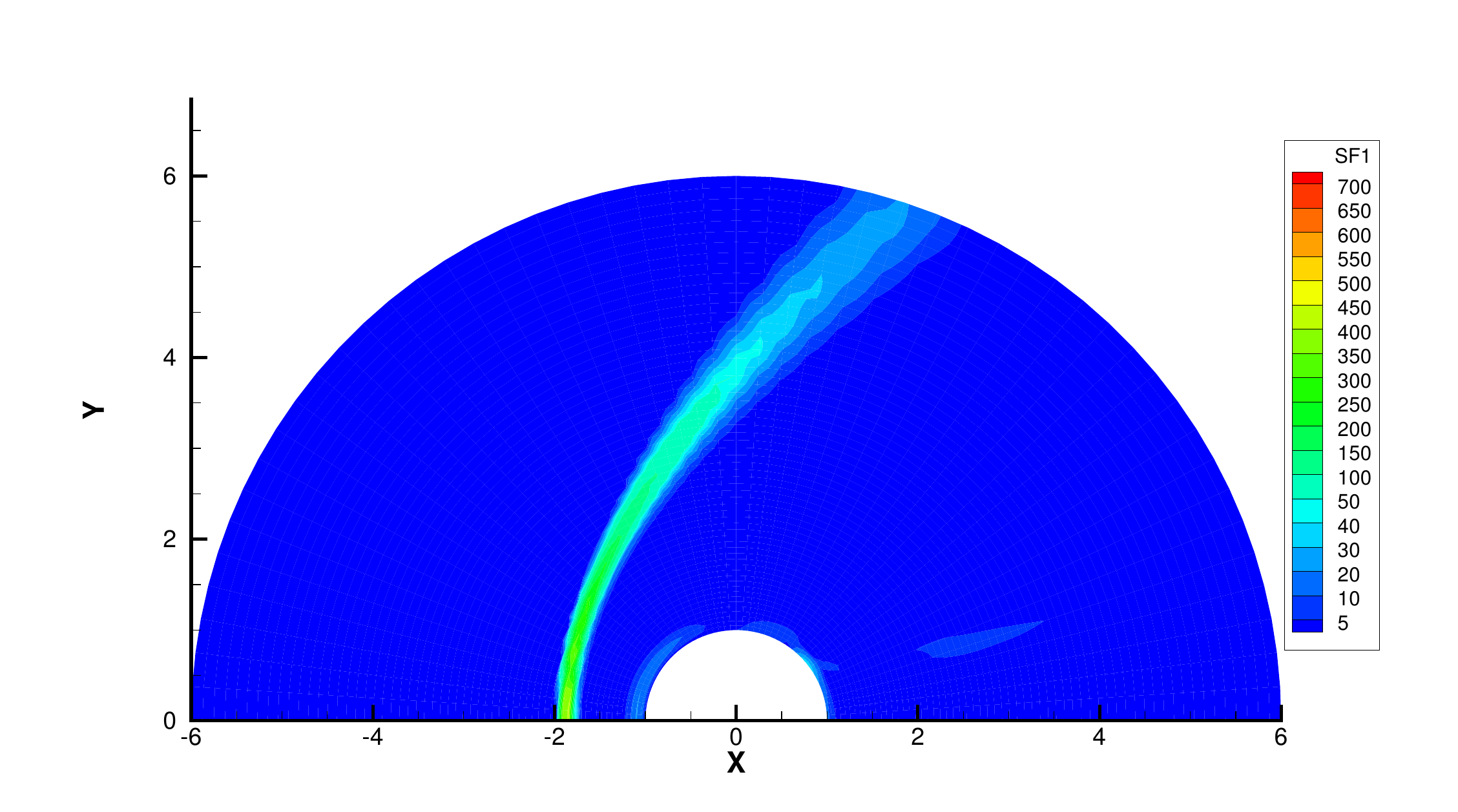}
	}
	\subfigure[$a_3$]{
		\includegraphics[width=7.5cm]{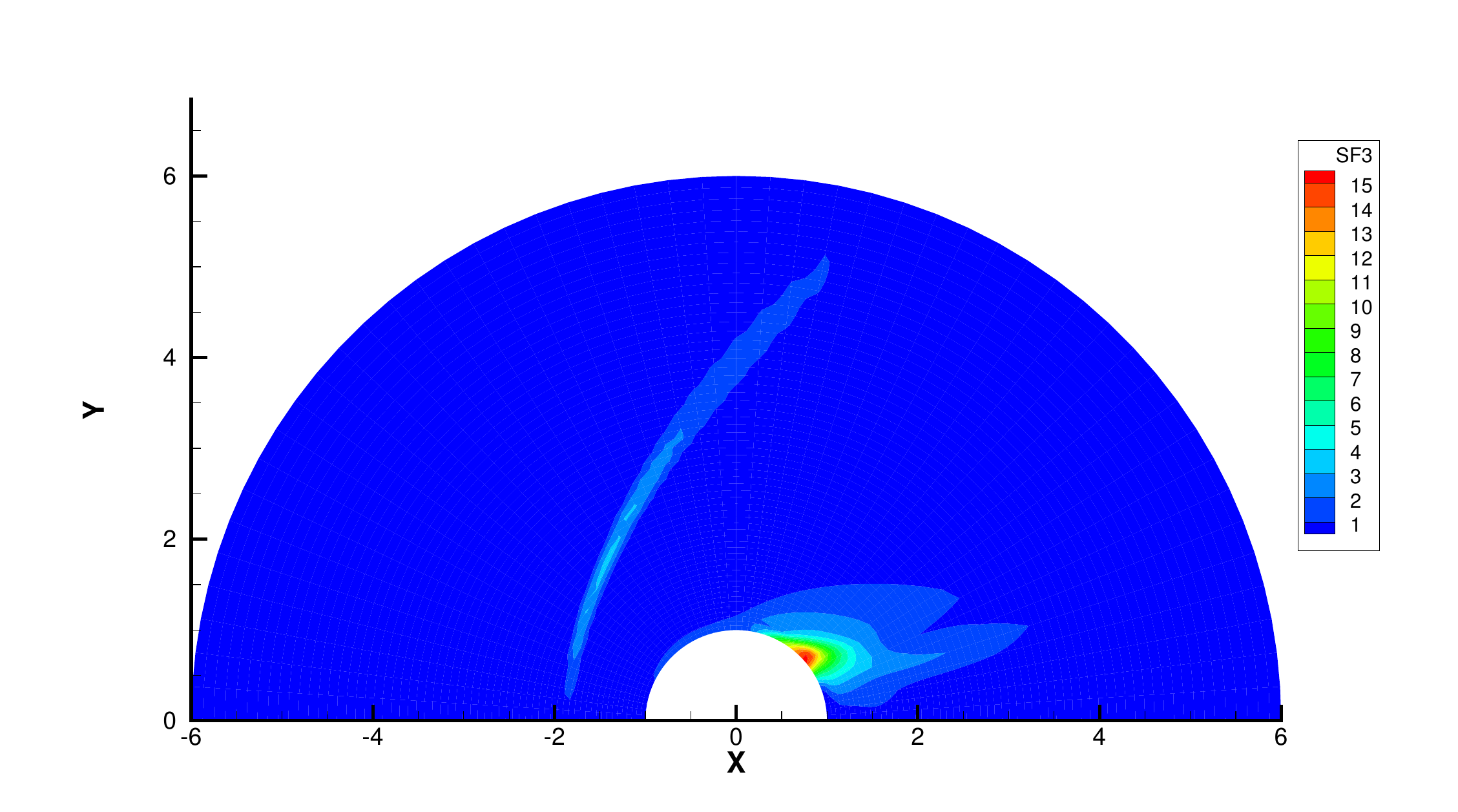}
	}
	\subfigure[Collision time]{
		\includegraphics[width=7.5cm]{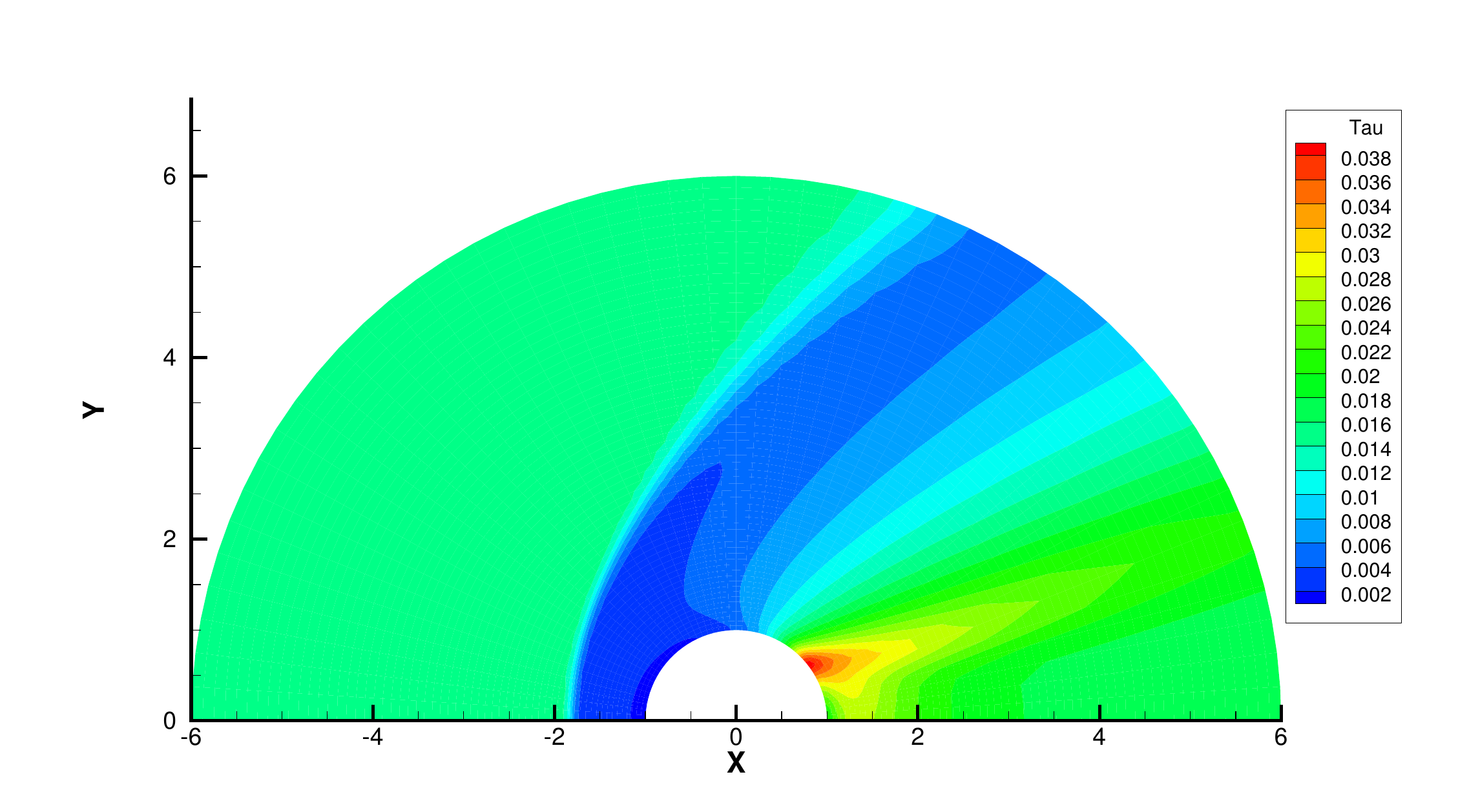}
	}
	\subfigure[Velocity adaptation]{
		\includegraphics[width=7.5cm]{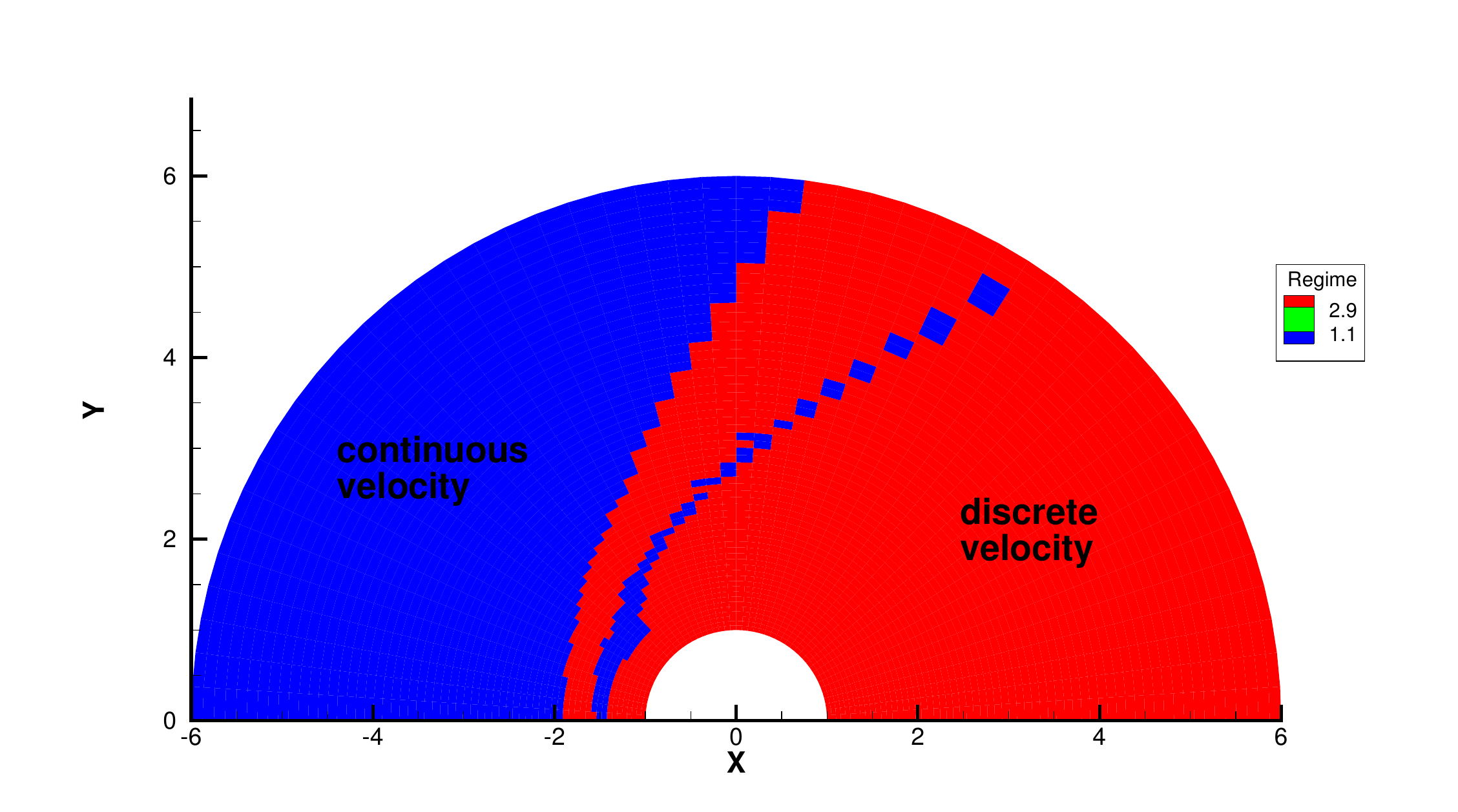}
	}
	\caption{Velocity space adaptation in the flow domain at $Kn=0.01$.}
	\label{pic:cylinder criterion2}
\end{figure}

Fig. \ref{pic:cylinder criterion1} and \ref{pic:cylinder criterion2} presents two components of spatial slope $\mathbf a$ used in the velocity-space switching criterion, the mean collision time and the adaptation of velocity space.
As can be seen, the shock wave and boundary are two sources for high gradients of flow variables, leading to the failure of Chapman-Enskog expansion and Navier-Stokes solutions.
Behind the cylinder, the low density wake leads to an increasing collision time, which is shown in Fig. \ref{pic:cylinder criterion1}c and Fig. \ref{pic:cylinder criterion2}c.
Therefore, a velocity adaptation is determined as shown in Fig. \ref{pic:cylinder criterion1}d and Fig. \ref{pic:cylinder criterion2}d .
The incoming flow as well as a small region between the bow shock and cylinder is computed with continuous velocity space, while the rest non-equilibrium region are simulated by the UGKS with discrete velocity space.
Due to increasing collision time and non-equilibrium flow dynamics, the region under discrete particle velocity space enlarges at $Kn=0.01$ than that at $Kn=0.01$.

Table 4 and 5 presents the computational cost of AUGKS, UGKS and GKS at $Kn=0.001$ and $Kn=0.01$ respectively.
With the current setup of physical mesh and velocity space, the continuous GKS solver is about 30 times faster than the UGKS,
and the AUGKS is about $3.3$ times faster than the original UGKS in this steady flow problem.
In the convergent steady state, there are about 3196 at $Kn=0.001$ and 1992 at $Kn=0.01$ out of 6000 total cells using continuous velocity space, and the corresponding memory burden is about $47\%$ and $67\%$ of the original UGKS.

\begin{table}  
	\label{table:cylinder1}
	\centering
	\begin{tabular*}{10cm}{llll}  
		\hline  
		& AUGKS  & UGKS & GKS \\
		\hline  
		CPU time (s) &36130.68  & 117371.67 & 2975.07 \\  
		Memory (KB)   &452508  & 857520 & 14652 \\  
		\hline 
	\end{tabular*}  
	\caption{CPU time and memory cost in the flow around circular cylinder at $Kn=0.001$.}  
\end{table}

\begin{table}  
	\label{table:cylinder2}
	\centering
	\begin{tabular*}{10cm}{llll}  
		\hline  
		& AUGKS  & UGKS & GKS \\
		\hline  
		CPU time (s) &22145.10  & 75510.33 & 2536.55 \\  
		Memory (KB)   &614542  & 856944 & 12636 \\  
		\hline 
	\end{tabular*}  
	\caption{CPU time and memory cost in the flow around circular cylinder at $Kn=0.01$.}  
\end{table}

\section{Conclusion}

The gas dynamics has intrinsic multiple scale nature due to the large variations of density and characteristic length scale of the flow structures. 
Based on scale-dependent time evolving solution of the Boltzmann model equation, a velocity-space adaptive unified gas kinetic scheme has been developed in this paper for the simulation of multiscale flow transport.
The current adaptive algorithm is based on a dynamic velocity-space transformation, where the particle velocity space is continuous in near-equilibrium region and discrete in non-equilibrium one.
A switching criterion for particle velocity space transformation is proposed based on the Chapman-Enskog expansion and then is validated through numerical experiments.
With a unified framework with the adaptation of particle velocity space only, the AUGKS needs no buffer zone for the connection between continuum and kinetic solutions.
This compact property leads to an effective method for multiscale flow simulation with unsteadiness and complex geometries.
Compared with single-velocity-space framework of the original UGKS, the AUGKS is more efficient and less memory demanding for multiscale flow computations.
The AUGKS provides a useful tool for non-equilibrium flow studies, and it can be further improved in the future with the combination of implicit and multigrid techniques \cite{zhu2016implicit,zhu2017unified}.

\section*{Acknowledgement}
The authors would like to thank Dr. Chang Liu and Dr. Lei Wu for the help on numerical implementation of the fast spectral method for the Boltzmann collision term.
The current research is supported by Hong Kong research grant council (16207715, 16211014, 16206617), and  National Science Foundation of China (11772281,91530319).

\bibliographystyle{unsrt}
\bibliography{tbxiao}

\end{document}